\documentclass[acmsmall]{acmart}

\AtBeginDocument{%
  \providecommand\BibTeX{{%
    \normalfont B\kern-0.5em{\scshape i\kern-0.25em b}\kern-0.8em\TeX}}}
\pdfoutput=1
\setcopyright{rightsretained}
\acmJournal{TKDD}
\acmYear{2021} \acmVolume{1} \acmNumber{1} \acmArticle{1} \acmMonth{1} \acmPrice{}\acmDOI{10.1145/3449361}

\input{preamble}
\begin{document}

\title[Demarcating Endogenous and Exogenous Opinions]{Demarcating Endogenous and Exogenous Opinion Dynamics: An Experimental Design Approach}

%\email{webmaster@marysville-ohio.com}
%\affiliation{%
%  \institution{Institute for Clarity in Documentation}
%  \streetaddress{P.O. Box 1212}

\author{Paramita Koley}
\email{paramita.koley@iitkgp.ac.in}
\affiliation{%
	\institution{Department of Computer Science and Engineering, IIT Kharagpur}
	\city{Kharagpur}
	\country{India}
	\postcode{721302}}

\author{Avirup Saha}
\affiliation{%
	\institution{Department of Computer Science and Engineering, IIT Kharagpur}
	\city{Kharagpur}
	\country{India}
	\postcode{721302}}
\email{avirupsaha@iitkgp.ac.in}

\author{Sourangshu Bhattacharya}
\affiliation{%
	\institution{Department of Computer Science and Engineering, IIT Kharagpur}
	\city{Kharagpur}
	\country{India}
	\postcode{721302}}
\email{sourangshu@cse.iitkgp.ac.in}

\author{Niloy Ganguly}
\affiliation{%
	\institution{Department of Computer Science and Engineering, IIT Kharagpur}
	\city{Kharagpur}
	\country{India}
	\postcode{721302}}
\email{niloy@cse.iitkgp.ac.in}

\author{Abir De}
\affiliation{%
	\institution{Department of Computer Science and Engineering, IIT Bombay}
	\city{Mumbai}
	\country{India}
	\postcode{400076}}
\email{abir@cse.iitb.ac.in}

\renewcommand{\shortauthors}{Koley, et al.}


\begin{abstract}
The networked opinion diffusion in online social networks (OSN) is often governed by the 
two genres of opinions--- \emph{endogenous} opinions that are driven by the influence of social contacts among
users, and \emph{exogenous} opinions which are formed by external effects like news, feeds etc.
Accurate demarcation of endogenous and exogenous messages offers an important cue to opinion
modeling, thereby enhancing its predictive performance.
In this paper, we design a suite of unsupervised classification methods
based on experimental design approaches, in which, we aim to select
the subsets of events which minimize different measures of mean estimation error. 
In more detail, we first show that these subset selection tasks are NP-Hard. Then we show that the
associated objective functions are weakly submodular, which allows us to cast efficient approximation algorithms
with guarantees. 
Finally, we validate the efficacy of our proposal  on various real-world datasets crawled from Twitter as well as diverse synthetic datasets. Our experiments range from validating prediction performance on unsanitized and sanitized events to checking the effect of selecting optimal subsets of various sizes. Through various experiments, we have found that our method offers a significant improvement in accuracy in terms of opinion forecasting, against several competitors. 

\end{abstract}



%\if{0}
\maketitle
\section{Introduction}
\label{sec:intro}

Research on understanding opinion dynamics, from both modeling and control perspectives, abounds in literature~\cite{clifford1973model,degroot1974reaching,c1,holme2006nonequilibrium,c6,c2,yildiz2010voting,c4,das2014modeling,de2014learning,c3,c5,c7,c8,nipsxx}, predominantly following two approaches. 
The first approach  is grounded on the concepts of statistical physics, it is barely data-driven and therefore shows poor predictive performance~\cite{clifford1973model,degroot1974reaching,c1,holme2006nonequilibrium,c6,c2,yildiz2010voting,c4,c3,c5,c7,c8},
The second class of models aims to overcome such limitations, by learning a tractable linear model from transient opinion dynamics~\cite{de2014learning,das2014modeling,nipsxx}.
%
% }
Barring the individual limitations of these existing approaches, they all have 
assumed the absence or lack of external effects, despite 
empirical evidences advocating the presence of such signals~\cite{icwsm,ex1,ex2,ex3,ex4}. 
{Since a social network is an open system encouraging both inward and outward flow of information, a continuous flux of external information is funneled to its users, via a gamut of sources like news, feeds, etc.} 
As a result, 
a networked opinion formation process that involves extensive interactive discussions among
 connected users, is also propelled by such external sources recommended to those users. 
{Therefore, at the very outset, we observe two families of opinions -- \emph{endogenous} opinions which evolve due to the influence from neighbors, and \emph{exogenous} opinions that are driven mostly by the externalities.
In most practical situations, the true labels of the posts (endogenous or exogenous) are not available.
{Therefore, an accurate unsupervised labeling of the posts has immense potential impact on opinion modeling --
boosting the predictive performance for a broad spectrum of applications like pole-prediction, brand sentiment estimation, etc.}
In this paper, our goal is to demarcate endogenous and exogenous messages and demonstrate the utility of our proposal from an opinion modeling viewpoint.

\subsection{Proposed approach} 
We initiate  investigating  the dynamics of organic opinion in the presence of exogenous actions, using a recent temporal point-process based model \ourm~\cite{nipsxx}. It allows users' latent \emph{endogenous} opinions to be modulated over time, by \emph{both endogenous and exogenous} opinions of their neighbours, expressed as sentiment messages (Section \ref{sec:model}).

Subsequently, in Section \ref{sec:learning} we propose \our\footnote{This paper is an extension of \cite{de2018demarcating} where the idea of \our was first introduced. However, it has been substantially refined and expanded in this paper.}, a suite of learning algorithms, based on experimental design methods that optimally demarcate the endogenous and exogenous opinions under various circumstances. 
In a nutshell,  in order to categorize messages, we aim to
select the set of events that comply with the organic dynamics with a high confidence, \ie\ a low variance of influence estimation.
To this end, we devise this problem as an inference task of the message category (endogenous or exogenous) by means of subset selection (\ie\ demarcating a subset of endogenous messages from the whole message stream).  
We find that this proposed inference problem can be formulated as an instance of a cardinality constrained 
submodular maximization problem. To solve this optimization problem, we present a 
greedy approach which, like an ordinary greedy submodular maximization algorithm, enjoys approximation bounds. However, since some of the optimization objectives we consider are only weakly submodular, they admit some special approximation bounds which have been proposed recently in \cite{hashemi2019submodular}. 

Finally, we perform experiments on various real datasets, crawled from Twitter about diverse topics (Section \ref{sec:expt}) and synthetic datasets, built over diverse networks (Section \ref{sec:expt_syn}) and show that \our can accurately classify  endogenous and exogenous messages, thereby helping to achieve a substantial performance boost in forecasting opinions. 

%\vspace*{-0.4cm}
\section{Related Work}
Opinion modeling and its applications have been widely studied in different guises
in many years. In this section, we review some of them, from two major perspectives--
(i) opinion dynamics modeling and (ii) opinion sensing. 

\xhdr{Opinion dynamics modeling}
Modeling the evolution process of opinion flow over networks, mostly follows
two approaches, based on (a) statistical physics and (b) data-driven techniques. The first type of models, e.g. Voter, Flocking, DeGroot, etc. 
is traditionally designed to capture  various regulatory real-life phenomena e.g. consensus, polarization, clustering, coexistence etc.~\cite{clifford1973model,yildiz2013binary,nvm,yildiz2010voting,
cod,bil,medeiros2006domain,mobilia2003does,anvoter,durrett1996spatial,degroot1974reaching,hegselmann2002opinion,bc1,bc2,bc3}. 
Voter model~\cite{clifford1973model} is a discrete opinion model, where opinions
are represented as nominal values, and copied from influencing
neighbors in every step. This underlying principle
is still a major workhorse for many discrete opinion models~\cite{yildiz2013binary,nvm,yildiz2010voting,
cod,bil,medeiros2006domain,mobilia2003does,anvoter,durrett1996spatial}. In contrast to these models,
Flocking and DeGroot are continuous opinion models. In Flocking model and its variations~\cite{hegselmann2002opinion,bc1,bc2,bc3}, a node $i$ 
having opinion $x_i$ first selects the set of neighbors $j$ with $|x_i-x_j|\leq \epsilon$, and then updates its own opinion by averaging these opinions.
DeGroot model~\cite{degroot1974reaching}, on the other hand, allows a user to update her opinion with the average opinions of \emph{all} her neighbors. In this model,
the underlying influence matrix is row stochastic, enforcing consensus for a strongly connected graph. 
The second class of models, e.g. Biased Voter, AsLM, SLANT, etc. aim to learn a tractable linear model from a temporal message stream reflecting transient opinion dynamics~\cite{das2014modeling,de2014learning,nipsxx,gupta21}. While a Biased Voter model~\cite{das2014modeling} unifies various aspects of DeGroot and Flocking models, AsLM~\cite{de2014learning} generalizes the DeGroot model by relaxing the structure of the influence matrix. 
In contrast to   these   models that ignore the temporal effects of messages (post-rate), SLANT~\cite{nipsxx} blends
the opinion dynamics along with the message dynamics, using a stochastic generative model. 
In contrast to the above modeling approaches, there exist abundant empirical studies \cite{centola2010spread,aral2011creating,bakshy2012role,bakshy2012social,friggeri2014rumor} which investigate various factors influencing information diffusion in social networks through large-scale field experiments.
However, all these approaches skirt the effect of externalities, which severely constrains their forecasting prowess.

\xhdr{Opinion sensing} 
Sensing opinions, or mining sentiments from textual data traditionally relies on sophisticated NLP based machineries. See~\cite{liu2012sentiment, pang2008opinion} for details. Both these monographs provide a comprehensive survey. 
In general, LIWC~\cite{liwcPaper} is widely considered as benchmark tool to compute sentiments from rich textual data. 
On the other hand, Hannak \etal\ developed a simple  yet effective method for sentiment mining from short informal text like tweets~\cite{hannak2012tweetin}, also used by
~\cite{de2014learning,nipsxx}. Recently, a class of works~\cite{li2015analyzing,hosseinia2017detecting,kc2016temporal,li2017bimodal} designs simple supervised strategies to sense opinion spams, and some of them~\cite{hosseinia2017detecting,kc2016temporal,li2017bimodal} also advocate the role of temporal signals in opinion spamming. Note that, exogenous opinions are fundamentally different from opinion spams. In contrast to a spam which is unsolicited and irrelevant to the discussion, an exogenous post is often relevant, yet just an informed reflection of some external news or feeds. 
Also, since spamminess of a message is its intrinsic property, it does not depend on the messages before it.  
However, an exogenous post when retweeted, can become endogenous. 
Furthermore, the opinion spam detection techniques rest on the principle of supervised classification that in turn requires labeled messages. 
However, in the context of networked opinion dynamics, the messages (tweets) come unlabeled, which renders the spam detection techniques practically inapplicable for such scenarios.

Finally we conclude this section with this note that our work closely resembles SLANT~\cite{nipsxx} and is built upon the modeling framework proposed by SLANT, but the major difference is that SLANT assumes the entire event stream as endogenous, whereas our work is motivated towards exploring various techniques for systematically demarcating the externalities from the heterogeneous event stream. Similarly, our proposed algorithms are closely influenced by recent progress in subset selection literature \cite{hashemi2019submodular} where authors deal with designing new alphabetical optimality criteria for quadratic models. However, our work is motivated towards investing the subset selection problem for linear models in a temporal setting.

% \vspace*{-0.3cm}

\section{Problem setup}\label{sec:model}

In what follows, we describe the scenario of a social network of users who post opinion-bearing messages. For ease of reference, we list a compendium of all important notations used in this section in Table \ref{tab:notations}.
\begin{table}
	\begin{tabular}{c l}
		\toprule
		Symbol & Meaning \\\midrule
		$\Gcal$ & A directed social network (such as Twitter) \\
		$\Vcal$ & Vertices of $\Gcal$ (users) \\
		$\Ecal$ & Edges of $\Gcal$ (follower-followee links) \\
		$\Ncal(u)$ & Set of users followed by $u$, subset of $\Vcal$ \\
		$\Ucal(t)$ & History of messages posted by all users until time $t$ \\
		$\Ucal_u(t)$ & History of messages posted by user $u$ until time $t$ \\
		$e_i$ & $i$th message in $\Ucal(t)$ \\
		$u_i$ & User posting the $i$th message in $\Ucal(t)$ \\
		$\zeta_i$ & Opinion/sentiment value of the $i$th message in $\Ucal(t)$ \\
		$t_i$ & Timestamp of the $i$th message in $\Ucal(t)$ \\
		$\Hcal(t)$ & Set of endogenous messages in $\Ucal(t)$ \\
		$m_i$ & Opinion/sentiment value of the $i$th endogenous message in $\Hcal(t)$ \\
		$\Hcal_u(t)$ & Set of endogenous messages posted by user $u$ in $\Ucal_u(t)$ \\
		$\Ccal(t)$ & Set of exogenous messages in $\Ucal(t)$, complementary to $\Hcal(t)$ \\
		$w_i$ & Opinion/sentiment value of the $i$th exogenous message in $\Ccal(t)$ \\
		$\Ccal_u(t)$ & Set of exogenous messages posted by user $u$ in $\Ucal_u(t)$, complementary to $\Hcal_u(t)$ \\
		$N_u(t)$ & Counting process for endogenous messages of user $u$, equal to $|\Hcal_u(t)|$ \\
		$M_u(t)$ & Counting process for endogenous messages of user $u$, equal to $|\Ccal_u(t)|$ \\
		$\Nb(t)$ & Set of counting processes $(~N_u(t)~)_{u \in \Vcal}$ \\
		$\Mb(t)$ & Set of counting processes $(~M_u(t)~)_{u \in \Vcal}$ \\
		$\lambda_u^*(t)$ & Intensity of $N_u(t)$, i.e. the endogenous message rate of user $u$ \\	
		$\lambdab^*(t)$ & Set of intensities $(~\lambda_{u}^*(t)~)_{u \in \Vcal}$ \\
		\bottomrule
	\end{tabular}
	\caption{List of important notations used in Section \ref{sec:model}.}
	\label{tab:notations}
\end{table}

We use two sources of data as input: 
(I). a directed social network $\Gcal=(\Vcal, \Ecal)$  of users with the connections between them (e.g. friends, following, etc), and 
(II). an aggregated  history $\Ucal(T)$ of the messages posted by these users during a given time-window $[0,T)$.
In this paper, we summarize each message-event $e_i \in \Ucal(T)$ using only three components, the user $u_i$ who has posted the message, 
the opinion or sentiment value $\zeta_i$ associated with the message, and the timestamp $t_i$ of the post. Therefore, $\Ucal(T):=\{e_i = (u_i, \zeta_i, t_i) | t_i < T\}$.
% More generally, we denote $\Ucal(t):=\{e_i = (u_i, m_i, t_i) | t_i < t\}$, as the set of collected messages until any timestamp $t$.
We also use the notation $\Ucal(t)$ to denote the set of messages collected until $t<T$ i.e. $\Ucal(t):=\{e_i = (u_i, \zeta_i, t_i) | t_i < t\}$.

In the spirit of \cite{nipsxx}, we assume that the history of events until time $t$ influences the arrival process of events after time $t$.
However, in a direct contrast to~\cite{nipsxx} which skirts the potential influence of externalities, 
we posit that the message events belong to two categories-- \textit{endogenous} and \textit{exogenous}. 
Whereas the arrivals of endogenous events are driven by the previous events in the network i.e. these are \emph{history-dependent}, 
exogenous events originate from external influence outside the given social network and are, therefore \emph{not history-dependent}.
Note that the distinction between endogenous and exogenous events is not directly observable from the data, but needs to be inferred from the characteristics of the event sequence.
To this end, we split the entire set of messages observed until time $t$, $\Ucal(t)$ into two complementary subsets, $\Hcal(t)$ and $\Ccal(t)$ representing the sets of endogenous and exogenous events respectively, with $\Ucal(t) = \Hcal(t) \cup \Ccal(t)$ \& $\Hcal(t) \cap \Ccal(t)$ = $\phi$. 
At a user level, we denote $\Hcal_u(t) = \{e_i = (u_i, m_i, t_i) | u_i = u\ \mbox{and}\, t_i<t\}$ as the collection of all endogenous messages with sentiments $m_i$, posted by user $u$ until time $t$. 
Similarly, $\Ccal_u(t)=\{e_i = (u_i, w_i, t_i) | u_i = u\ \mbox{and}\, t_i<t\}$ denotes the history of exogenous messages posted by user $u\in \Vcal$ with sentiments $w_i$ until time $t$. Finally we write $\Ucal_u(t)=\Hcal_u(t)\cup \Ccal_u(t)$, as the history gathering both types of messages posted by user $u$, until time $t$.
Therefore, $\cup_{u\in\Vcal}\Hcal_u(t)=\Hcal(t)$, $\cup_{u\in\Vcal}\Ccal_u(t)=\Ccal(t)$, and $\cup_{u\in\Vcal}\Ucal_u(t)=\Ucal(t)$.
Note that, for clarity we denote $m_i$ and $w_i$ as endogenous and exogenous sentiments respectively, while $\zeta_i$ denotes opinion of any type. 
However, both types of sentiments belong to identical domain.
%Therefore, $\cup_{u\in\Vcal}\Hcal_u(t)=\Hcal(t)$.\\
To model the endogenous message dynamics, we represent the message times by a set of counting processes denoted as a vector $\Nb(t)$, in which the $u$-th entry, $N_{u}(t) \in \cbr{0} \cup \ZZ^+$, counts the number of endogenous messages user $u$ 
posted until time $t$, i.e. $N_u(t)=|\Hcal_u(t)|$.
Then, we characterize the message rates with the conditional intensity function 
%\vspace{-1mm}
\begin{equation}
\EE[d\Nb(t)\, |\,  \Ucal(t)] = \lambdab^*(t) \, dt,
%\vspace{-0.4mm}
\end{equation}
where $d\Nb(t):=\rbr{~dN_{u}(t)~}_{u \in \Vcal}$ counts the endogenous messages per user in the interval $[t, t+dt)$ and $\lambdab^*(t) := (~\lambda_{u}^*(t)~)_{u \in \Vcal}$ 
denotes the  user intensities that depend on the history $\Ucal(t)$. 

Note that we assume the endogenous events do not depend on their own history $\Hcal(t)$ only, but rather on the combined history $\Ucal(t)$ of both endogenous and exogenous events. Hence, every exogenous post influences the subsequent endogenous events in the same manner as the previous endogenous events. This is because a recipient user cannot distinguish between exogenous or endogenous posts made by her neighbors. 

In order to represent the arrival times of the exogenous message set $\Ccal(t)$, we introduce an additional counting process $\Mb(t)$ that describes the rate of generation of exogenous events, in which the $u$-th entry, $M_{u}(t) \in \cbr{0} \cup \ZZ^+$, counts the number of exogenous messages user $u$ posted until time $t$, i.e. $M_u(t)=|\Ccal_u(t)|$.
Note that, we do not aim to model the dynamics of exogenous events, since their source is not known to us.

\subsection{Opinion dynamics in absence of exogenous actions}
For clarity, we briefly discuss the proposal by De \etal~\cite{nipsxx}, that ignores the effect of exogenous messages.
The user intensities $\lambda_u^*(t)$ are generally modeled using multivariate Hawkes Process~\cite{Liniger2009}. We denote the set of users that \emph{u follows} by $\Ncal(u)$.
In absence of exogenous actions, \ie, when $\Ucal(t)=\Hcal(t)$, we have: 
\begin{equation}
\lambda^{*}_u(t) = \mu_u + \sum_{v \in   \Ncal(u)} b_{vu} \sum_{e_i \in \Hcal_{v}(t)} \kappa(t - t_i).
% = 
% \mu_u + \sum_{v \in u \cup \Ncal(u)} b_{vu}~(\kappa(t) \star dN_v(t)), 
\label{eq:intensity-multidimensional-hawkes}
\end{equation}
Here, the first term, $\mu_u \geqslant 0$, captures the posts by user $u$ on her own initiative, and the second term, with $b_{vu} \geqslant 0$, reflects the influence of previous posts on her intensity (self-excitation).
The users'{} latent opinions are represented as a history-dependent, multidimensional stochastic process $\mathbf{x}^{*}(t)$:
%\vspace*{-0.4mm}
\begin{equation}
%\vspace*{-1mm}
x^{*}_u(t) = \alpha_u + \sum_{v \in \Ncal(u)} a_{vu} \sum_{e_i \in \Hcal_{v}(t)} m_i g(t - t_i)
%\vspace*{-0.8mm}
% =\alpha_u + \sum_{v \in \Ncal(u)} a_{vu}~(g(t) \star (m_v(t) dN_v(t))), 
\label{eq:opinion} 
\end{equation}
%\vspace*{-0.1mm}
where the first term, $\alpha_u \in \RR$, models the original opinion of a user $u$  and
the second term, with $a_{vu} \in \RR$, models updates in user $u$'{}s opinion due to the influence from previous messages of her neighbours. Here, $\kappa(t)=e^{-\nu t}$ and $g(t) = e^{-\omega t}$ (where $\nu,\ \omega \geqslant 0$) denote exponential triggering kernels, which model the decay of influence over time.
%, where the $u$-th entry, $x^{*}_{u}(t) \in \RR$, represents the opinion of user $u$ at time $t$ and the sign $^{*}$ means that it may depend on the history $\Hcal(t)$. 
Finally, when a user $u$ posts a message at time $t$, the \emph{message sentiment $m$ reflects the expressed opinion} which is sampled from a distribution $p(m | x^*_u(t))$. 
Here, the sentiment distribution $p(m|x^* _u(t))$  is assumed to be normal, \ie \ $p(m | x_u(t)) = \Ncal(x_u(t), \sigma_u)$.
\subsection{Opinion dynamics with exogenous events}

In this section, we model the effect of exogenous events, $\Ccal(t)$, on the latent endogenous opinion process $x^*(t)$ and the endogenous rate $\lambda^*(t)$. Recall that $\Ncal(u)$ denotes the set of users that \emph{u follows}.
We present the dynamics of latent opinion $x^{*}_u(t)$ of user $u$, in the presence of exogenous messages in the following equation.
\begin{align}
	%\vspace*{-2mm}
	&\hspace*{-0.3cm} x^{*}_u(t) = \alpha_u +\hspace*{-0.3cm} \sum_{v \in \Ncal(u)}\hspace*{-0.2cm} a_{vu} \Big(\hspace*{-0.3cm}\sum_{e_i \in \Hcal_{v}(t)}\hspace*{-0.2cm} m_i g(t - t_i)+\hspace*{-0.3cm} \sum_{e_i \in \Ccal_{v}(t)}\hspace*{-0.3cm} w_i g(t - t_i)\Big)\label{eq:opinionCon}
	%\vspace*{-2mm}
\end{align}
where, the last term captures signals from exogenous posts. Similarly,
the endogenous message rate $\lambda_u ^* (t)$ of a user $u$ evolves as,
\begin{align}
	%\vspace*{-4mm}
	&\hspace*{-0.3cm} \lambda^{*}_u(t) = \mu_u +\hspace*{-0.3cm} \sum_{v \in \Ncal(u)}\hspace*{-0.2cm} b_{vu} \Big(\hspace*{-0.3cm}\sum_{e_i \in \Hcal_{v}(t)}\hspace*{-0.2cm} \kappa(t - t_i)+\hspace*{-0.3cm} \sum_{e_i \in \Ccal_{v}(t)}\hspace*{-0.3cm} \kappa(t - t_i)\Big).\label{eq:intensityCon}
	%\vspace*{-2mm}
\end{align}
Note that same parameters, $a_{vu}$ and $b_{vu}$, are used to model the effect of endogenous and exogenous processes, on both opinion and message dynamics. The above equation can be equivalently written as:
%\vspace*{-0.2cm}
\begin{align}
%\vspace*{-0.2cm}
	&\hspace*{-0.3cm} \xb^{*}(t) = \alphab + \hspace{-0.1cm}\int_0 ^t\hspace*{-0.2cm} g(t-s)\Ab\big[\mb(s)\odot d\Nb(s)+\wb(s)\odot d\Mb(s)\big]\label{eq:opConX} \\
	&\hspace*{-0.25cm}\lambdab^{*}(t) = \mub + \int_0 ^t \Bb \kappa(t-s) \big[d\Nb(s)+ d\Mb(s)\big].\label{eq:mesConX}
\end{align}
%	\vspace*{-0.1cm}
 Here $\Ab=(a_{vu})\in \RR^{|\Vcal|\times|\Vcal|}$, $\Bb=(b_{vu})\in \RR^{|\Vcal|\times|\Vcal|} _+$, $\xb^*(t)=(x^*_u(t))_{u\in\Vcal}$. Similarly we define $\lambdab^*(t), \ \mb(s),  \ \wb(s) $.
Furthermore, the exogenous intensity  is given by: $\EE[d\Mb(t)|\Ucal(t)]=\etab(t)$. We do not aim to model $\etab(t)$. %However, we do utilize it during opinion shaping in section \ref{sec:shaping}. 

By defining, $\Pb(t):=\Nb(t)+\Mb(t)$,
as the counting process associated with the combined history $\Ucal(t)=\Hcal(t)\cup\Ccal(t)$ of both endogenous and exogenous events, we further simplify
Eqs.~\eqref{eq:opConX} and  ~\eqref{eq:mesConX} as,
%\vspace*{-0.1cm}
\begin{align}
	&\hspace*{-0.2cm} \xb^{*}(t) = \alphab + \hspace{-0.1cm}\int_0 ^t\hspace*{-0.2cm}  g(t-s)\Ab\big[\zetab(s)\odot d\Pb(s)\big]\label{eq:oopConX}\\
	&\hspace*{-0.2cm}%\vspace*{-0.4cm}
	\lambdab^{*}(t) = \mub + \int_0 ^t \Bb  \kappa(t-s) d\Pb(s).\label{eq:omesConX}
\end{align}
%\vspace*{-0.4cm}
%\\

\section{Demarcation of endogenous and exogenous messages}
\label{sec:learning}
In this section, we propose a novel technique for demarcating endogenous messages $\Hcal(T)$ and exogenous messages $\Ccal(T)$ from a stream of unlabelled messages $\Ucal(T)$ gathered during time $[0,T)$. % 
Then, based on the categorized messages, we find the optimal parameters $\alphab$, $\mub$, $\Ab$ and $\Bb$ by solving  a maximum likelihood estimation (MLE) problem. 
From now onwards, we would write $\Ucal(T)$, $\Hcal(T)$, $\Ccal(T)$ as $\UU$, $\HH$ and $\Ccal_T$ to lighten the notations.
Hence, succinctly, the problem can be stated as follows:
\begin{enumerate}
 \item identify a subset $\HH\subseteq\UU$ of endogenous events
 \item find the optimal (maximum-likelihood) parameters $\alphab$, $\mub$, $\Ab$ and $\Bb$ based only on $\HH$
\end{enumerate}

\subsection{\ourx: Our proposed approach}
Now, we attempt to design an unsupervised learning algorithm to isolate the endogenous events $\HH$ and exogenous events $\Ccal_T$ from the stream of unlabeled 
sentiment messages $\UU$, which is equivalent to assigning each event $e\in \UU$ into either $\HH$ or $\Ccal_T$. 
This is achieved by extracting the set of events that comply with the endogenous dynamics with high confidence that in turn is indicated by a low variance of estimated parameters.

In more detail, given a candidate set of endogenous events  $\HH$, the opinion parameters $\Ab, \alphab$ can be estimated by maximizing  the likelihood of endogenous opinions $m_i$,  $\sum_{i}\log p(m_{i}|x^*_{u_i}(t_i))$,
\ie, minimizing the following,  
\begin{align}
\hspace*{-0.1cm}\underset{\Ab,  \alphab}{\text{min}}\hspace*{-0.1cm} & \hspace*{-0.1cm}\sum_{\substack{\ e_i\in \HH \\ u\in \Vcal}}\hspace*{-0.2cm}\sigma^{-2} \Big(m_i - \alpha_u -\hspace*{-0.1cm}\int_{0}^{t_i}\hspace*{-0.2cm} g(t-s)(\zetab(s)\odot d\Pb(s))^T \Ab_{u}\Big)^2 \Big.\nn\\
&+c||\Ab||_F ^2+c||\alphab||_2 ^2. \label{eq:loss1}
%\vspace*{-0.2cm}
\end{align}
Here, the first term is derived using the Gaussian nature of $p(m|x^* _u(t))$ and the last two are the regularized terms.
The optimal parameters  ($\hat{\Ab},\ \hat{\alphab}$) depend on the candidate set of endogenous messages $\HH$.  
To this end, we compute the estimation covariance as,
\begin{align}
\covm(\Hcal_T):=\EE(\thetahat-\thetab)(\thetahat-\thetab)^T, \ \thetab:=\text{vec}([\Ab, \ \alphab]).
\end{align}
%%%%%%%%%%%%%%%%%%%%%
Here the expectation is taken over the noise process induced while getting the message sentiment $m_i$, from the opinion $x_{u_i} ^*(t_i)$ according to the distribution $p(m_i|x^* _{u_i}(t_i))$.
Before proceeding further, we want to make a clarification that we exclude intensity parameters ($\mu$ and $B$) in covariance estimation as their MLE estimation do not offer closed form solution, making it mathematically inconvenient. 
Prior to going into the selection mechanism of
$\HH$, we first look into the expression of covariance matrix $\covm$ in the Lemma~\ref{lem:covv}. Note that, the inference problem given by Eq.~\eqref{eq:loss1} is that of regularized least squares estimation, and so the covariance matrix for the optimal parameters can be derived in a closed form given  in
the following:
%
% \vspace*{-0.2cm}
\begin{lemma}\label{lem:covv}
For a given endogenous message-set $\HH$,
\begin{align}\label{eq:covv}
\covm(\Hcal_T)=\diag_{u\in \Vcal}\big(c\Ib+\sigma^{-2}\sum_{e_i\in \Hcal_T}\phib^u _i\phib _i ^{uT}\big)^{-1}
\end{align}
%where, $\phib^u _i=\mathbbm{1}_{u_i=u}\big[\int_{0} ^{t_i} g(t-s)\zetab(s)\odot d\Pb(s),\ 1].$ $ \mathbbm{1}_X$ is the indicator function with respect to $X$.
where, 
\begin{equation}
\phib^u _i =
\begin{cases}
\text{vec}([\int_{0} ^{t_i} g(t-s)\zetab(s)\odot d\Pb(s), 1]) & u_i=u \\
0 & u_i\neq u
\end{cases}
\end{equation}
\end{lemma}
%  \vspace*{-0.3cm}
The proof of this lemma is given in the Appendix (Section \ref{sec:proof1}).

Our  objective is to identify $\HH$, given its size $N_\Hcal$, so that $\covm(\Hcal_T)$ is small.
Such a demarcated message-set $\HH$ would then follow endogenous opinion dynamics more faithfully than its complement $\UU\cp\HH$.
In order to compute the best candidate for $\Hcal_T$, we need to minimize a suitable function $\Omega_X(\Hcal_T)$ which is some measure of $\covm(\Hcal_T)$.

In accordance with the alphabetical design criteria of A-optimality, D-optimality, E-optimality, and T-optimality used by  \cite{hashemi2019submodular,chaloner1995bayesian}, we define, 
%\vspace*{-0.3cm}
\begin{align}\label{eq:fdef}
 &\Omega_A(\Hcal_T):=\tr\left[\covm(\Hcal_T)\right]\\
 &\Omega_D(\Hcal_T):=\tr\left[\log\covm(\Hcal_T)\right]=
 \log \left[ \mbox{det}(\covm(\Hcal_T)) \right]\\
 &\Omega_E(\Hcal_T):=\lambda_{max}\left[\covm(\Hcal_T)\right] \\
 &\Omega_T(\Hcal_T):=-\tr\left[\covm(\Hcal_T)^{-1}\right]
 %\vspace*{-0.5cm}
\end{align}
 where $\log \covm$  is the matrix logarithm of $\covm$, and $\lambda_{max}\left[\covm(\Hcal_T)\right]$ refers to the maximum eigenvalue of $\covm(\Hcal_T)$. 
 These functions $\Omega_X(\HH)$, where $X\in\{A,D,E,T\}$ can be viewed as complexity measures of $\covm(\HH)$~\cite{hashemi2019submodular} that make them good candidates for minimizing $\covm$.
 Hence, 
 by defining 
 $ f_X(\Hcal_T):=-\Omega_X(\HH)$, where $X\in\{A,D,E,T\}$,
 we pose the following optimization problem to obtain the best cardinality constrained candidate set $\HH$:
\begin{align}\label{eq:suboh}
& \underset{\HH \in \Ucal_T}{\text{maximize}}f_X(\HH)\nn\\
& \text{subject to, } |\HH|=N_\Hcal
\end{align}

Normally such a cardinality constrained subset selection problem would be NP-Hard \cite{williamson2011design,krause2014submodular}. Hence, we will rely on a greedy heuristic for maximizing $f_X$ (Algorithm~\ref{alg}), that, we would show later, gives an $(1-1/e)$ approximation bound.
Before going to that, we first specify two properties defined for any set function $h(V)$ in general (Definition~\ref{def:charx}) . 
We would show that, $f_X$ specifically enjoys these properties, thereby affording an approximation guarantee from the proposed simple greedy algorithm.

\begin{definition}
\label{def:charx}
A multidimensional set function $h(V)$ in a set argument $V\subseteq U$, is said to be
\begin{enumerate}
 \item \textbf{submodular}, if for any set $V\subseteq\barx{V}$,  $x\not\in\barx{V}$
      \begin{equation}
	h(V\cup x)-h(V)\geq h(\barx{V}\cup x)-h(\barx{V})
      \end{equation}
       In addition, if for all sets $V\subseteq U$, $h(V)$ can be expressed as a linear function of weights of individual set elements, i.e.
      \begin{equation}
       h(V)=w(\phi) + \sum_{x\in V} w(x) 
      \end{equation}
      for some weight function $w:U\rightarrow \mathbb{R}$, then $h(V)$ is said to be \textbf{modular}.
 \item \textbf{weakly submodular}, if for any set $V\subseteq\barx{V}$,  $x\not\in\barx{V}$, the two quantities
	
	\begin{equation}
	c_h = \max_{V,\barx{V},x} \frac{h(\barx{V}\cup x)-h(\barx{V})}{h(V\cup x)-h(V)}
	\end{equation}
	
	and
	
	\begin{equation}
	\epsilon_h = \max_{V,\barx{V},x} (h(\barx{V}\cup x)-h(\barx{V}))-(h(V\cup x)-h(V))
	\end{equation}
	
	are bounded. $c_h$ and $\epsilon_h$ are called the multiplicative and additive weak submodularity constants, respectively.
 
\end{enumerate}

\end{definition}

\begin{theorem}[\textbf{Characterizing $f_X$}]
\label{thm:subm}
Let $\Vcal(\HH)$ be the set of users of the message set $\HH$.
\begin{enumerate}
 \item $f_X(\HH)$ is \textbf{monotone} in $\HH$ $\forall X\in\{A,D,E,T\}$.
 \item $f_A(\HH)$ and $f_E(\HH)$ are \textbf{weakly submodular} in $\HH$.
 \item $f_D(\HH)$ is \textbf{submodular} in  $\HH$.
 \item $f_T(\HH)$ is \textbf{modular} in  $\HH$.
 
\end{enumerate}

\end{theorem}

\textsc{Proof Idea: }
The key to the proof of monotonicity relies on mapping the given set-function $f_X$ to a suitably chosen continuous functions $g(p)$ so that, $g(1)>g(0)$ implies the monotonicity of $f_X$. Noting that $f_X$ is linear, the rest of the proof follows from the properties of the A, D, E and T optimality criteria presented in \cite{hashemi2019submodular} and the citations therein.
The complete proof is given in the Appendix (Sec.~\ref{sec:proof2})

\subsubsection{Maximization of $f_X(\HH)$}
Since $f_X$ is  (weakly) submodular in $\HH$, it can be maximized by a traditional greedy approach adopted for maximizing submodular functions of a single set~\cite{subm}. The maximization routine is formally shown in Algorithm~\ref{alg}.
At each step, it greedily adds an event $e$ to $\HH$ sequentially, by maximizing the marginal gain
$f_X(\HH\cup\{e\})-f_X(\HH)$ (step 5, Algorithm~\ref{alg}) until $|\HH|$ reaches $N_\Hcal$.
\begin{lemma}[\textbf{Solution quality for $f_D$ and $f_T$}]
 	\label{lem:bound}
Algorithm~\ref{alg} admits a $(1-1/e)$  approximation bound for $f_D(\HH)$ and $f_T(\HH)$.	
\end{lemma}	
This result is due to submodularity and monotonicity of the functions $f_D(\HH)$ and $f_T(\HH)$.
It has been shown in \cite{nemhauser1978analysis} that such a greedy algorithm for maximizing a monotone and submodular function admits a $(1-1/e)$ approximation bound.

\begin{lemma}[\textbf{Solution quality for $f_A$ and $f_E$}]
 	\label{lem:bound2}
 	Let $c_{f_A}$ and $\epsilon_{f_A}$ be the multiplicative and additive weak submodularity constants (see Definition \ref{def:charx}) for $f_A$ and $c_{f_E}$ and $\epsilon_{f_E}$ be those for $f_E$. By Theorem \ref{thm:subm}, these quantities exist and are bounded provided $\Vcal(\HH)$. Let $(\HH^{Ag})$ and $(\HH^{Eg})$ be the subsets obtained by maximizing $f_A$ and $f_E$ respectively, and $(\HH^{A*})$ and $(\HH^{E*})$ be the optimal subsets achieving the maximum of $f_A$ and $f_E$ respectively.
 	Then,
 	\begin{align}
 	 f_A(\HH^{Ag})\ge (1-e^{-1/c_A})f_A(\HH^{A*})\\
 	 f_E(\HH^{Eg})\ge (1-e^{-1/c_E})f_E(\HH^{E*})
 	\end{align}
	where $c_A=\max\{c_{f_A},1\}$ and $c_E=\max\{c_{f_E},1\}$.
	Also,
	\begin{align}
	 f_A(\HH^{Ag})\ge (1-\frac{1}{e})(f_A(\HH^{A*})-(N_\Hcal-1)\epsilon_{f_A})\\
	 f_E(\HH^{Eg})\ge (1-\frac{1}{e})(f_E(\HH^{E*})-(N_\Hcal-1)\epsilon_{f_E})
	\end{align}
	where $N_\Hcal$ is the required number of endogenous events that is fed as an input to Algorithm~\ref{alg}.
%Algorithm~\ref{alg} admits an $(1-1/e)$  approximation bound for $f_D(\HH,\OO)$ and $f_T(\HH,\OO)$.	
\end{lemma}	
This result is due to weak submodularity and monotonicity of the functions $f_A(\HH)$ and $f_E(\HH)$. This result directly follows from Proposition 2 of \cite{hashemi2019submodular}.

Note that, one can find fast adaptive algorithms as an alternative to standard greedy for submodular function maximization. For $f_D$ being submodular monotone, recent fast adaptive algorithm proposed in \cite{breuer2020fast} easily applies for $f_D$. Though recent fast adaptive techniques in submodular maximization do not apply to weakly submodular functions in general, it has been shown that $f_A$ satisfies $\gamma$-differential submodularity \cite{qian2019fast} and thereby DASH, a recent fast adaptive technique proposed by \cite{qian2019fast} can be applied for maximizing $f_A$. However, in practice, we find  the performance obtained by both fast adaptive algorithms for maximizing $f_A$ as well as $f_D$ are inferior to standard greedy in some cases, despite speeding up the demarcation process. Thereby we have added a detailed comparative evaluation of standard greedy and adaptive algorithms as additional results in supplementary (see Section~\ref{sec:appen_expt}).

\begin{algorithm}[!t]                      % enter the algorithm environment
	 \smaller
	\caption{$\bm{\Upsilon}$=\ourx($f_X,N_\Hcal$, $\UU$)}          % give the algorithm a caption
	\label{alg:findSN}     
	%     \begin{multicols}{1}% and a label for \ref{} commands later in the document
	\begin{algorithmic}[1]\label{alg}                    % enter the algorithmic environment
		% \STATE\textbf{Input: }{$f,N_\Ocal,N_\Hcal$ } \\
		% \STATE\textbf{Output: }{$\HH,\OO$}\\
		\STATE \textbf{Initialization}:
		\STATE \text{$\HH \leftarrow \emptyset,\ \CT\leftarrow\UU$}
		\STATE \textbf{General subroutine}:
		\WHILE{$|\HH|<N_\Hcal$}
		%\IF{$|\Ocal|<N_\Ocal$}
		%\STATE /*\textttx{Choose $e$ in a greedy manner}*/
		%\STATE $e\leftarrow \text{arg }{\text{max}}_{e}f_X(\Hcal_T\cup\{e\})-f_X(\HH)$ 
		%\STATE $\CT\leftarrow \CT\cp \{e\}$
		%		\STATE /*\textttx{Update endogenous message-set}*/
		%\STATE $\HH\leftarrow\HH\cup\{e\}$
		%\ENDIF 
		%	\STATE /*\textttx{The user budget is reached. \textttx{$|\OO|=N_\OO$}\newline So, select only messages from now on.}*/
		\STATE $e\leftarrow \text{arg }{\text{max}}_{e\in \Ucal_T}f_X(\Hcal_T\cup\{e\})-f_X(\HH)$   
		\STATE $\CT\leftarrow \CT\cp \{e\}$
		 \STATE /*\textttx{Update endogenous message-set}*/
				\STATE $\HH\leftarrow\HH\cup\{e\}$
		\ENDWHILE
		\STATE $\bm{\Upsilon}=(\HH,\CT)$
% 		\STATE$\Ical\leftarrow\Vcal\cp\OO$
		\STATE \textbf{return} $\bm{\Upsilon}$.
\end{algorithmic}
\end{algorithm}
%

%\vspace*{-0.2cm}
%\setlength{\textfloatsep}{-0pt}% Remove \textfloatsep
\begin{algorithm}[h]   
% enter the algorithm environment
	 \smaller
	\caption{Parameter Estimation}          % give the algorithm a caption
% 	\label{alg:findSN}     
	%     \begin{multicols}{1}% and a label for \ref{} commands later in the document
	\begin{algorithmic}[1]\label{est}                    % enter the algorithmic environment
		\STATE\textbf{Input: }{$N_\Hcal$, $\UU$ } \\
		 \STATE\textbf{Output: }{$(\alphab^*, \mub^*,\Ab^*,\Bb^*)$}\\
		\STATE/*\cmnt{First find the endogenous messages}*/
		\STATE ($\HH,\CT$)=\ourx($f_X,N_\Hcal$, $\UU$)
		\STATE/*\cmnt{Estimate parameters over only $\HH$}*/
		\STATE $(\alphab^*, \mub^*,\Ab^*,\Bb^*)=\argmax \Lcal(\alphab,\mub,\Ab,\Bb|\HH)$
		\STATE \textbf{return} $\alphab^*, \mub^*,\Ab^*,\Bb^*$.
\end{algorithmic}
\end{algorithm}

%\vspace{1cm}
The event-set $\HH$ thus obtained would be used next to estimate all the parameters $\Ab,\mub,\alphab,\Bb$ (See Algorithm~\ref{est}) by maximizing $\Lcal(\alphab,\mub,\Ab,\Bb|\HH)$ which is given by 
% \begin{align}
% \vspace*{-1cm}
\begin{equation}\label{eq:LL}
 \Lcal(\alphab,\mub,\Ab,\Bb|\HH)=\sum_{e_i\in\HH} p(m_{i}|x^* _{u_i}(t_i))+\sum_{e_i\in \HH}  \log (\lambda_{u _i}(t_i))-\sum_{u\in \Vcal} \int_0 ^T \lambda^* _u(s)ds
\end{equation}

% \vspace*{-1cm}
% \end{align}
Since $\Lcal$ is a concave function, one can maximize this efficiently. We adopt the method given by the authors in~\cite{nipsxx}, which can accurately compute the parameters. In conclusion, the above procedures yield four distinct methods for demarcating the endogenous and exogenous dynamics, viz. \oura, \ourd, \ourd and \ourt according to the optimality criterion applied. 
\subsection{Discussion on the various optimalities} \label{discussion_optimality}
Since $f_A$ and $f_E$ are only weakly submodular while $f_D$ and $f_T$ enjoy full submodularity, \oura and \oure may, in theory, achieve poorer performance than \ourd and \ourt. 
However, \cite{hashemi2019submodular} have noted that, if the difference between the minimum and maximum information of individual observations is small, E-optimality is nearly submodular and under these conditions, \oure is expected to find a good (informative) subset. 
Furthermore \cite{chamon2017approximate,hashemi2019submodular} have also noted that the behavior of both A-optimality and E-optimality approaches that of a submodular function if the highest SNR of the observations is relatively small and in this case \oura and \oure should perform well. 
%See Theorems 5 and 6 of \cite{hashemi2019submodular} for mathematical definitions of information and SNR associated with a single observation, respectively.
Also, \cite{chamon2017approximate,hashemi2019submodular} have noted that even when the SNRs are large, when the observations are not too correlated, greedy design for A and E optimality achieves good results. 
In this context, we note that Proposition 1 of \cite{bian2017guarantees} provides a lower bound of the submodularity ratio of A-optimality in terms of the spectral norm $||\HH||$ of the observations $\HH$.
From there it follows that if the spectral norm $||\HH||$ is low, then the lower bound of the submodularity ratio approaches 1, i.e. the behaviour of A-optimality approaches that of a submodular function.

We also note that the modular nature of $f_T(\HH)$ implies that each $x\in\HH$ contributes independently to the function value. 
Consequently, the optimization of $f_T$ is easily achieved by simply evaluating $f_T$ for each individual event, sorting the result, and then choosing the top $N_\Hcal$ individual events from the sorted list to obtain the best subset $\HH$.

Finally, as noted by \cite{hashemi2019submodular}, under certain conditions A-optimality and D-optimality have more intuitive interpretations than E-optimality and T-optimality, A-optimality being related to the mean-square-error (MSE) and D-optimality being related to maximization of entropy of model parameters. 

\section{Experiments With Real Data}
\label{sec:expt}
In this section, we provide a comprehensive evaluation
of the four variants of $\our$ on a diverse set of real datasets. 
Since the category of being exogenous or endogenous is latent to a message event in most public datasets, our proposals cannot be tested in terms of their classification error.
Hence, we resort to measure the utility of our methods in terms of their impact on the predictive power of the underlying endogenous model. To that aim, we address the following research questions:
\begin{enumerate} 
 \item How do the variants of $\our$ compare against the competitive baselines in terms of the predictive accuracy of the trained endogenous model?
 \item How does the pre-specified value of the fraction of exogenous messages $\gamma$ impact the predictive accuracy?
 \item Do the proposed methods have any positive impact on the long term forecasting task?
 \item How do they perform well on a curated test set, which only contains endogenous messages?
  \item How does their performance vary across different sizes of training set?
\end{enumerate}

\subsection{Datasets}

We consider five real datasets (summarized in Table~\ref{tab:datasets}) corresponding to various real-world events, collected from Twitter.
They are tweets about a particular story. Specifically, we have:
\begin{enumerate}
 \item \textbf{\barca~\cite{cheshire}:} Barcelona winning the La-liga, from May 8 to May 16, 2016.
\item \textbf{\british~\cite{cheshire}:} British national election from May 7 to May 15, 2015.
\item \textbf{\jaya~\cite{cheshire}:} Verdict for the corruption-case against Jayalalitha, an Indian politician, from May 6 to May 17, 2015
\item \textbf{\juv~\cite{cheshire}} Champions League final in 2015, between Juventus and Real Madrid, from May 8 to May 16, 2015.
\item \textbf{\twitter~\cite{nipsxx}:} Delhi assembly elections, from 9th to 15th of December 2013
\end{enumerate}

Along with other statistics, the last column ($\bar{r}$) in Table~\ref{tab:datasets} indicates average absolute correlations of the observation matrix for each dataset. We  observe that the average correlations are quite small (in range of 0.001-0.006) despite high SNR value (in range of 30-50 dB), which justifies the application of \oura or \oure on these datasets as the greedy design of A and E optimality should achieve good performance for such cases even though they are only weakly submodular (see Section~\ref{discussion_optimality}).

For all datasets, we follow a very standard setup for both network construction and message sentiment computation~\cite{de2014learning,nipsxx,icml17}. 
We built the follower-followee network for the users that posted related tweets using the Twitter rest API\footnote{\scriptsize \url{https://dev.twitter.com/rest/public}}. 
Then, we filtered out users that posted less than 200 tweets during the account lifetime, follow less than 100 users, or have less than 50 followers. 
For each dataset, we compute the sentiment values of the messages using a popular sentiment analysis toolbox~\cite{hannak2012tweetin}. Here, the sentiment takes values $m\in[-1,1]$ and we consider the sentiment polarity to be simply $\text{sign}(m)$. Note that, while other sentiment analysis tools~\cite{liwcPaper} can be used to extract sentiments from tweets, we appeal to~\cite{hannak2012tweetin} due to two major reasons-- its ability of accurately extract sentiments from short informal texts like tweets, and its wide usage in validating data-driven opinion models~\cite{de2014learning,nipsxx}.

\subsection{Evaluation protocol}\label{sec:eval_protocol}
The temporal stream of sentiment messages is split into training and test sets, assigning the first 90\% of the total number of messages to the training set. The training set $\UU$, collected until time $T$, is categorized into endogenous $\HH$ and exogenous messages $\Ccal_T$, and model parameters are estimated over the classified $\HH$. 
During categorization, we took a range of pre-specified values of $|\HH|/|\UU|$, the pre-specified fraction of organic messages. 
%However, we assumed $\OO=\Vcal$ to extract the endogenous dynamics from all users for all of the \ourx\  methods. 
%
Finally, using this estimated model, we forecast the sentiment value $m$ for each message in the test set given the history up to \T hours before the time of the message as $\hat{m} = E_{\Hcal_t \backslash \Hcal_{t-\T}}[x^*_u(t) | \Hcal_{t-\T}]$ that we compute using an efficient simulation method given by~\cite{nipsxx,coevolve}. 
\begin{table}[!t]
	\begin{center}
		\begin{tabular}{|c|c|c|c|c|c|c|}
			\hline
			Datasets & $|\mathcal{V}|$ & $|\mathcal{E}|$ & $|\mathcal{U}_T|$ & $\EE[m]$ & $\rho[m]$ & $\bar{r}$ \\ \hline \hline    
			\barca & 703 & 4154 & 9409 & 0.109 & 0.268 & 0.0026\\
			\british  & 231 & 1108 & 1584 & -0.096 & 0.232 & 0.0062\\
			%    GTwitter & 947 & 5126 & 13203 & 0.482 & 0.238 \\
			\jaya & 1059 & 17452 & 10691 & 0.062 & 0.266 & 0.0016 \\
			\juv & 703 & 4154 & 7431 & 0.562 & 0.224 & 0.0041\\
			%    MsmallTwitter & 679 & 3284 & 17850 & 0.490 & 0.245 \\
			\twitter & 548 & 5271 & 20026 & 0.016 & 0.178 & 0.0018 \\
			%    VTwitter & 567 & 2443 & 14016 & 0.447 & 0.281 \\
			\hline

		\end{tabular}
		\caption{Statistics of real datasets.}
		\label{tab:datasets}
		%\vspace*{-0.4cm}
	\end{center}
\end{table}

\subsection{Baselines}\label{sec:baselines}
We compare the four variants of \our with four unsupervised event classification techniques, borrowed from robust regression literature as well as various outlier detection techniques. Finally, we compare a representative of \our with three best performing baselines for each experiment. 
% We summarize the list of algorithms in ~Table\ref{tab:algorithms}.
%

\xhdr{Huber regression~\cite{tsakonas2014convergence}} Here, we apply Huber penalty in our learning objective, which follows from the underlying assumption that a subset of the samples are outliers. 
	\begin{align}
		\hspace*{-0.1cm}\underset{\Ab,  \alphab}{\text{min}}\hspace*{-0.1cm} \hspace*{-0.1cm}\sum_{\substack{\ e_i\in \UU}}\hspace*{-0.2cm}\rho_{h} \Big(m_i - \alpha_u -\hspace*{-0.1cm}\int_{0}^{t_i}\hspace*{-0.2cm} g(t-s)(\zetab(s)\odot d\Pb(s))^T \Ab_{u}\Big) \Big.
		\label{eq:huber_reg}
		%\vspace*{-0.2cm}
	\end{align}
	where $\rho_h\colon\Rb \rightarrow \Rb$ is defined as $\rho_h(u)=u^2$ if $|u|\leq c/2$, otherwise $c|u|-c^2/4$.
	
 \xhdr{Robust lasso~\cite{nasrabadi2011robust}} 
Here, we define $\epsilon_i$ as a measure of exogenous behavior of an event $e_i$. Such a measure is matched with the training error using a mean square loss, which is further penalized by an $L_1$ regularizer on $\bm{\epsilon}=(\epsilon_i)_{i\in\Ucal_T}$. Such a regularizer controls the fraction of exogenous messages identified by this model. During our implementation, we tune it so that the fraction of exogenous messages is close to $\gamma$. 
	\begin{align}
		\hspace*{-0.1cm}\underset{\Ab,  \alphab, \ob}{\text{min}}\hspace*{-0.1cm} & \hspace*{-0.1cm}\sum_{\substack{\ e_i\in \UU}}\hspace*{-0.2cm}  \Big(m_i - \alpha_u -\hspace*{-0.1cm}\int_{0}^{t_i}\hspace*{-0.2cm} g(t-s)(\zetab(s)\odot d\Pb(s))^T \Ab_{u}-o_i\Big)^2 \Big.+
		  c_1 \Big( ||\Ab||_1+||\alphab||_1 \Big)+c_2||\ob||_{1}. \nn
		%\vspace*{-0.2cm}
	\end{align}
\xhdr{Robust hard thresholding~\cite{bhatia2015robust}} 
Instead of minimizing different measures of variance, such a method directly solves the training error of the endogenous model using a hard thresholding based approach.
\begin{align}
		\hspace*{-0.1cm}\underset{\Ab,  \alphab,|\HH| \geq (1-\gamma)|\UU|}{\text{min}}\hspace*{-0.1cm} & \hspace*{-0.1cm}\sum_{\substack{\ e_i\in \HH \\ u\in \OO}}\hspace*{-0.2cm}\frac{1}{|\UU|} \Big(m_i - \alpha_u -\hspace*{-0.1cm}\int_{0}^{t_i}\hspace*{-0.2cm} g(t-s)(\zetab(s)\odot d\Pb(s))^T \Ab_{u}\Big)^2 \Big.
		+c \Big(||\Ab||_F ^2+||\alphab||_2 ^2 \Big).  \nn
		%\vspace*{-0.2cm}
	\end{align}

\xhdr{Soft thresholding} This method is designed by assuming the presence of unlimited error signals on a limited number of data points and alternates between $(\alpha,A)$ and ${\{o_i\}}_{\forall i \in \HH}$ for solving the following objective. 	
	\begin{align}
		\hspace*{-0.1cm}\underset{\Ab,  \alphab, \ob}{\text{min}}\hspace*{-0.1cm} & \hspace*{-0.1cm}\sum_{\substack{\ e_i\in \UU }}\hspace*{-0.2cm}\frac{1}{|\UU|} \Big(m_i - \alpha_u -\hspace*{-0.1cm}\int_{0}^{t_i}\hspace*{-0.2cm} g(t-s)(\zetab(s)\odot d\Pb(s))^T \Ab_{u}-o_i\Big)^2 \Big. 
		 +c_1\Big(||\Ab||_F ^2+||\alphab||_2 ^2 \Big)+c_2||\ob||_{1}. \nn
		%\vspace*{-0.2cm}
	\end{align}
%For each of meth

\xhdr{\ourm ~\cite{nipsxx}} In this baseline, we have used all the samples for parameter estimation purpose, without any filtering.

\subsection{Evaluation metrics}
We measure the performance of our methods and the baselines using the prediction errors of the correspondingly trained endogenous model.
Specifically, we use 
(i) the \emph{mean squared error (MSE)} between the actual sentiment value ($m$) and the estimated  sentiment value ($\hat{m}$), \ie, $\EE[(m-\hat{m})^2]$. 
and (ii) the \emph{failure rate (FR)} which is the probability that the polarity of actual sentiment ($m$) does not coincide with the polarity of predicted opinion ($\hat{m}$), \ie, $\PP(\sgn(m) \neq \sgn(\hat{m}))$ to measure the predictive error of the resulting endogenous model.

\begin{table}[!h]
\centering
		\resizebox{0.85\textwidth}{!}{
		\begin{tabular}{|c|c|c|c|c||c|c|c|c|c|}
			\hline
            &  \multicolumn{9}{c|}{ \textbf{Mean squared error:} $\EE(m-\hat{m})^2$ }\\\hline
            &  \multicolumn{4}{c|}{\our} & \multicolumn{5}{c|}{Competitive methods}\\\hline
			Datasets &  A & D & E  & T  & Hard  & Huber & Lasso & Soft & Slant \\ \hline
			\barca  &  0.038 & \textbf{0.037} & 0.038 & 0.038 & 0.038 & 0.039 & 0.039 & 0.039 & 0.040 \\\hline
			\british &  0.054 & 0.053 & 0.054 & 0.051 & \textbf{0.049} & 0.055 & 0.054 & 0.054 & 0.055 \\\hline
			\jaya  &  \textbf{0.069} & 0.071 & 0.071 & 0.071 & 0.070 & 0.076 & 0.074 & 0.074 & 0.073 \\\hline
			\juv  &  0.066 & \textbf{0.056} & 0.060 & 0.071 & 0.069 & 0.060 & 0.073 & 0.074 & 0.076 \\\hline
			\twitter  &  \textbf{0.035} & 0.039 & 0.0 & 0.036 & 0.037 & 0.040 & 0.040 & 0.039 & 0.042 \\ \hline  \hline
			&  \multicolumn{9}{c|}{ \textbf{Failure Rate:} $\PP(\text{sign}(m)\neq \PP(\text{sign}(\hat{m}))$ }\\\hline
			\barca  &  0.120 & 0.113 & 0.121 & 0.122 & \textbf{0.108} & 0.119 & 0.115 & 0.117 & 0.126 \\\hline
            \british &  0.169 & 0.176 & 0.169 & 0.188 & \textbf{0.163} & 0.213 & 0.213 & 0.201 & 0.188 \\\hline
			\jaya  &  0.207 & 0.210 & 0.208 & \textbf{0.204} & 0.214 & 0.211 & 0.210 & 0.211 & 0.215 \\\hline
			\juv  &  0.114 & \textbf{0.079} & 0.096 & 0.122 & 0.090 & 0.080 & 0.086 & 0.086 & 0.098 \\\hline
			\twitter  &  \textbf{0.137} & 0.147 & 0.0 & 0.144 & 0.141 & 0.152 & 0.157 & 0.152 & 0.160 \\ \hline
		\end{tabular}
	}
		\caption{Sentiment prediction performance for five real-world datasets for {all competing methods} for a fixed $\gamma = 0.2$. For each message $m$ in test set, we predict its sentiment value given the history up to \T= 4 hours before the time of the message. For the \T hours, we predict the opinion stream using a sampling algorithm. Mean squared error and failure rate have been reported. We observe that the variants of \our generally perform better than the baselines on all datasets. Among the baselines, \ourR performs comparably with \our on some of the datasets presented here.}
		\label{tab:forecast}
		%\vspace*{-0.4cm}
\end{table}

\subsection{Results}
\label{sec:results:real}

\xhdr{Comparative analysis}\label{sec:perf_eval} 
Here we aim to address the research question (1). More specifically, we compare the endogenous model obtained using our method against the baselines.
 Table~\ref{tab:forecast}
summarizes the results, which reveals the following observations.
(I) Our method outperforms other methods. %, which is due to the fact that it performs better than any other method.
(II) $\our_{A}$ and $\our_D$ perform best among the variants of \our. This is because $\our_E$ and $\our_T$ often suffer from poor training due to the computational inefficiency of eigenvalue optimization in $\our_E$ and the trace optimization of inverse matrices in $\our_T$. (III) The robust regression with hard thresholding performs best across the baselines, which together with the superior performance of our methods indicate that the hard thresholding based methods are more effective than those based on soft thresholding in the context of unsupervised demarcation of the opinionated messages. 	
Here we want to clarify that Table~\ref{tab:forecast} presents a complete comparative evaluation of all the baselines along with all variants of \our whereas, in subsequent experiments, we present only the three best performing baselines for clarity.

\begin{figure}[h!]
	
	\centering
	\subfloat{ 	\includegraphics[scale=0.75]{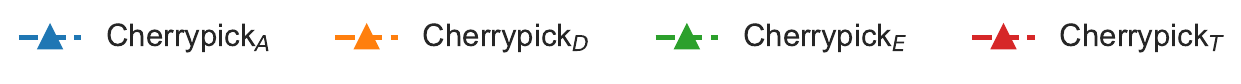}}
	\vspace{-3mm}
	
	\subfloat[\barca]{ 	\includegraphics[width=0.30\textwidth]{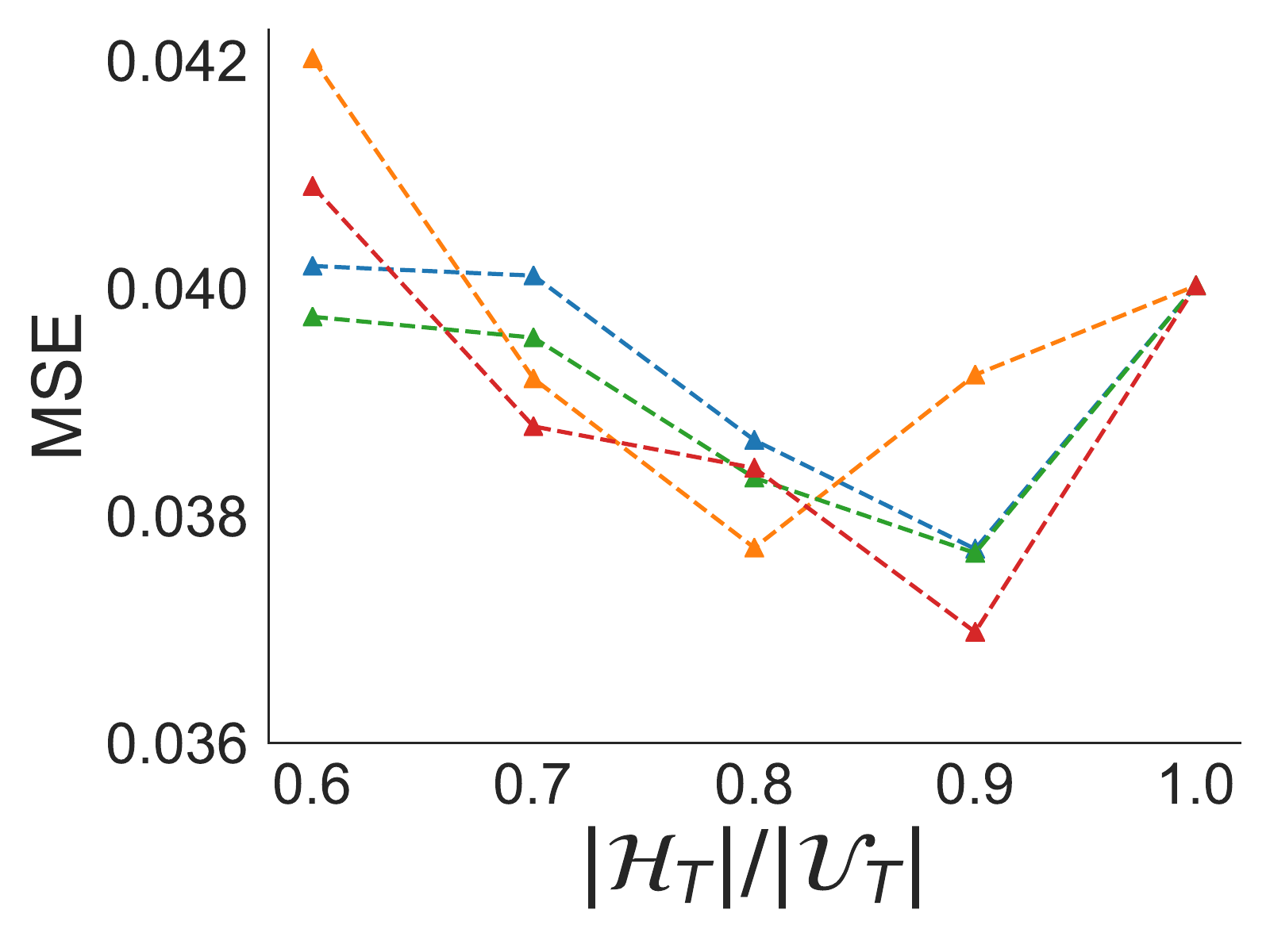}}\hspace*{0.2cm}
	%\subfloat[\british]{ 	\includegraphics[width=0.30\textwidth]{FIG_new/exp2_british_election_MSE.pdf}}\hspace*{0.2cm}
	%\subfloat{ 	\includegraphics[width=0.30\textwidth]{FIG_new/exp2_GTwitter_MSE.pdf}}
	\subfloat[\jaya]{ 	\includegraphics[width=0.30\textwidth]{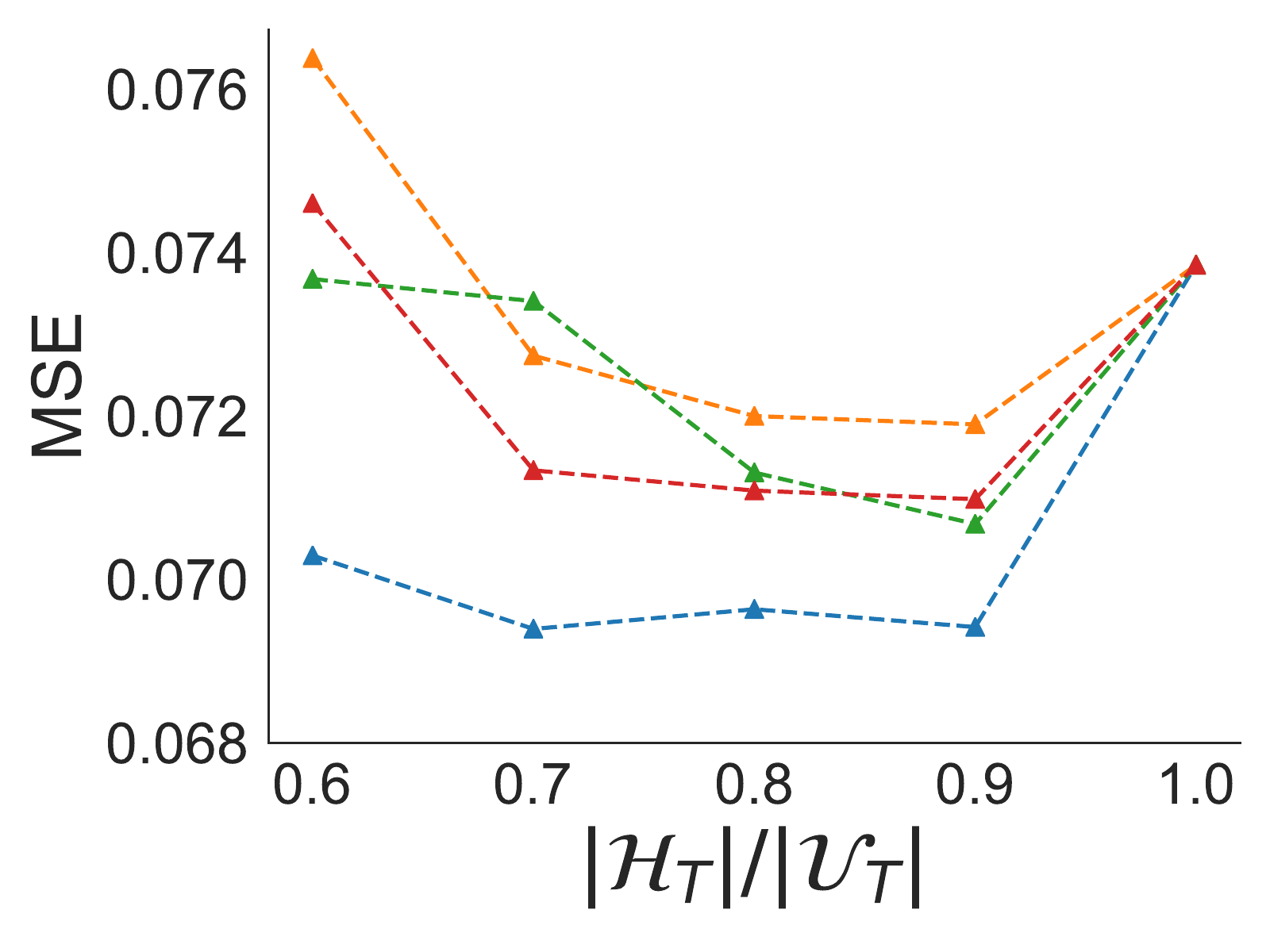}}\hspace*{0.2cm}
	\subfloat[\twitter]{ 	\includegraphics[width=0.30\textwidth]{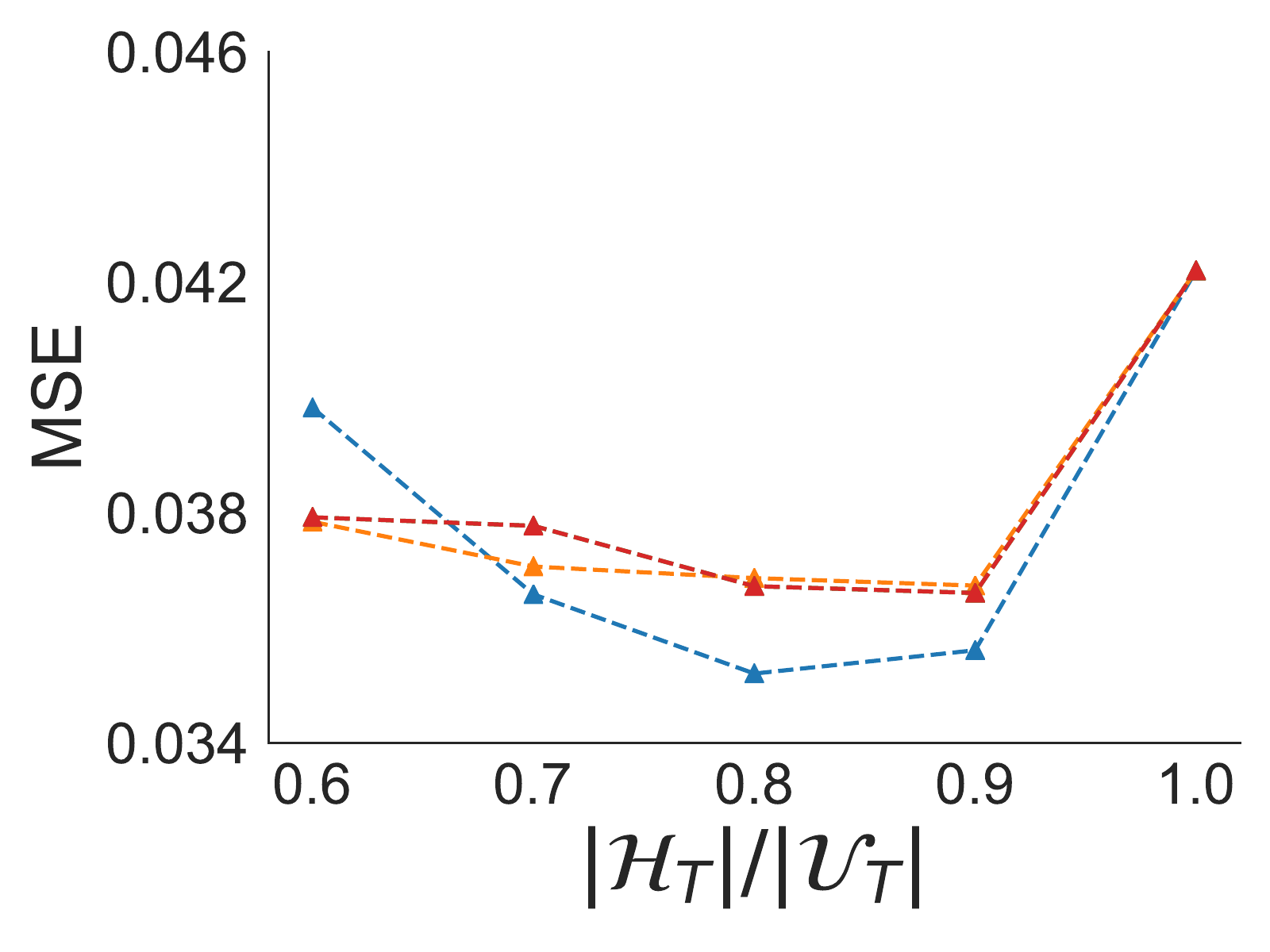}}\hspace*{0.2cm}

	%\subfloat[Juv]{ 	\includegraphics[width=0.30\textwidth]{FIG_new/exp2_JuvTwitter_MSE.pdf}}
	
	\vspace{-3mm}
	
	\subfloat[\barca]{\includegraphics[width=0.30\textwidth]{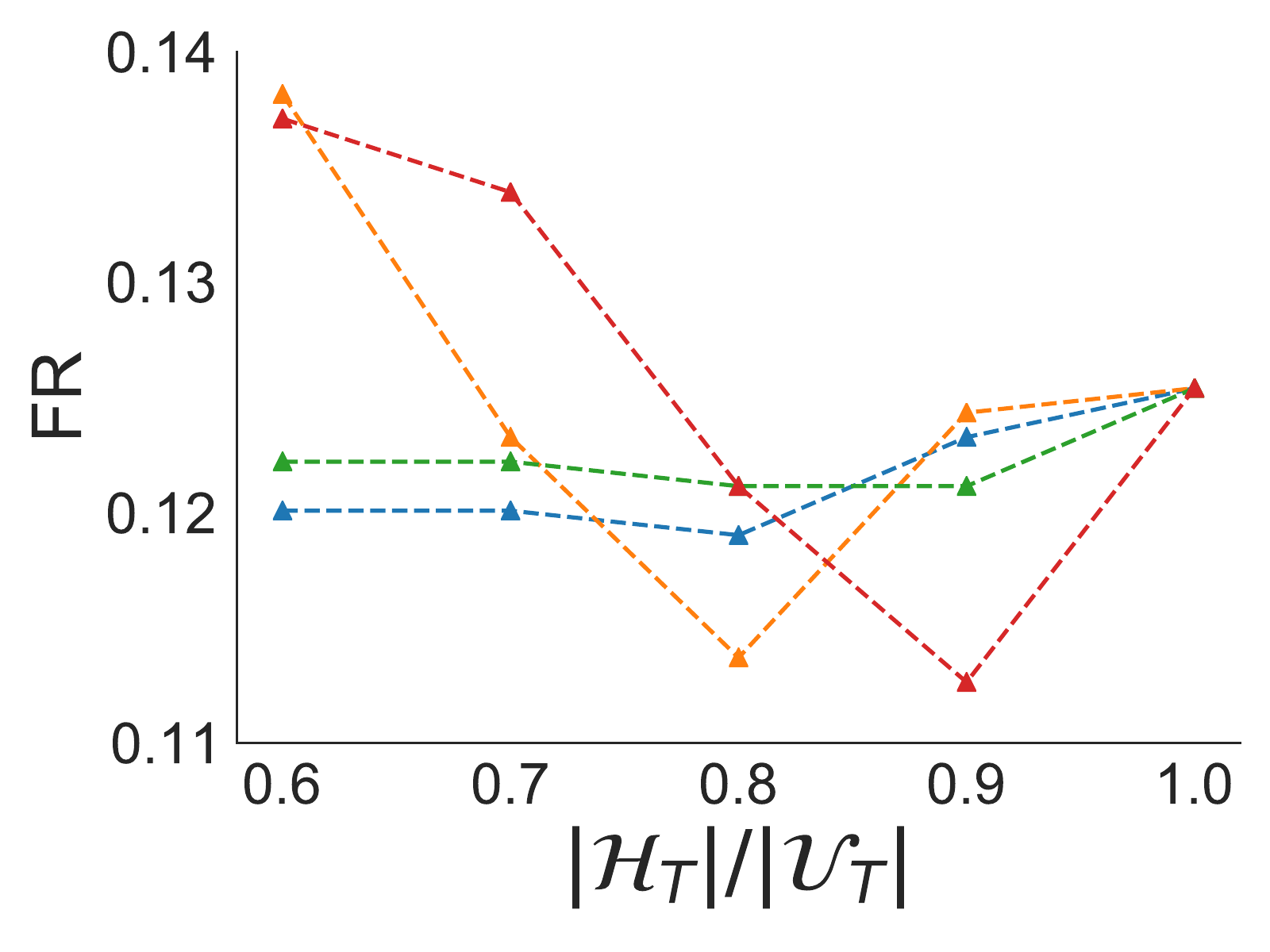}}\hspace*{0.2cm}
	%\subfloat[\british]{ 	\includegraphics[width=0.30\textwidth]{FIG_new/exp2_british_election_FR.pdf}}\hspace*{0.2cm}
	%\subfloat{ 	\includegraphics[width=0.30\textwidth]{FIG_new/exp2_GTwitter_FR.pdf}}
	\subfloat[\jaya]{\includegraphics[width=0.30\textwidth]{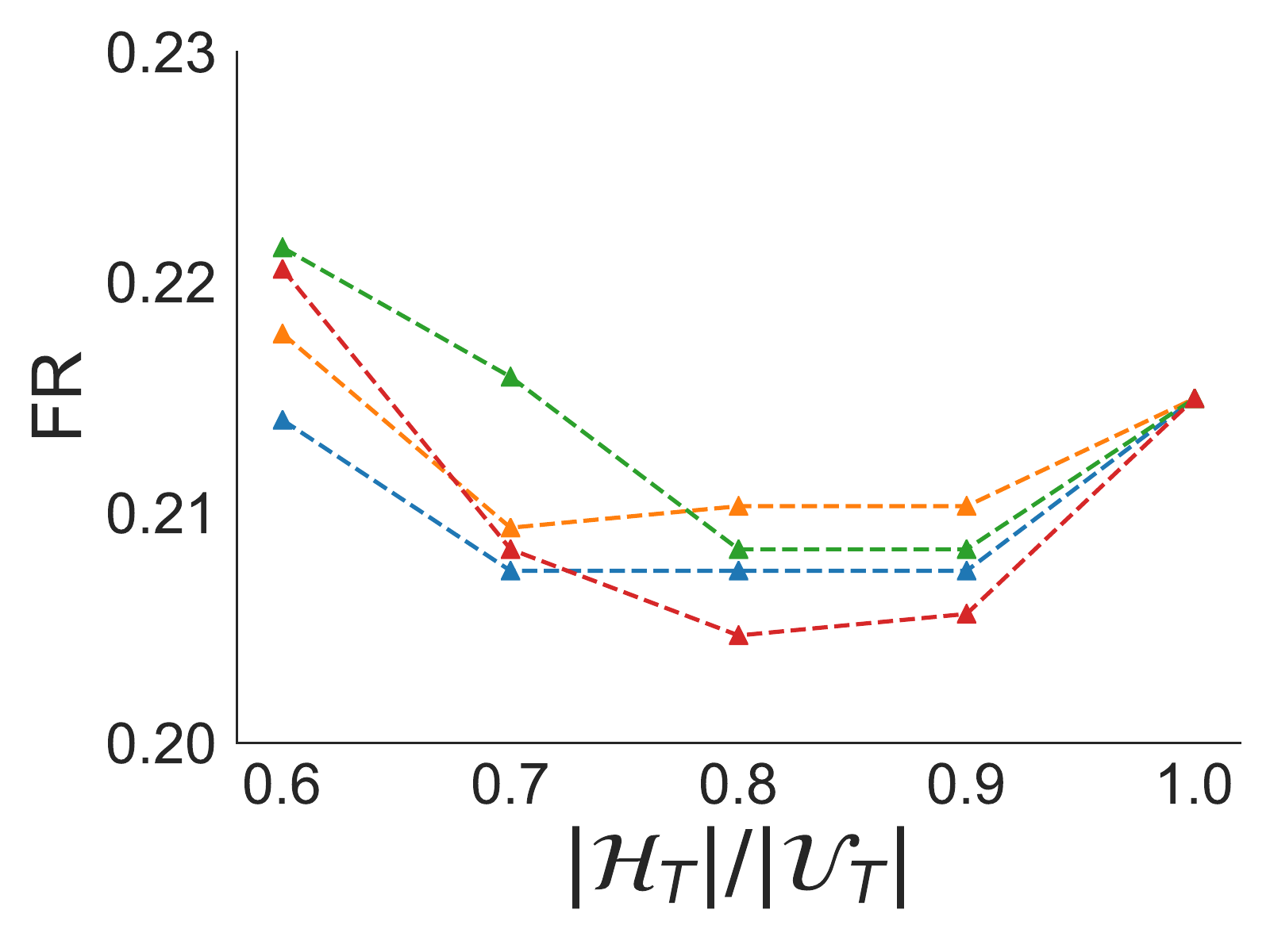}}\hspace*{0.2cm}
	%\vspace{-3mm}
	%\subfloat[Juv]{ 	\includegraphics[width=0.30\textwidth]{FIG_new/exp2_JuvTwitter_MSE.pdf}}
	%\subfloat{ 	\includegraphics[width=0.30\textwidth]{FIG_new/exp2_MsmallTwitter_MSE.pdf}}
	% \subfloat{ 	\includegraphics[width=0.30\textwidth]{FIG_new/real_vs_ju_703_MSE.pdf}}
	%\subfloat{ 	\includegraphics[width=0.30\textwidth]{FIG_new/exp2_VTwitter_MSE.pdf}}
	%\vspace{-3mm}
	%\subfloat[Juv]{ 	\includegraphics[width=0.30\textwidth]{FIG_new/exp2_JuvTwitter_FR.pdf}}
	%\subfloat{ 	\includegraphics[width=0.30\textwidth]{FIG_new/exp2_MsmallTwitter_FR.pdf}}
	% \subfloat{ 	\includegraphics[width=0.30\textwidth]{FIG_new/real_vs_ju_703_FR.pdf}}
	\subfloat[\twitter]{\includegraphics[width=0.30\textwidth]{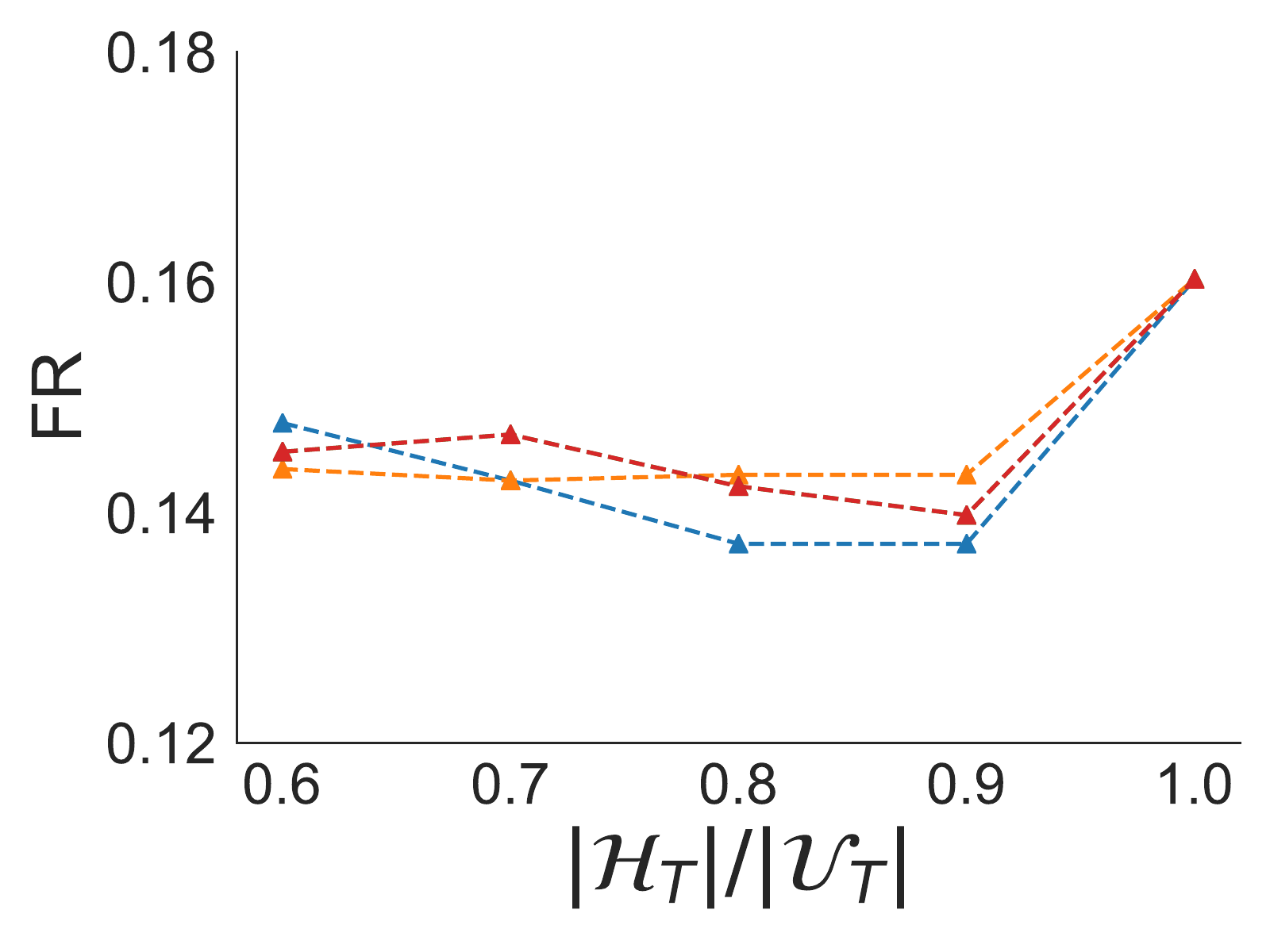}}\hspace*{0.2cm}
	%\subfloat{ 	\includegraphics[width=0.30\textwidth]{FIG_new/exp2_VTwitter_FR.pdf}}
	\vspace{-3mm}   
	\caption{Performance variation with size of endogenous subset for \barca, \jaya and \twitter datasets where the size of endogenous subset of data is varied from 50\% to 100\%. Time span \T has been set to 4 hours. Mean squared error and failure rate have been reported. We observe that our method performs better for the range of size of endogenous subset within 70\% to 90\% of the whole data whereas performance deteriorates when endogenous subset size is set too small like 60\% or too high like 100\% of the whole data. }
	\label{fig:varWithgamma}
\end{figure}

\xhdr{Variation of performance with the fraction of outliers ($\gamma$)}  
Next, we address the research question (2).
Figure \ref{fig:varWithgamma}  describes the variation of forecasting performance for different values of $\gamma$ \ie\ the pre-specified fraction of outliers, for \barca, \jaya and \twitter datasets. In this experiment, $\gamma$ is varied from $0.4$ to $0$, and demarcation methods are used to remove outlier according to a given $\gamma$, followed by parameter estimation over the refined set. 
As this experiment is mainly intended for showing the effect of the parameter $\gamma$ in our methods on predictive performance, 
%	the critical role of the $\gamma$ in our methods, 
	baselines are omitted here. 
For this experiment, the time span has been fixed to  $4$ hours.
Here we observe, 
%as we decrease $\gamma$, 
as we start refining the event set, the prediction performance improves. But if we increase the value of $\gamma$ beyond 
around 0.4, the performance drops,   
%the forecasting error first decreases and then increases, 
strongly suggesting an optimum number of outliers present in the training data.
If we set a high value of $\gamma$, our methods misclassify many regular events as outliers, while a small value
of $\gamma$ ignores their effects. 

\xhdr{Forecasting performance}
Next, we address the research question (3).
In particular, we compare the forecasting performance of \our against the baselines.
Figure \ref{fig:forecast} shows the forecasting performance with respect to variation of \T, across various representative datasets, for the best performing variant of \our 
and three best performing baselines (best according to MSE at \T = 4  hours), where $\gamma=0.2$. 
We make the following observations. (I) \our outperforms the baselines for the majority of the cases. (II) Generally, \ourR performs better among the baselines. (III) The performance deteriorates as we predict further in the future for all the methods, but the performance stabilizes after a while.

\begin{figure}[h!]
	
	\centering
	\subfloat{ 	\includegraphics[width=12cm,keepaspectratio]{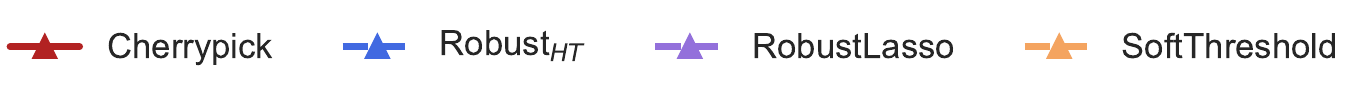}}
	\vspace{-3mm}
	
	\hspace{-8mm}\subfloat{ 	\includegraphics[width=0.20\textwidth]{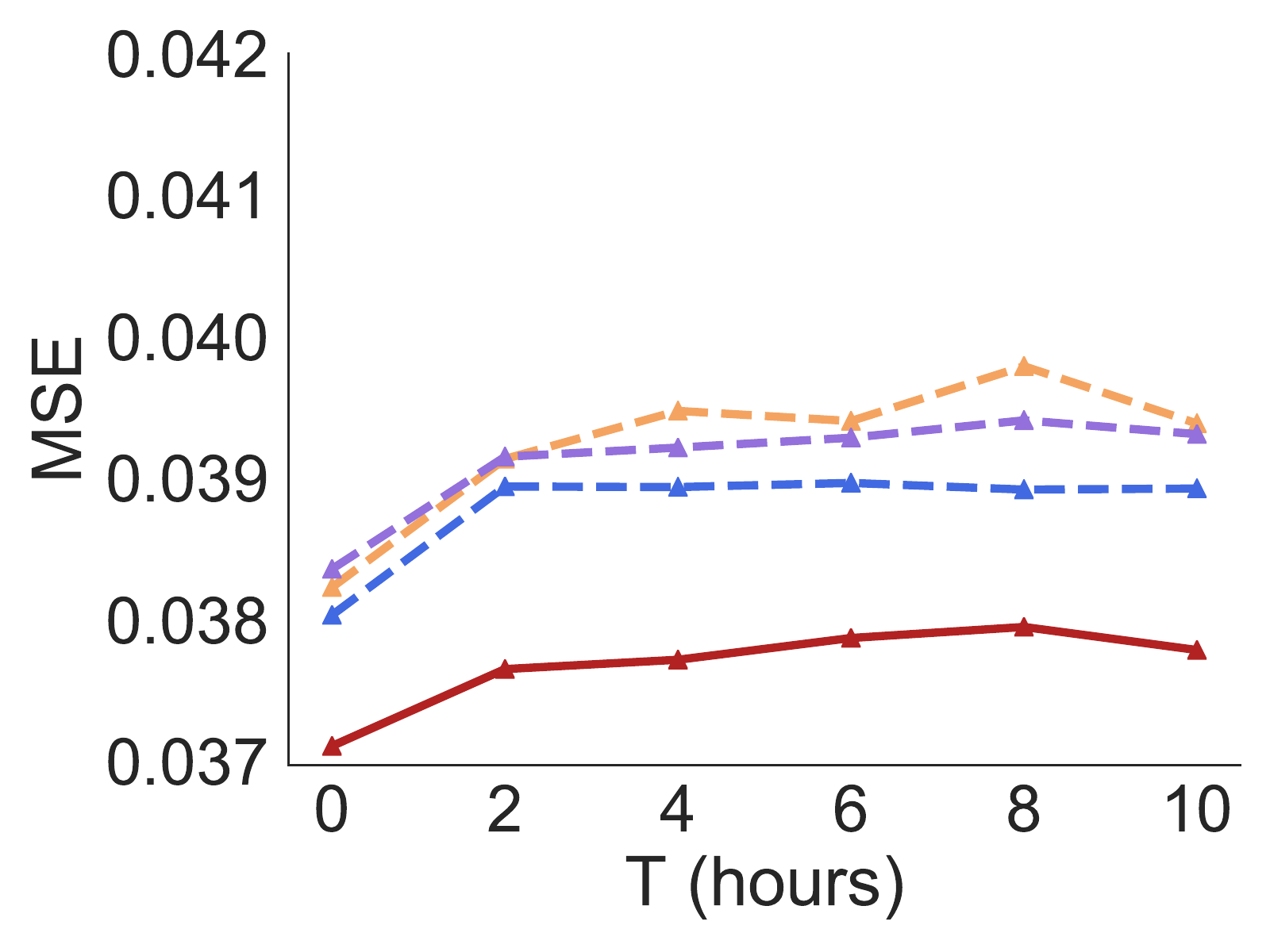}}
	\subfloat{ 	\includegraphics[width=0.20\textwidth]{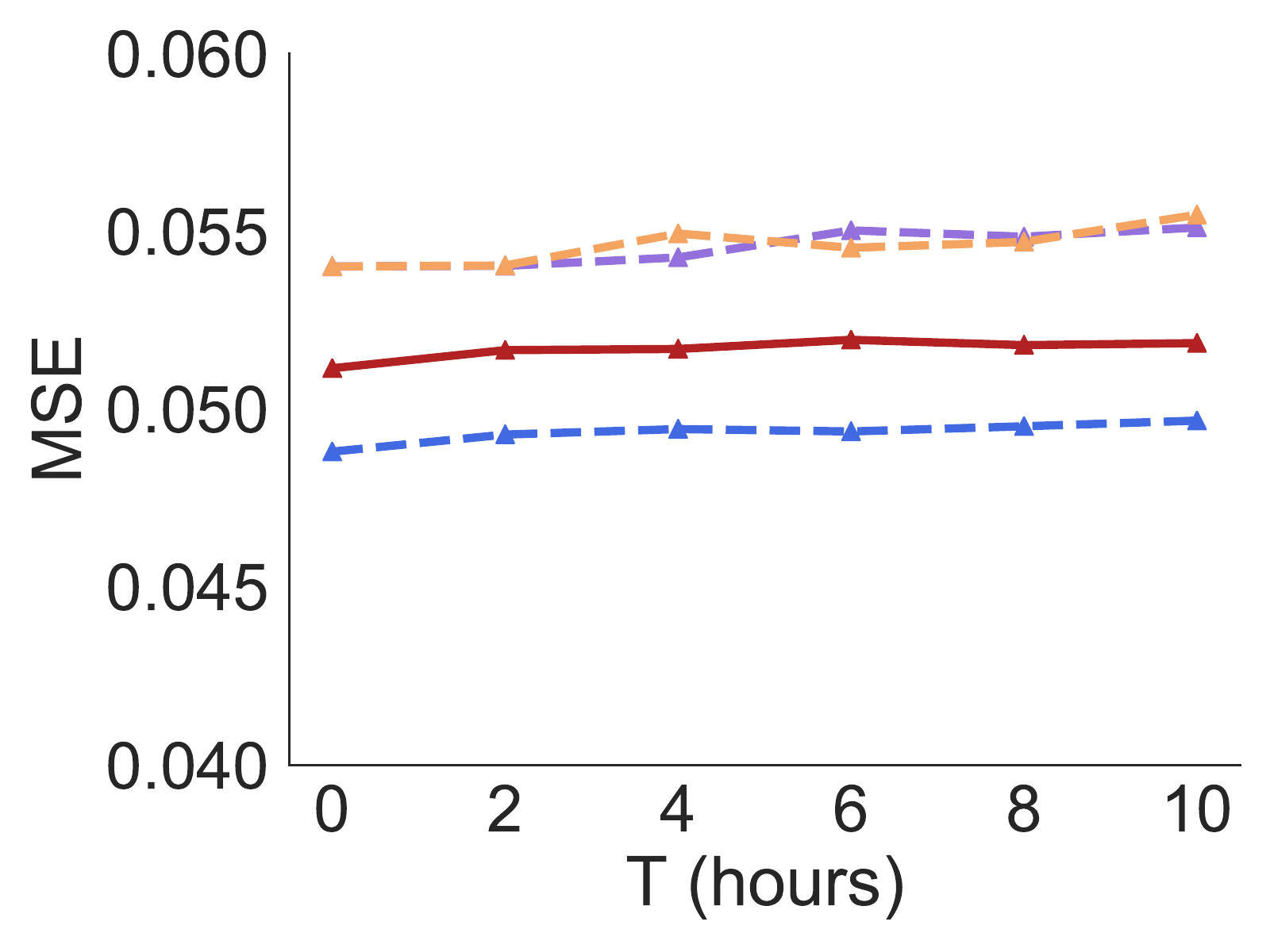}}
	%\subfloat{ 	\includegraphics[scale=0.20]{FIG_new/exp1_GTwitter_MSE.pdf}}
	\subfloat{ 	\includegraphics[width=0.20\textwidth]{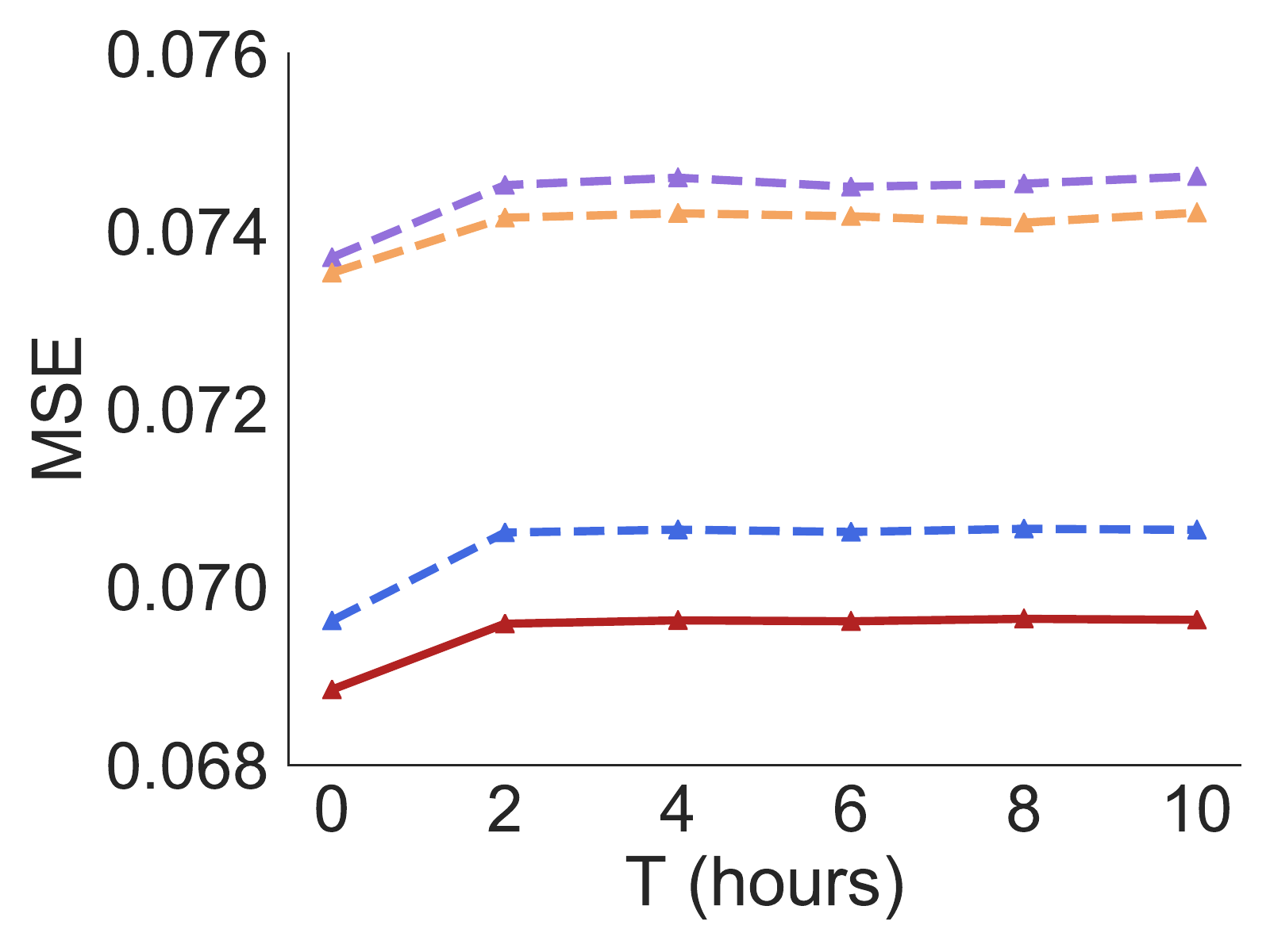}}
	\subfloat{ 	\includegraphics[width=0.20\textwidth]{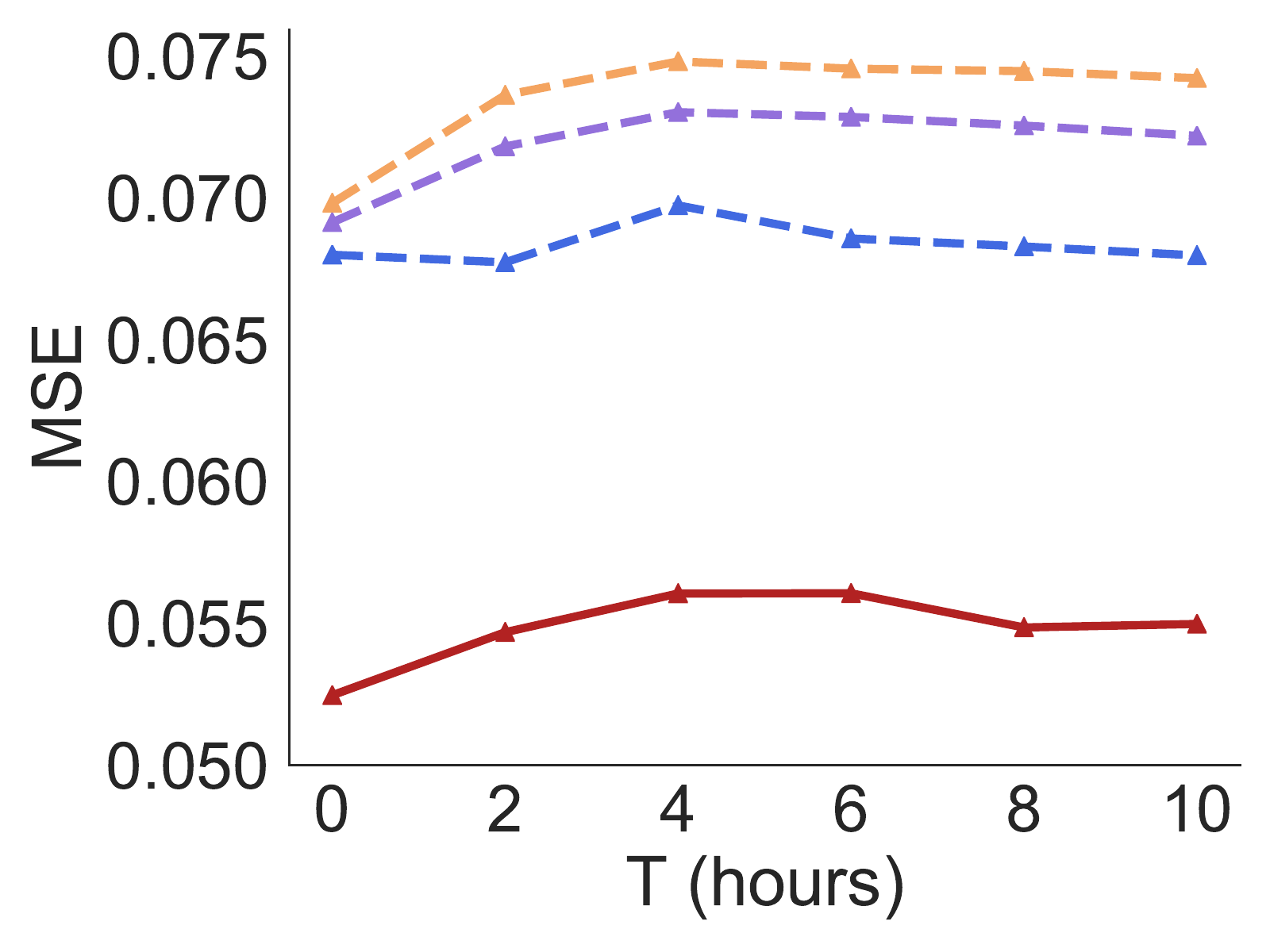}}
	%\subfloat{ 	\includegraphics[scale=0.20]{FIG_new/exp1_MsmallTwitter_MSE.pdf}}
	\subfloat{ 	\includegraphics[width=0.20\textwidth]{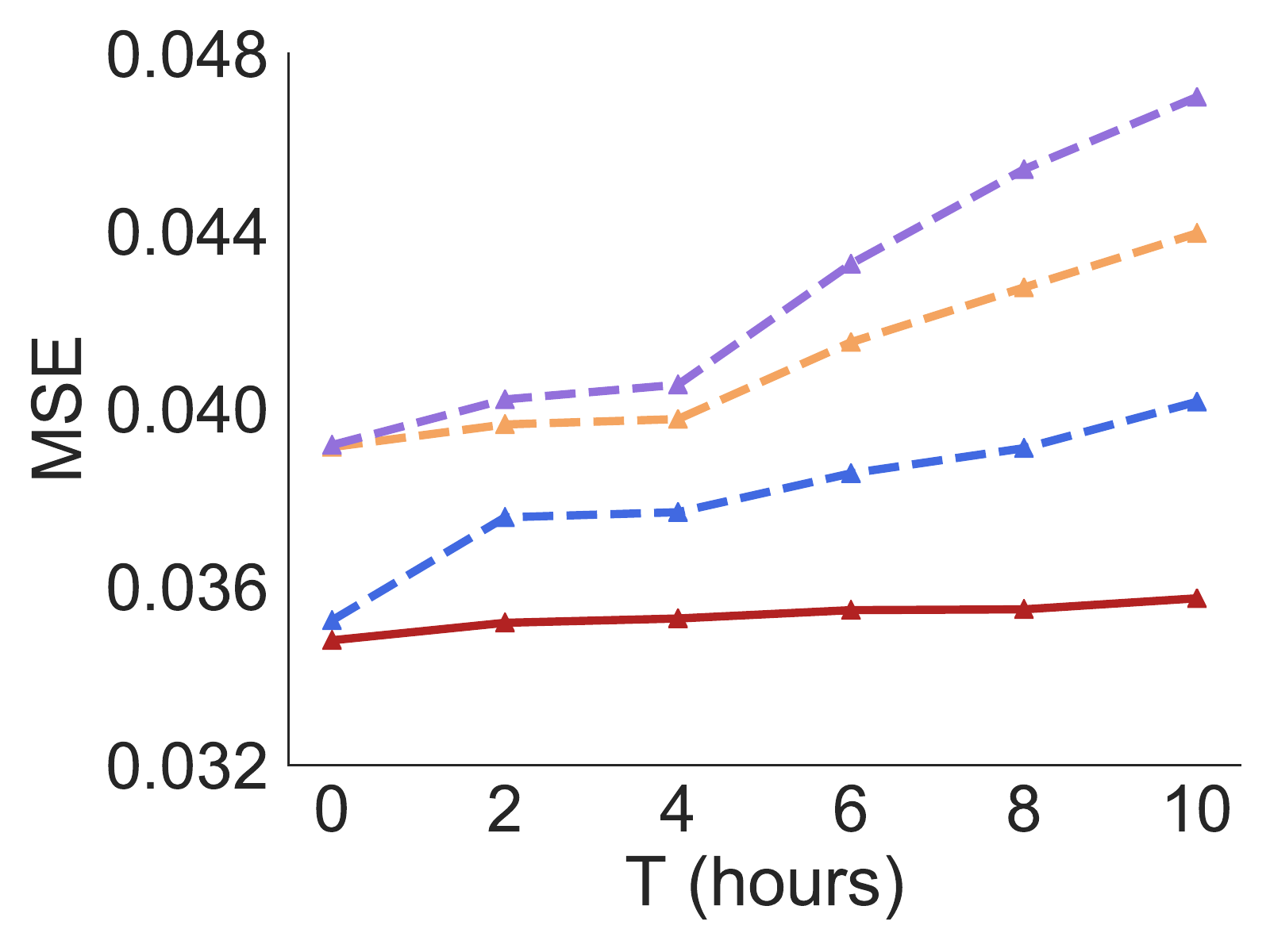}}
	%\subfloat{ 	\includegraphics[scale=0.20]{FIG_new/exp1_VTwitter_MSE.pdf}}
\\	
	\hspace{-8mm}	\subfloat[\barca]{ 	\includegraphics[width=0.20\textwidth]{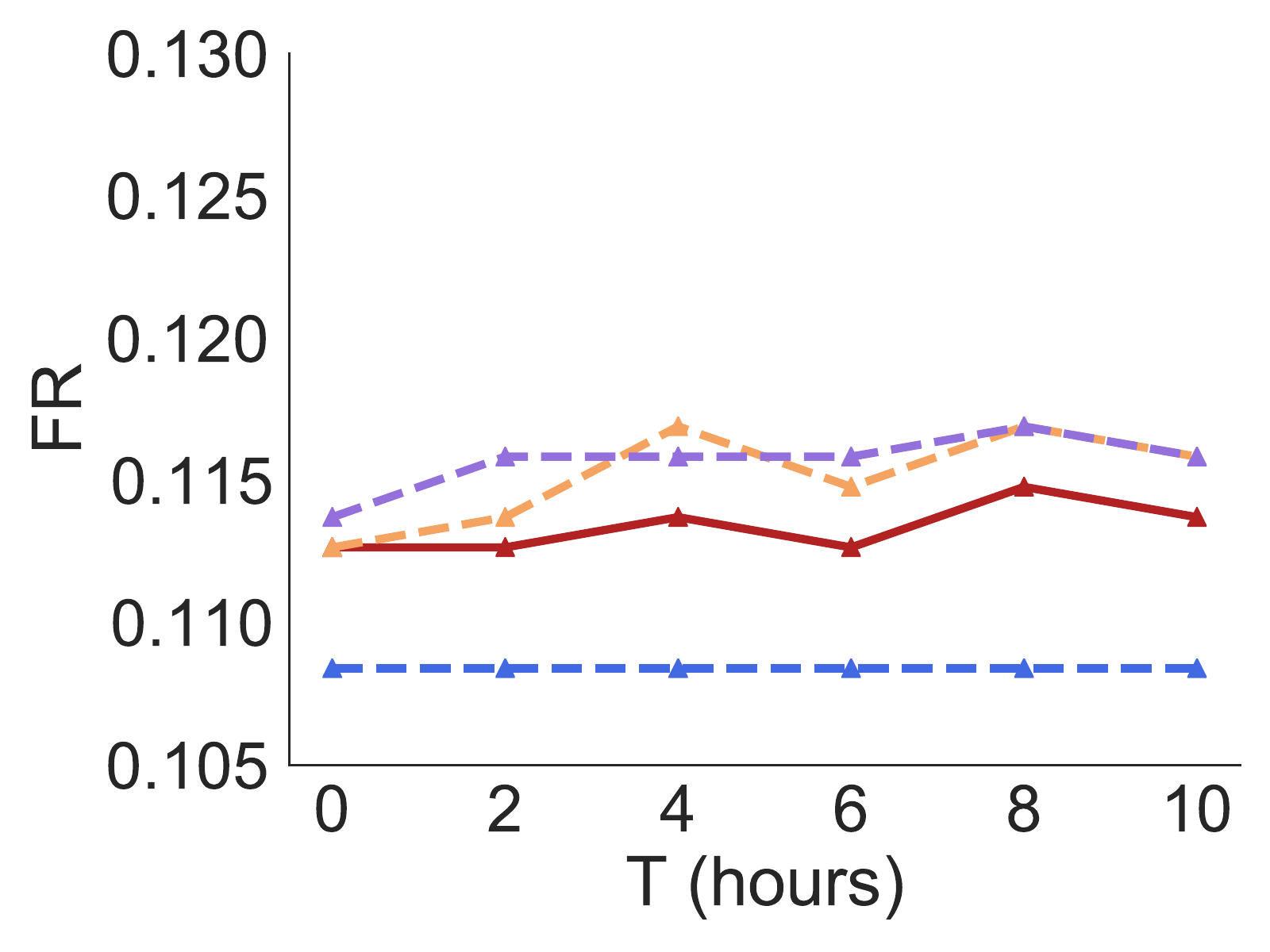}}
	\subfloat[\british]{ 	\includegraphics[width=0.20\textwidth]{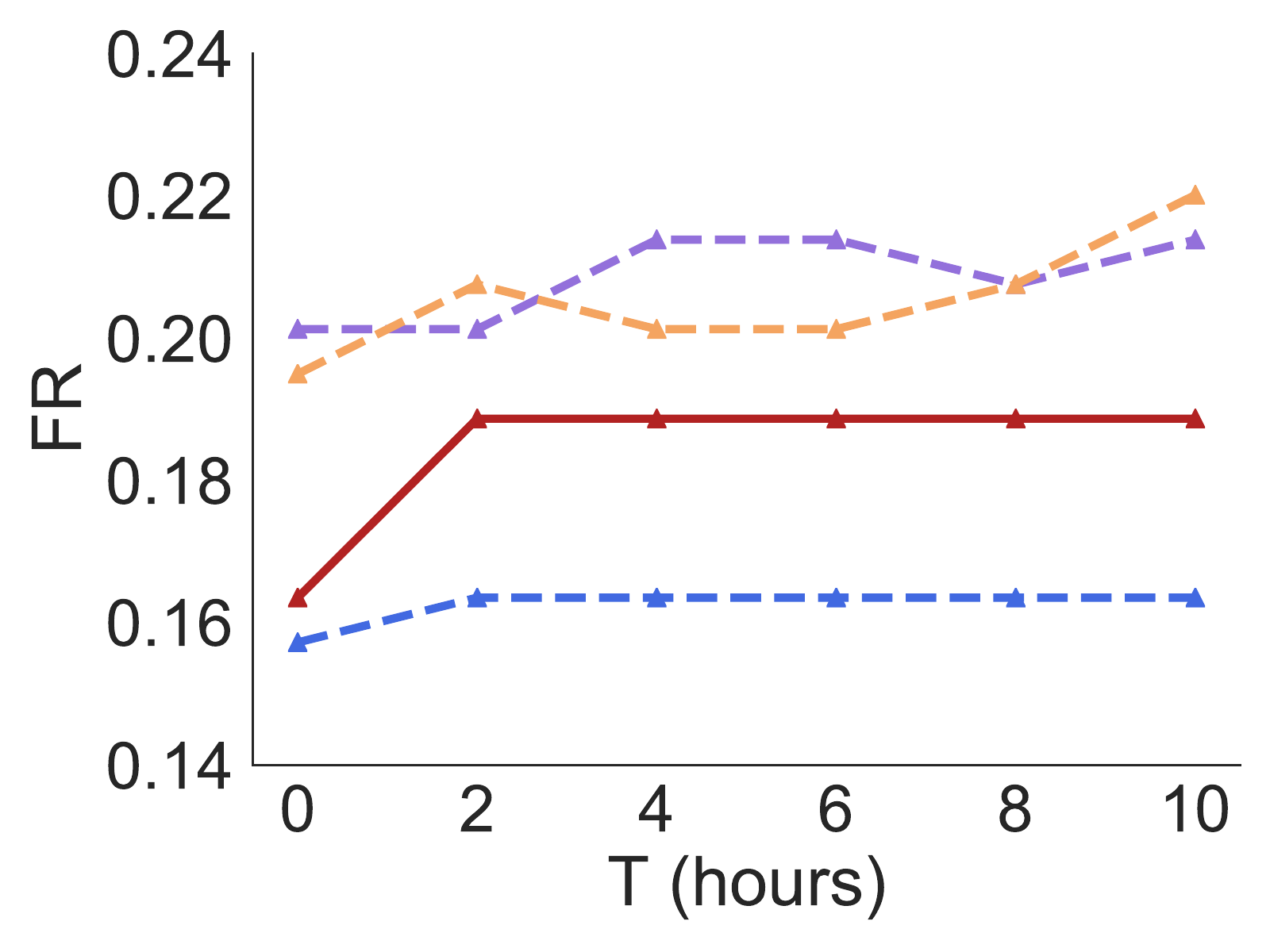}}
	%\subfloat{ 	\includegraphics[scale=0.20]{FIG_new/exp1_GTwitter_FR.pdf}}
	\subfloat[\jaya]{ 	\includegraphics[width=0.20\textwidth]{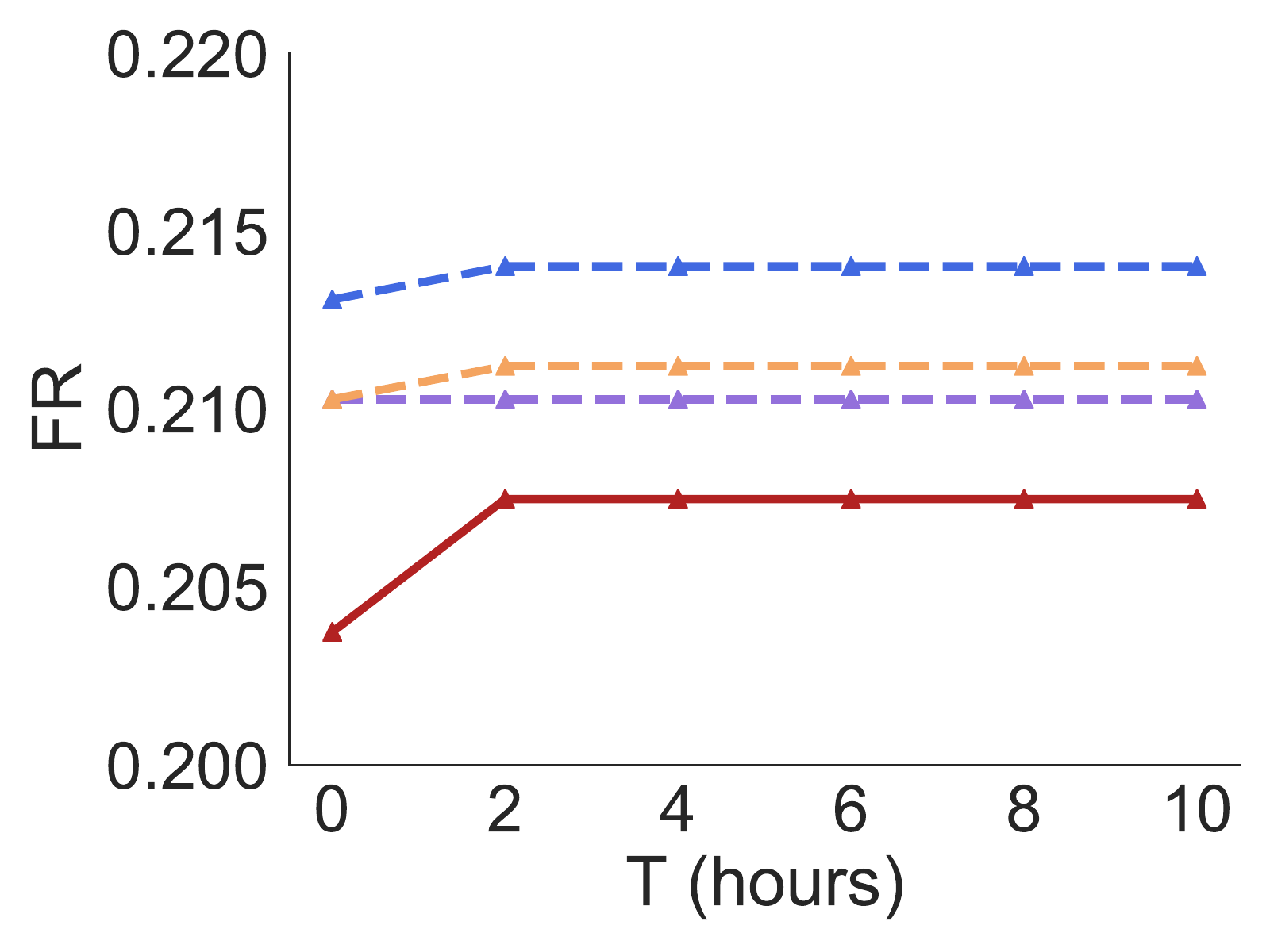}}
	%\vspace{-3mm}
	%\vspace{-3mm}
	\subfloat[\juv]{ 	\includegraphics[width=0.20\textwidth]{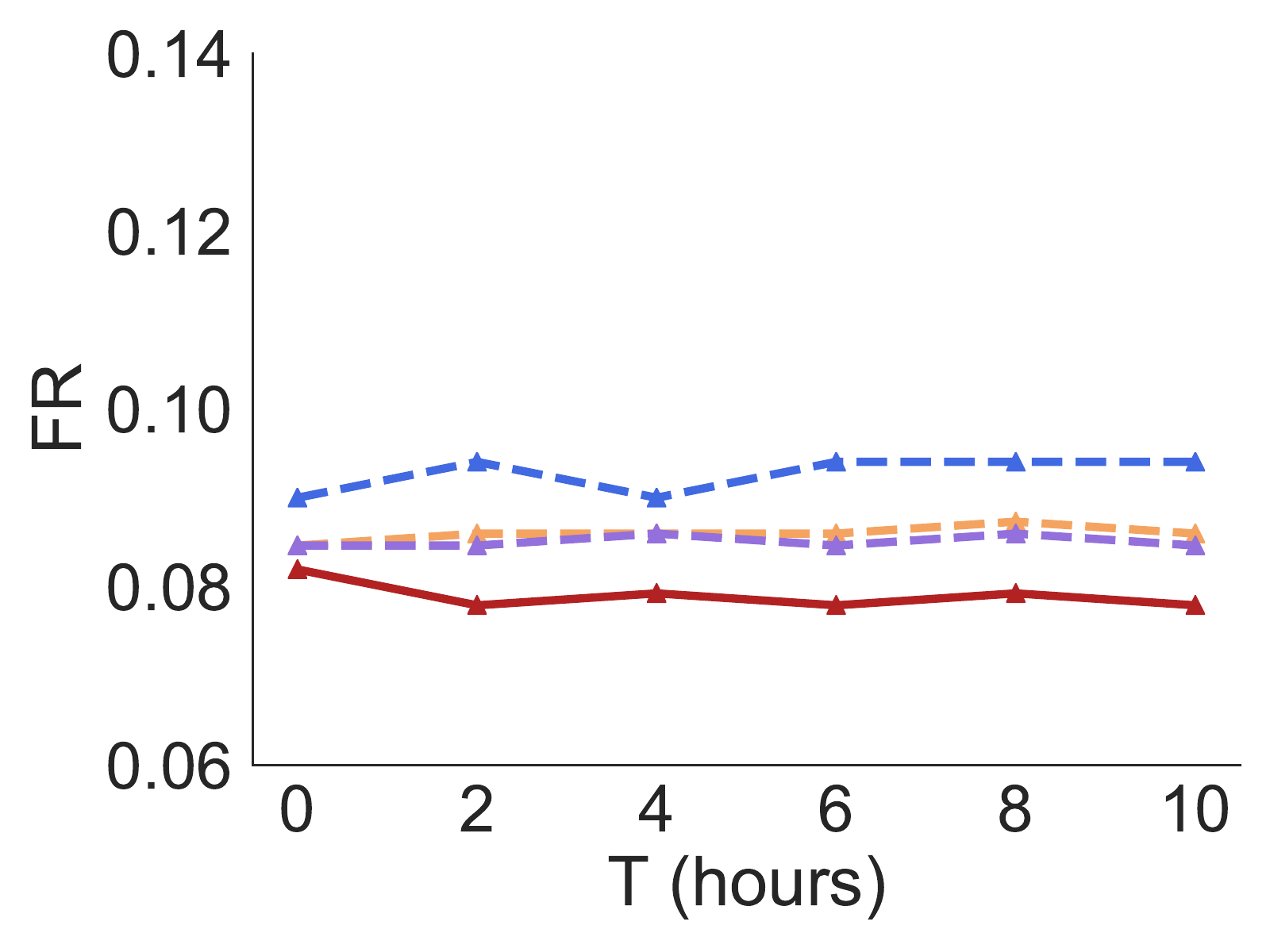}}
	%\subfloat{ 	\includegraphics[scale=0.20]{FIG_new/exp1_MsmallTwitter_FR.pdf}}
	\subfloat[\twitter]{ 	\includegraphics[width=0.20\textwidth]{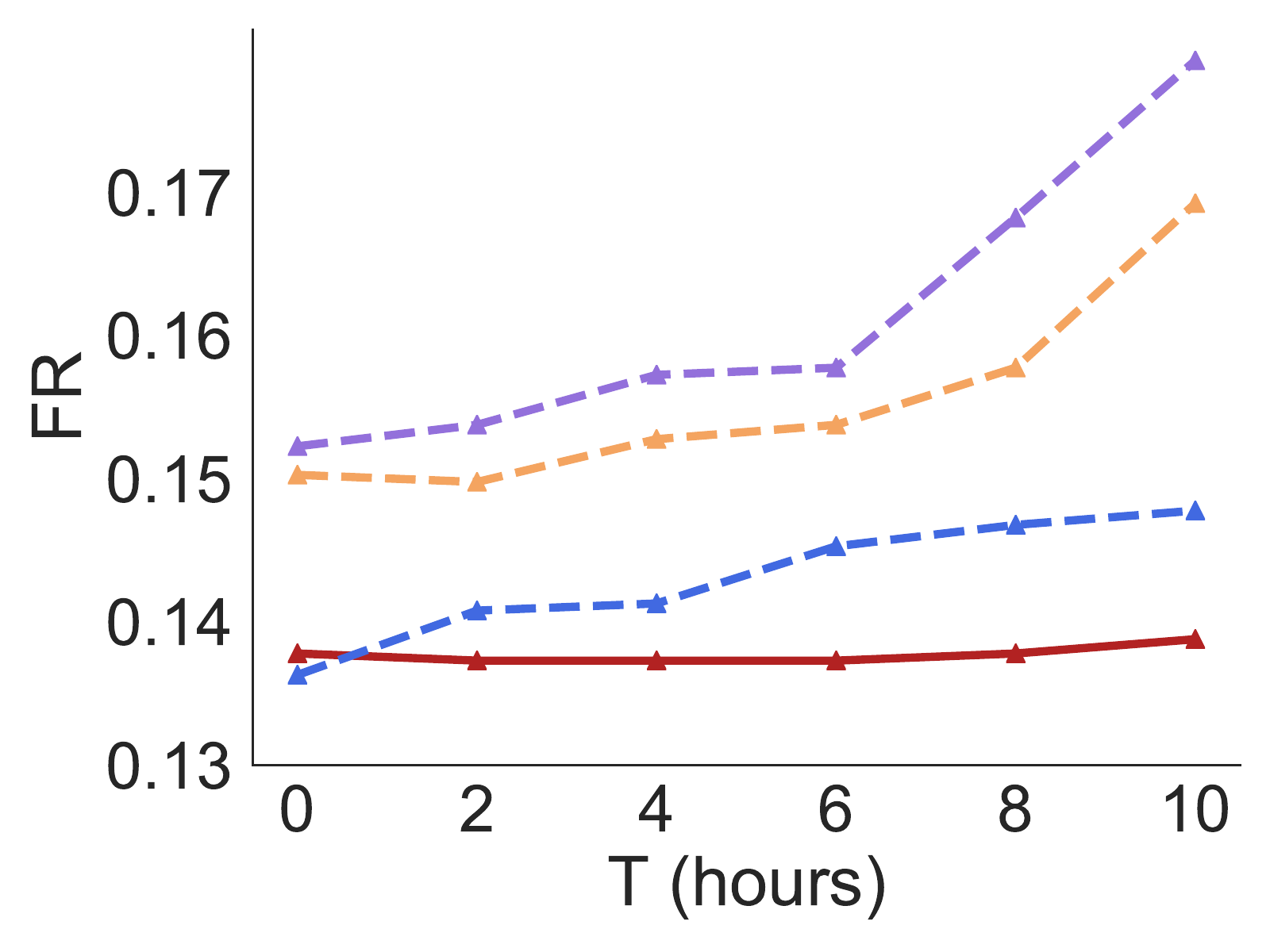}}
	%\subfloat{ 	\includegraphics[scale=0.20]{FIG_new/exp1_VTwitter_FR.pdf}}
	\vspace{3mm}
	
	% \subfloat{ 	\includegraphics[scale=0.30]{FIG_new/legend.png}}
	\caption{Sentiment prediction performance for five real-world datasets for the best performing variant of \our 
%		\as{Please clarify what you mean by ``one representative'': best performing variant?} 
		and three best performing baselines for a fixed $\gamma = 0.2$. For each message $m$ in test set, we predict its sentiment value given the history up to \T hours before the time of the message. For the \T hours, we predict the opinion stream using a sampling algorithm. Mean squared error and failure rate have been reported.
	We observe \our generally performs better than the baselines on all datasets. Among the baselines, \ourR performs comparably with \our on some of the datasets presented here.}
	\label{fig:forecast}
\end{figure}

\xhdr{Effect of sanitizing test data} 
Next, we address the research question (4).
To that aim,  we remove the outliers from the test data by refining entire datasets using variants of \our and then check the prediction performance of the previously computed model (in the experimental settings of Section \ref{sec:perf_eval}) over the refined test set. 
We compare the prediction error over this sanitized test set with the prediction error over the unrefined test set and report the improvement.
We report the prediction error in Table~\ref{tab:sanTest} (improvements over the error on the unrefined test set are given in brackets).
As observed from the results, for the majority of the cases, refining the test set improves prediction performance, confirming the presence of outliers in the test set which the estimated model will not be able to predict well.
Depending on the type and nature of the datasets, the error reduction varies, reaching its highest for \juv\ and lowest for \barca. Among our methods, we observe \oura and \ourd to perform better than the rest in terms of error reduction. 
In summary, the effectiveness of \our is more prominent after demarcating the test set. % effective

\begin{table}[h!]	
\centering
\resizebox{0.7\textwidth}{!}{\begin{tabular}{|c|c|c|c|c|c|}
			\hline
			Datasets & \oura & \ourd & \oure &  \ourt \\\hline
			& \multicolumn{4}{c|}{MSE} \\\hline
			\barca & 0.038 (1.78\%) & 0.044 (-16.\%) & 0.039 (-1.1\%) & 0.042 (-10.\%) \\\hline
			\british & 0.057 (-4.5\%) & 0.053 (1.48\%) & 0.051 (6.56\%) & 0.062 (-19.\%) \\\hline
			\jaya & 0.068 (2.04\%) & 0.069 (3.89\%) & 0.071 (0.51\%) & 0.074 (-4.4\%) \\\hline
			\juv & 0.059 (11.9\%) & 0.042 (24.8\%) & 0.057 (6.43\%) & 0.059 (17.1\%) \\\hline
			\twitter & 0.033 (6.41\%) & 0.039 (1.15\%) & 0.0 (100.\%) & 0.036 (1.41\%) \\\hline
			& \multicolumn{4}{c|}{FR} \\\hline
			\barca & 0.744 (0.29\%) & 0.140 (-23.\%) & 0.740 (1.52\%) & 0.650 (-7.1\%) \\\hline
			\british & 0.314 (13.9\%) & 0.168 (4.59\%) & 0.252 (28.4\%) & 0.140 (28.3\%) \\\hline
			\jaya & 0.685 (0.32\%) & 0.204 (2.94\%) & 0.701 (-0.4\%) & 0.694 (-0.0\%) \\\hline
			\juv & 0.062 (22.4\%) & 0.071 (10.9\%) & 0.051 (20.4\%) & 0.051 (43.8\%) \\\hline
			\twitter & 0.474 (19.2\%) & 0.143 (2.70\%) & 0.0 (100.\%) & 0.480 (0.83\%) \\\hline

			\hline
		\end{tabular}
		}
		\caption{ Mean squared error and failure rate of \ourx on all the datasets after demarcating exogenous events from the test set. Error reduction (from error reported on the entire test set) is given in brackets. After demarcating the test set, in the majority of the cases, error reduction is positive, indicating that the error on an unfiltered test set gives an overestimation of the true error. 
			\label{tab:sanTest}}

\end{table}

\xhdr{Variation of performance with training set size} 
Next, we answer the research question (5).
Specifically, we evaluate the efficacy of our approaches over varying training set sizes as follows. We use a subset of training data, varying from initial $50\%$ of the total training set to entire $100\%$ for estimating the parameters and test the estimated model on the same test data used in all the experiments, which is the last $10\%$ data of the entire stream of events. 
Figure \ref{fig:vary_train} summarizes the results over three datasets (\barca, \british and \juv) out of the five real datasets, which reveals the following observations. (I) The performance deteriorates with decreasing training set size for all methods. (II) Our methods generally perform better compared to the baselines over varying training set sizes showing the effectiveness of the demarcation technique. (III) For \british where the total number of events is quite small compared to the rest, \ourR performs better than the variants of \our, indicating hard thresholding to be quite effective for demarcation.
(IV) The performance of \our for smaller sample size, relative to its competitors, indicates its stability and robustness.

\begin{figure}[!t]
	\centering
	\subfloat{ 	\includegraphics[scale=0.75]{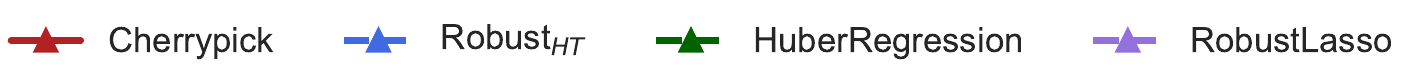}}
	\vspace{-3mm}
	\subfloat[\barca]{\includegraphics[width=0.30\textwidth]{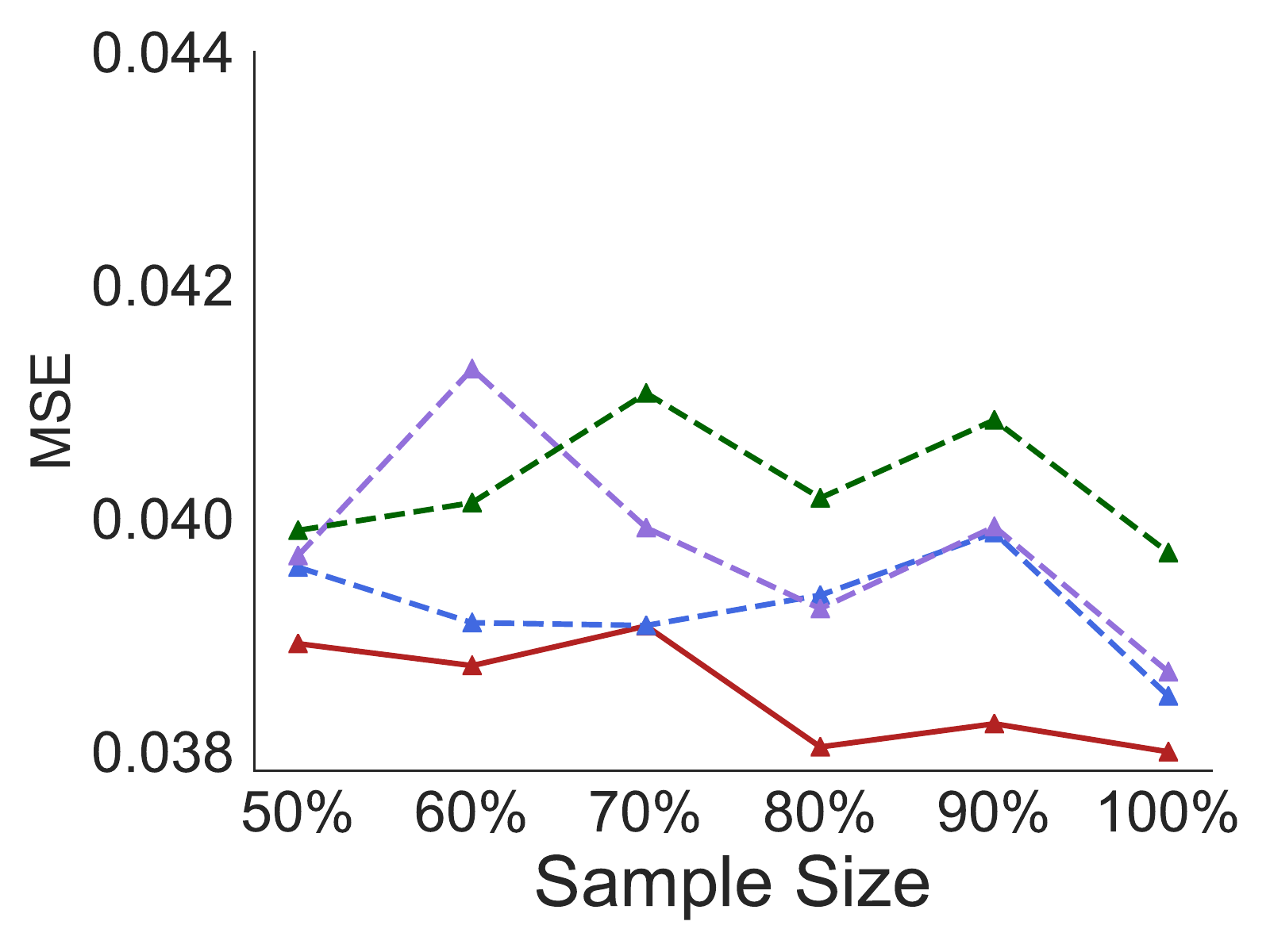}}\hspace*{0.2cm}
	\subfloat[\british]{\includegraphics[width=0.30\textwidth]{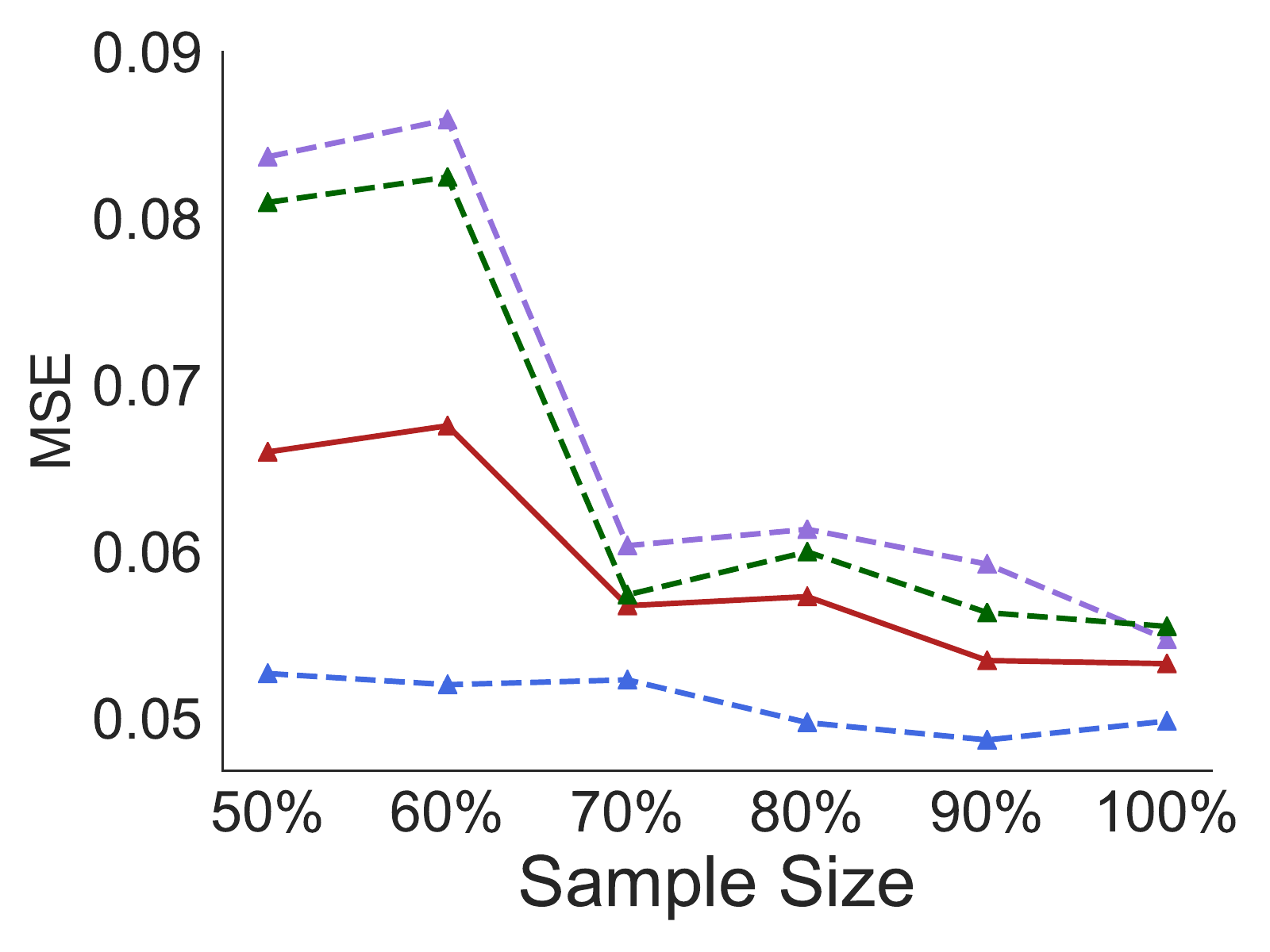}}\hspace*{0.2cm}
	\subfloat[\juv]{\includegraphics[width=0.30\textwidth]{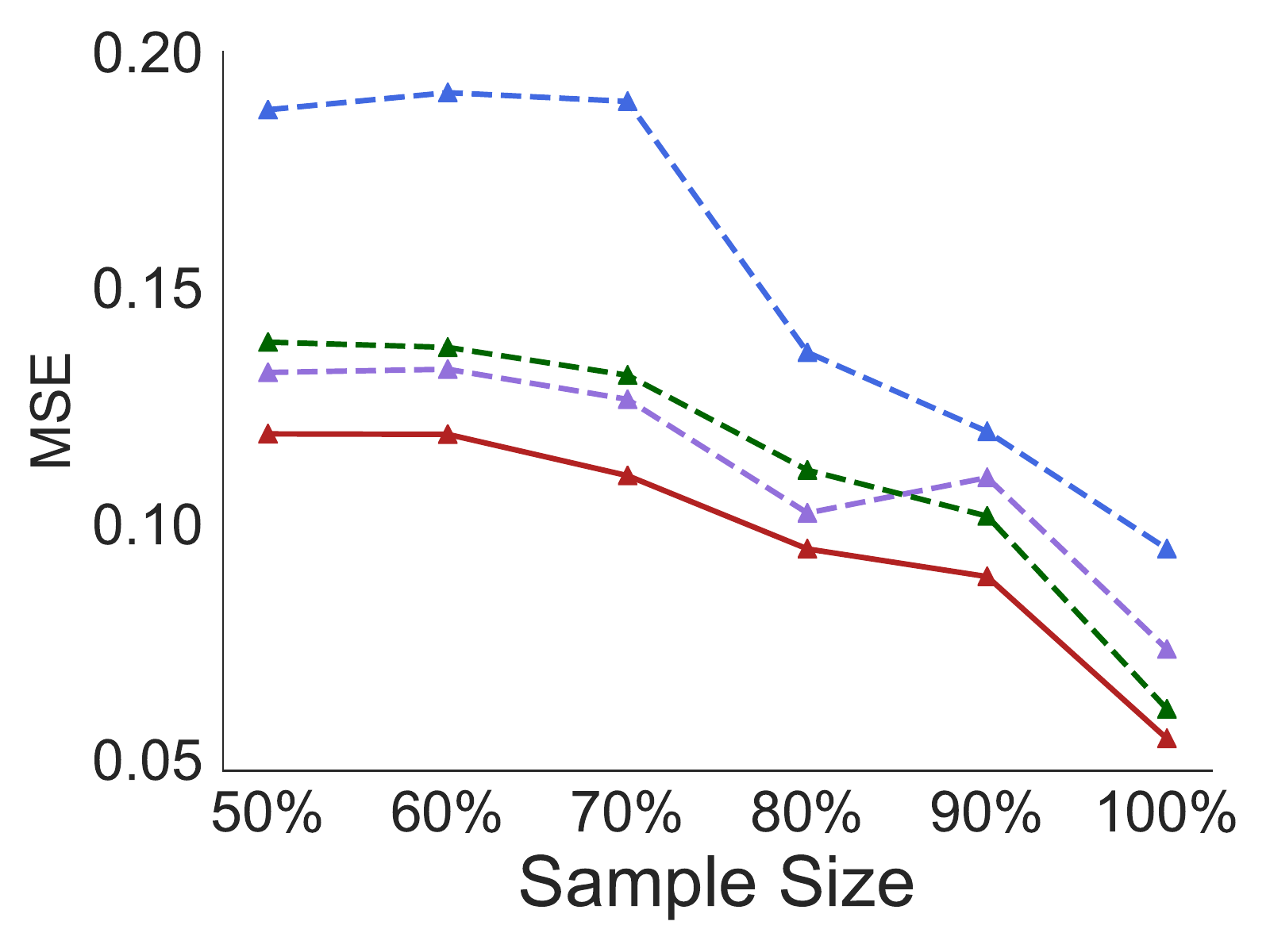}}\hspace*{0.2cm}
	\vspace{-3mm}
	
	\subfloat[\barca]{\includegraphics[width=0.30\textwidth]{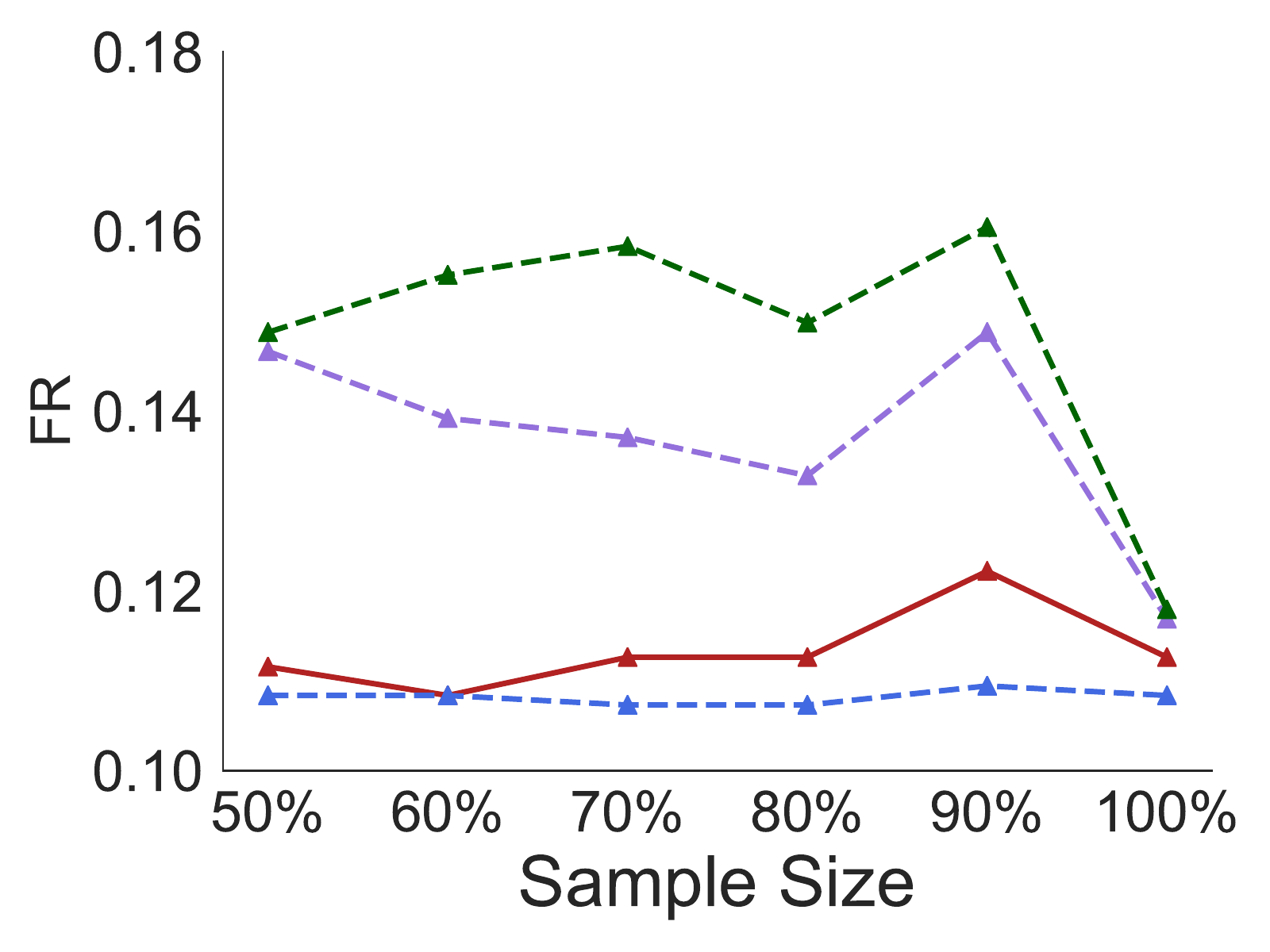}}\hspace*{0.2cm}
	\subfloat[\british]{\includegraphics[width=0.30\textwidth]{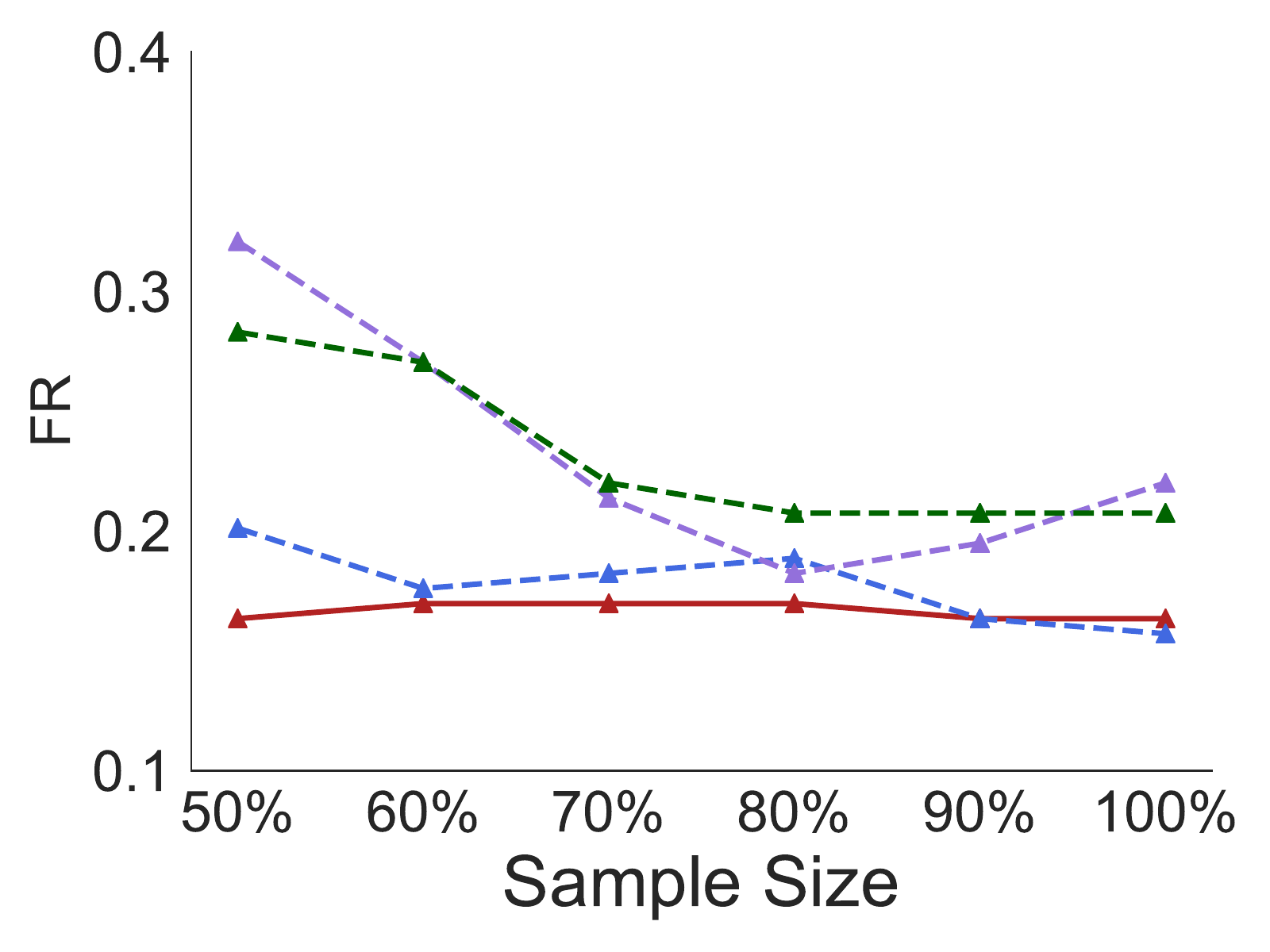}}\hspace*{0.2cm}
	\subfloat[\juv]{\includegraphics[width=0.30\textwidth]{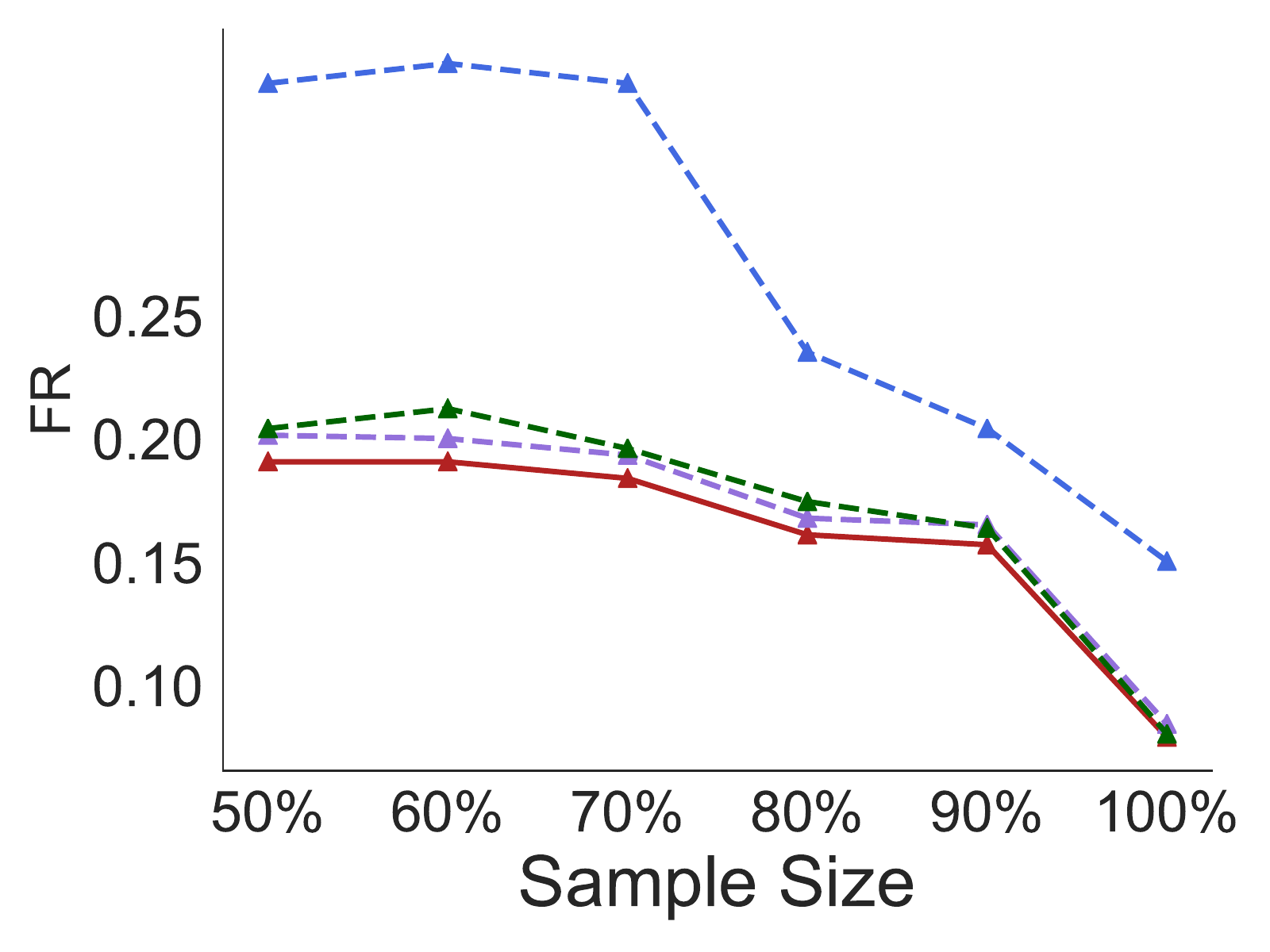}}\hspace*{0.2cm}
	\vspace{-3mm}
	\caption{Mean squared error and failure rate on three datasets \barca, \british and \juv for varying size of training data from 50\% to 100\% of total training data (initial 90\% data of entire event collection) for the best performing variant of \our 
%		\as{Please clarify what you mean by ``one representative'': best performing variant?} 
		and three best performing baselines. Time span for future prediction \T is set to 4 hours and $\gamma$ = 0.2. For all methods, error reduces with increasing training set size. Generally on the entire range, \our performs better than the baselines.}
	\label{fig:vary_train}
\end{figure}

\section{Experiments with synthetic data}
\label{sec:expt_syn}

In this section, we provide a comparative evaluation
of the four variants of $\our$ with baselines on a set of three synthetic datasets. 
To that aim, we address the following research questions here:
\begin{enumerate} 
	\item How do the variants of $\our$ compare in terms of parameter estimation performance across different sizes of training set?
	\item How does their predictive performance vary across different levels of noise present in the data?
\end{enumerate}

\subsection{Dataset generation}
For constructing the synthetic datasets, we have generated the following networks, each with $512$ nodes to use as input to the opinion model (SLANT)~\cite{nipsxx}.
\begin{itemize}
%	\item Kronecker Homophily network$\colon$ In this network, the nodes tend to link to nodes with similar degree by setting high value entries in the diagonal of the parameter matrix (parameter matrix [0.96, 0.3; 0.3, 0.96]).
%	\item Kronecker Heterophily network$\colon$ Here the nodes tend to link to nodes with different degrees and that is done by setting high-value entries off the diagonal (parameter matrix [0.3, 0.96; 0.96, 0.3]). 
	\item Kronecker Core-Periphery networks$\colon$ In this Kronecker network, the core is very well connected and there are comparatively fewer nodes in the periphery, sparsely connected with the core (parameter matrix [0.9, 0.5; 0.5, 0.3]).
	\item Kronecker Random networks: This Kronecker network has been generated using the parameter matrix [0.5, 0.5, 0.5, 0.5]. 
%	\item Kronecker Hierarchical networks$\colon$ \as{There should be some description here.}
	\item Barabasi Albert$\colon$ This is a scale-free network where the network grows iteratively with this preferential attachment property that the more well-connected nodes are more likely to receive new links.
\end{itemize}

The message stream is generated for each of the networks by simulating the message sampling algorithm proposed in ~\cite{nipsxx}. Each user starts with a latent opinion ($\alpha_i$) as well as non-zero opinion influence parameters corresponding to directed edges of the network ($A_{ij}$) sampled from a zero-mean unit variance Gaussian distribution. The rate of intensity ($\mu_i$), as well as the intensity influence parameters ($B_{ij}$), are uniformly sampled from the range $[0,1]$. Opinions are sampled from unit variance Gaussian with the mean set as the latent opinion of the user. Using multivariate Hawkes, the events are generated. While generating event times, each event is marked as exogenous with 20\% probability and if marked exogenous, it is sampled by a distinct distribution (described later).
%no longer sampled from latent opinion, instead it is sampled using from Gaussian with $0.1$ variance and mean sampled separately from a zero-mean-unit-variance Gaussian. 
The kernel used in exponential decay of the opinion ($\omega$) is $1000$ and the kernel for the intensity ($\mu$) is $10$. 

\subsection{Baselines}\label{sec:synthetic_baselines}

We compare our proposed approaches with the same baselines introduced in Section \ref{sec:baselines}.

\subsection{Evaluation protocol}\label{sec:synthetic_eval_protocol}

The evaluation protocol used is the same as in the case of real datasets (Section \ref{sec:eval_protocol}).

\subsection{Metrics}

We measure the performance of our methods and the baselines using the sentiment prediction errors and parameter estimation error of the correspondingly trained endogenous model, depending on corresponding experiment. 
Specifically, we measure the sentiment prediction error using the \emph{mean squared error (MSE)} between the actual sentiment value ($m$) and the estimated  sentiment value ($\hat{m}$), \ie, $\EE[(m-\hat{m})^2]$ and the parameter estimation error using the \emph{mean squared error (MSE)} between the estimated ($\bar{x}$)and true opinion parameters ($x$), i.e., $E[(x-\bar{x})^2 ]$.

\subsection{Results} 
\xhdr{Variation of performance with sample size} 
Here, we investigate the research question (1). More specifically, to understand the effect of sample size on parameter estimation, we vary 
the number of events per node from 20 to 400.
$20\%$ of the messages are perturbed as exogenous events and they are sampled from a Gaussian distribution with $0.1$ variance and mean sampled separately from a zero-mean-unit-variance Gaussian.

Figure \ref{fig:synTrain} summarizes the results for all variants of \our along with three best performing baselines, which reveals following observations. (I) \ourd and \ourt are able to boost parameter estimation performance across a wide range of training set size, along with showing their robustness and stability with decreasing sample size. Hoewever, \oura and \oure are observed to perform poorly over the entire range.(II) \ourR shows very comparable performance with best performing variants of \our. 
We can conclude that \ourd and \ourt are able to identify more useful samples for accurately estimating the parameters compared to \oura and \oure.  This is because \oura and \oure suffer from their weak submodularity property, which renders them disadvantageous due to high SNR of the datasets (40-50 dB) \cite{chamon2017approximate}.  

\begin{figure}[!h]
	
	\centering
	\subfloat{ \includegraphics[scale=0.75]{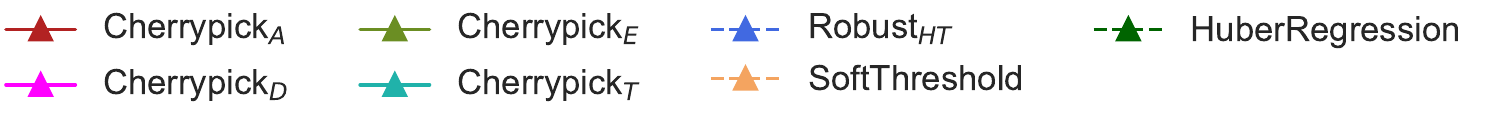}}
	\vspace{-3mm}
	
	\subfloat[\alberta]{ 	\includegraphics[width=.25\textwidth]{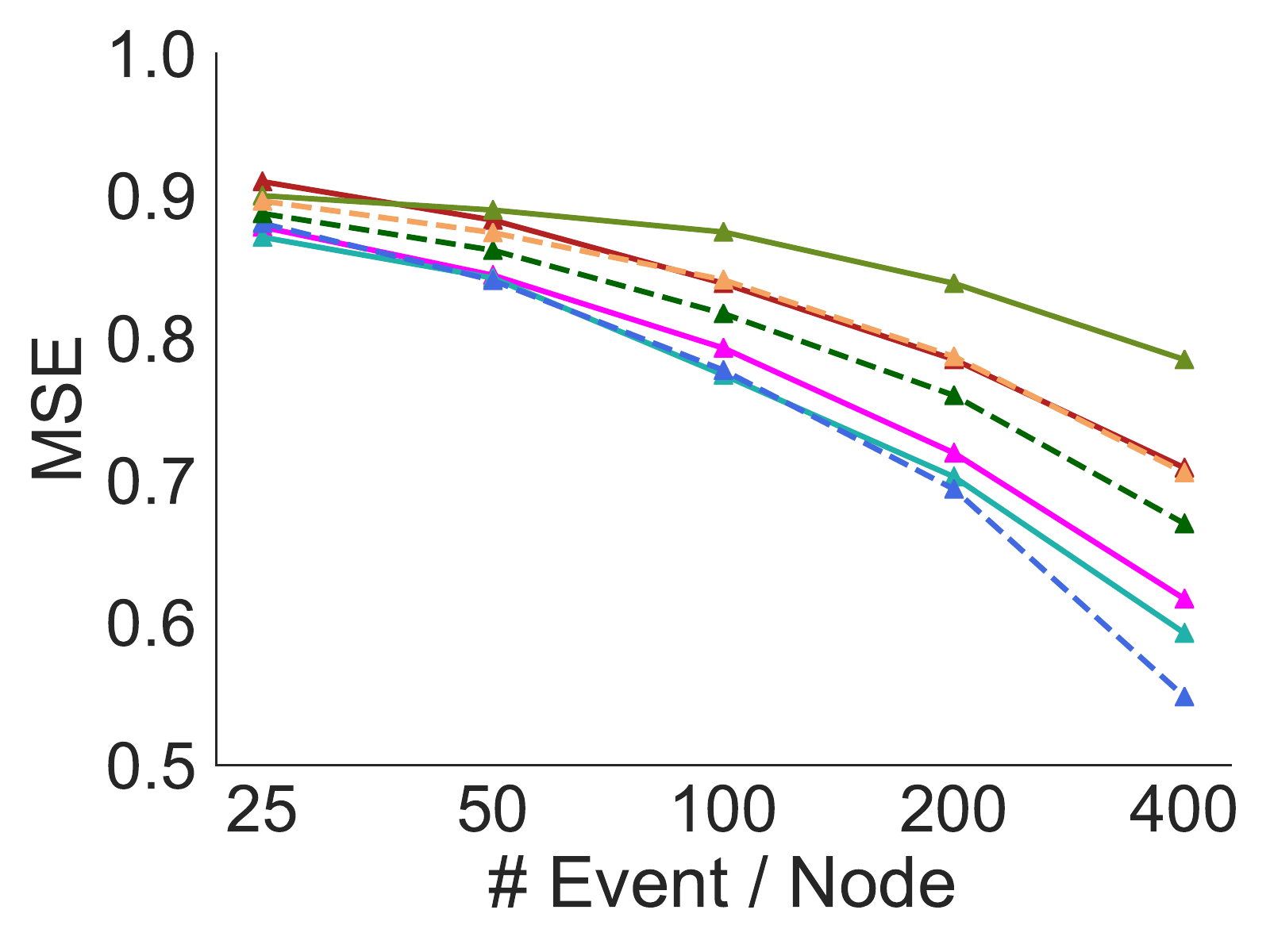}}\hspace{1cm}
	\subfloat[\kcp]{ 	\includegraphics[width=0.25\textwidth]{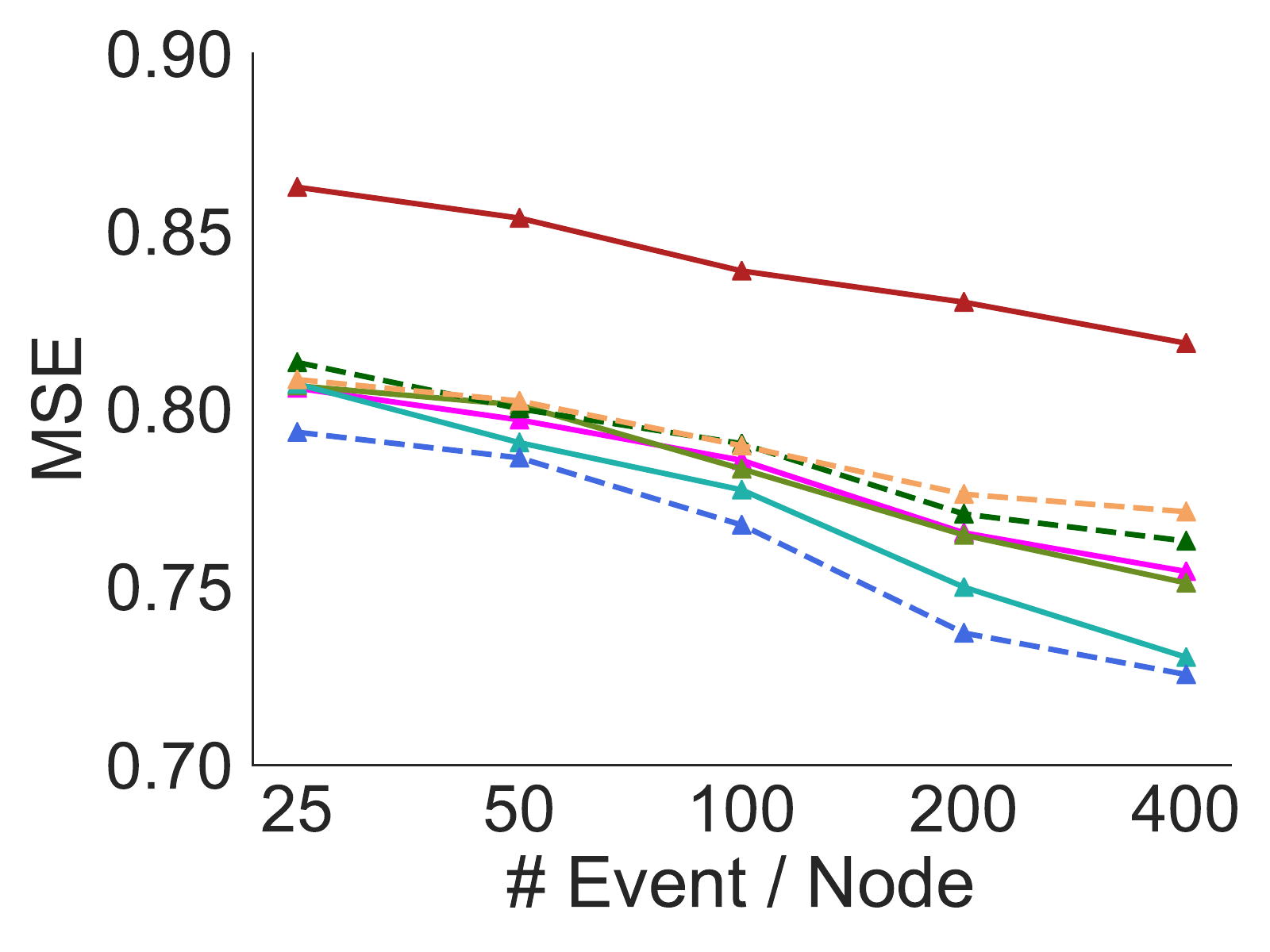}}\hspace{1cm}
%	\subfloat[\khier]{ 	\includegraphics[width=0.25\textwidth]{FIG_new/exp5_i_hier_k_MSE.pdf}}
%	\vspace{3mm}
	\subfloat[\krnd]{\includegraphics[width=0.25\textwidth]{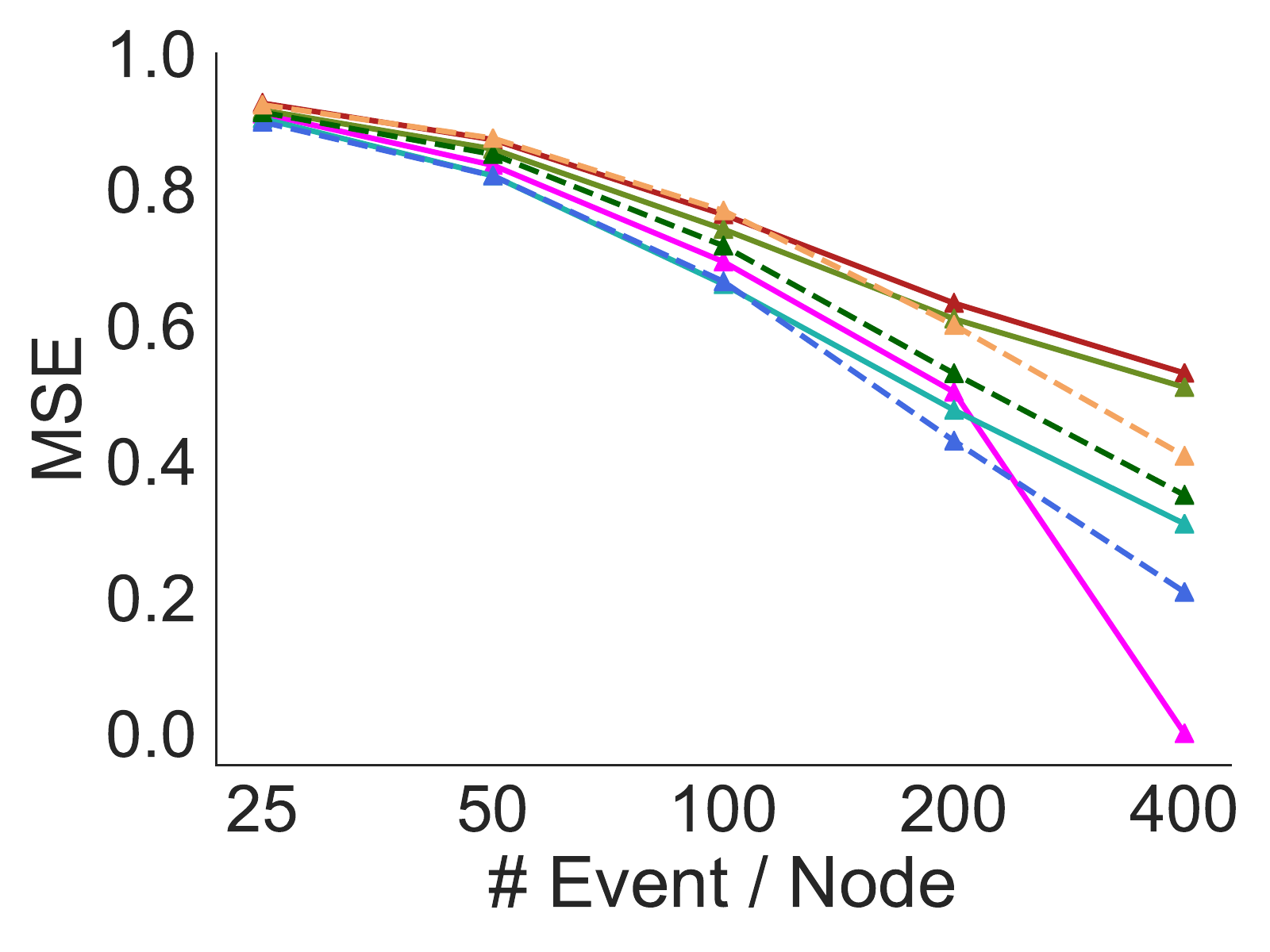}}\hspace{1cm}
%	\subfloat[\khomo]{\includegraphics[width=0.25\textwidth]{FIG_new/exp5_m_homo_k_MSE.pdf}}\hspace{1cm}
%	\subfloat[\khetero]{\includegraphics[width=0.25\textwidth]{FIG_new/exp5_r_hetero_k_MSE.pdf}}
	\vspace{3mm}
	
	\vspace{-3mm}
	\caption{Mean squared error for parameter estimation on three synthetic datasets against varying training set size. Best performing variants of \our always achieve comparable (if not better) performance with best of the baselines. }
	\label{fig:synTrain}
\end{figure}
\xhdr{Variation of performance with noise} 
Here we address research question (2). In particular, to have a better understanding of the effect of noise intensity in the performance of \our, we gradually increase the noise intensity in the message stream in all synthetic datasets and report the sentiment prediction performance of all competing methods. 
%Here we compare our proposed demarcation methods with baselines while .  
$30000$ events are sampled for each network and $20$\% of them are perturbed by adding noises of increasing intensity. The noise is sampled from Gaussian with variance set at $0.05$ and mean varying in the range  $[0.5, 0.75, 1.0 , 1.5, 2.0, 2.5 ]$. 

Figure \ref{fig:synPred} summarizes the results for all variants of \our along with three best performing baselines, where it shows, (i) as prediction error increases sharply with increasing noise for all the methods, \ourd or \ourt outperform or perform comparably with the  baselines, (iii) interestingly \ourR performs comparably with \ourx in majority of the cases. As we already mentioned, for all the cases, the performance of \oura or \oure suffers because their weak submodularity property which renders them ineffective due to high SNR of the corresponding datasets (40-50 dB) \cite{chamon2017approximate}.  %\as{Mention that this is due to weak submodularity of \oura and \oure. Verify by SNR and correlation analysis.}

\begin{figure}[!h]
	\centering
	
	\subfloat{\includegraphics[scale=0.75]{FIG_new/exp5_legend_short.pdf}}
	\vspace{-3mm}
	
	\subfloat[\alberta]{\includegraphics[width=0.25\textwidth]{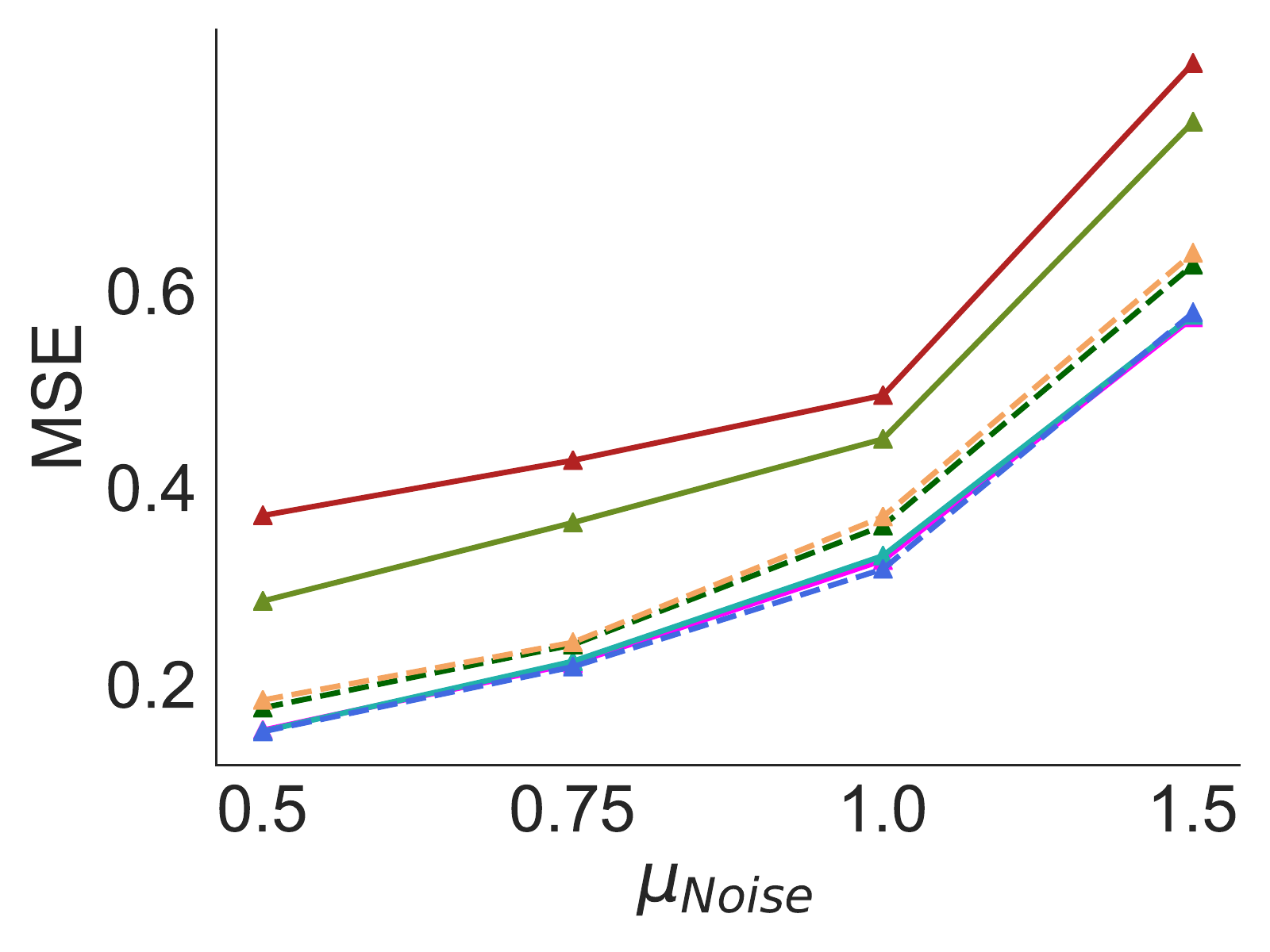}}\hspace{1cm}
	\subfloat[\kcp]{\includegraphics[width=0.25\textwidth]{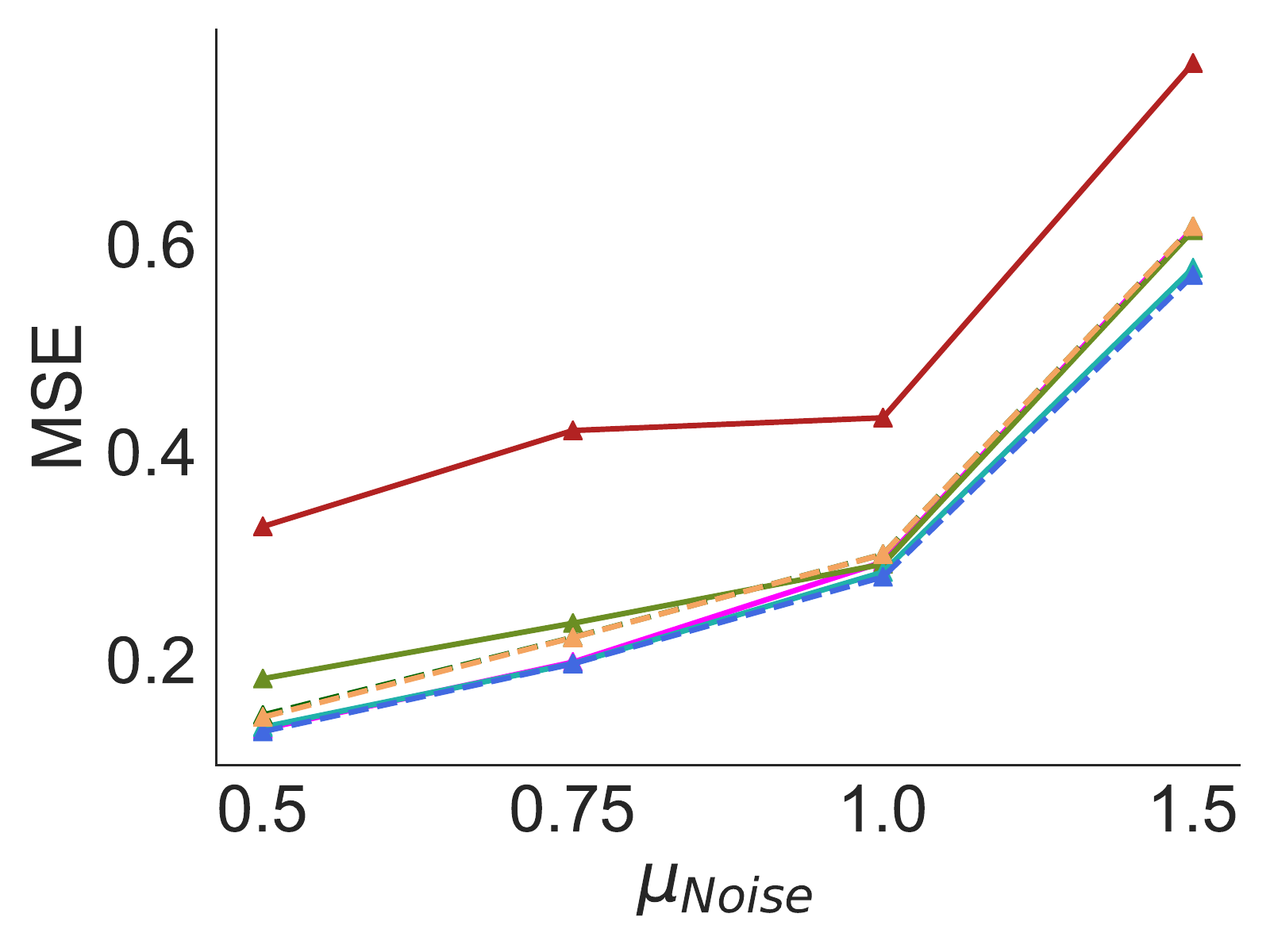}}\hspace{1cm}
%	\subfloat[\khier]{\includegraphics[width=0.25\textwidth]{FIG_new/exp6_i_hier_k_MSE.pdf}}\hspace{1cm}
%	
%	\vspace{3mm}
	\subfloat[\krnd]{\includegraphics[width=0.25\textwidth]{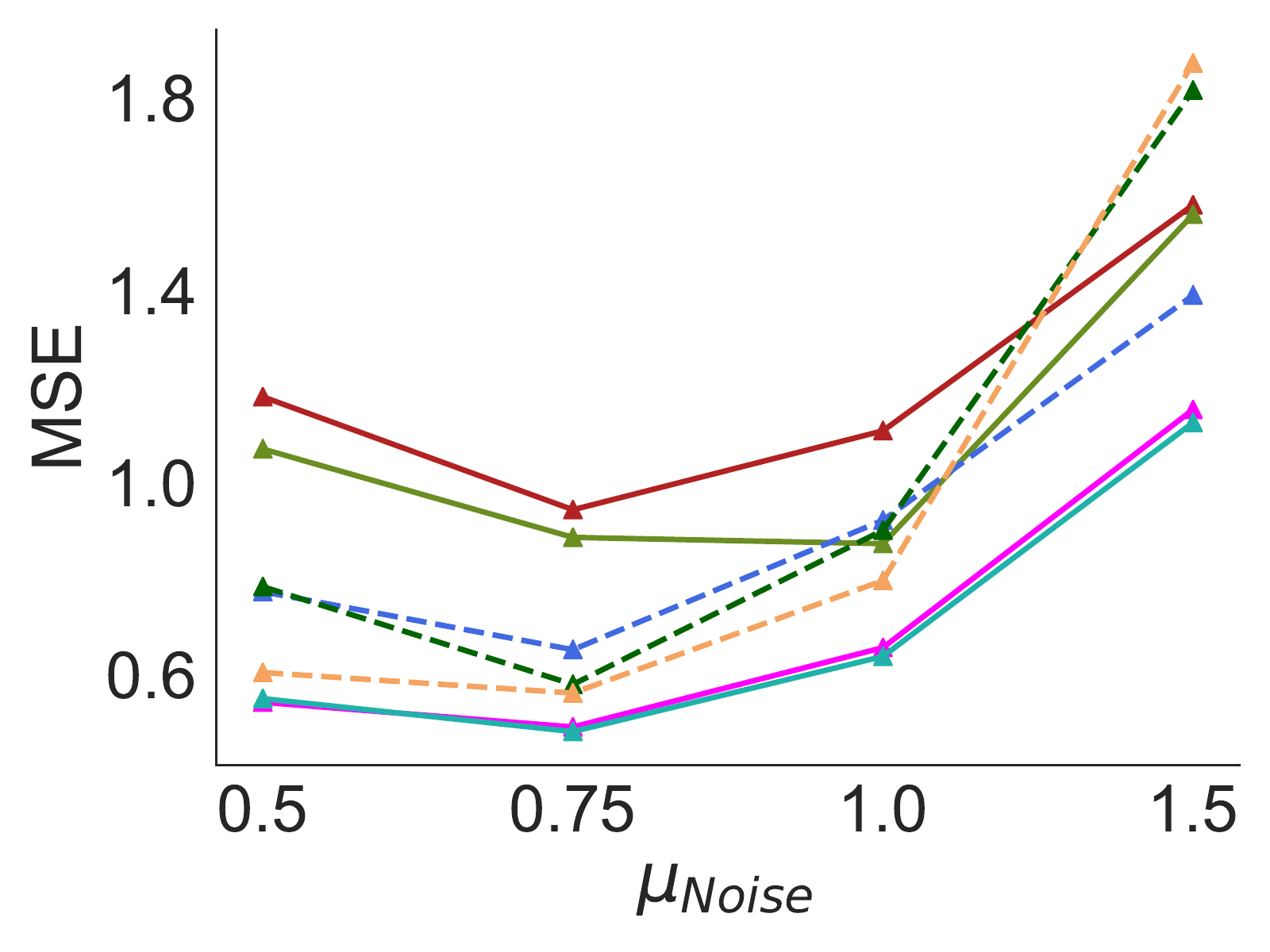}}\hspace{1cm}
%	\subfloat[\khomo]{\includegraphics[width=0.25\textwidth]{FIG_new/exp6_m_homo_k_MSE.pdf}}\hspace{1cm}
%	\subfloat[\khetero]{\includegraphics[width=0.25\textwidth]{FIG_new/exp6_r_hetero_k_MSE.pdf}}
	\vspace{3mm}
	
	\vspace{-3mm}
	\caption{Sentiment prediction error in terms of mean squared error on three synthetic datasets against varying intensity of noise. Best performing variants of \our degrade gracefully in comparison with the baselines for most of the cases.}
	\label{fig:synPred}
\end{figure}

\vspace*{-0.2cm}
\section{Conclusion}
The principal contribution of this paper lies in emphatically establishing the dual nature of message flow over online social network: injection of exogenous opinions and influence-based dynamics, internal to the network. The realization helps us to propose \ourx, a set of novel learning methodologies to demarcate endogenous and exogenous opinions and illustrate their utility by analyzing their performance from an opinion modeling perspective.
In \ourx, to this aim, we formulated the message classification problem as a submodular optimization task in the set of messages, which we solved using an efficient greedy algorithm. 
Our proposed techniques are very easy to implement, extremely scalable (particularly \oura\ and \ourt) and quite effective in serving their purpose, showing their superiority over the baselines which were designed by robust regression literature.
Finally, on various real datasets crawled from Twitter as well as synthetic datasets, we showed that our proposals consistently outperform various outlier removal algorithms in terms of predictive performance. The superior performance is even more remarkable considering the fact that we train our system on smaller (but relevant) amounts of data. 
%

%\begin{acks}
%To Robert, for the bagels and explaining CMYK and color spaces.
%\end{acks}
%\fi 
\bibliographystyle{ACM-Reference-Format}
\bibliography{demarcate}

\appendix
\appendix
\onecolumn
%{\huge Supplementary material}
\section{Proofs of results}% in Section~\ref{lem:covv}}
\label{sec:appen}

\subsection{Proof of Lemma~\ref{lem:covv}}\label{sec:proof1}
For any $u\in\Vcal$, define $\thetab_u=[\Ab_u;\alpha_u]$. We observe the loss (Eq.~\ref{eq:loss1}) associated with only user $u$ is a regularized least squares loss, \ie,
\begin{align}
\hat{\thetab}_u=\hspace*{-0.1cm}\underset{\thetab_u}{\text{min}}\hspace*{-0.1cm} & \hspace*{-0.1cm}\sum_{e_i\in \HH}\hspace*{-0.2cm}\sigma^{-2} \Big(m^*_{u}(t_i) -\phib_i ^{uT} \thetab_u\Big)^2 +c||\thetab_u||_2 ^2, 
\end{align}
%%%
\begin{align*}
\hat{\thetab}_u&=\Big(c\Ib+\Vxx\Big)^{-1}\sigma^{-2}\sum_{e_i\in\HH}m_u(t_i)\phib^u_i\nn\\
&=\Big(c\Ib+\Vxx\Big)^{-1}\Vxx \thetab_u + \Big(c\Ib+\Vxx\Big)^{-1}\sum_{e_i\in \HH}\phib^u _i\frac{\epsilon(t_i)}{\sigma^2}
\end{align*}
The equality follows from the fact that, \ $m_u(t_i)=\phib^{uT} _i \thetab_u+\epsilon(t_i) $
\begin{align*}
\hat{\thetab}_u-\thetab_u&=-c\Big(c\Ib+\Vxx\Big)^{-1} \thetab_u + \Big(c\Ib+\Vxx\Big)^{-1}\sum_{e_i\in \HH}\phib^u_i\frac{\epsilon(t_i)}{\sigma^2}
\end{align*}
Then the covariance product is given in the following:
\begin{align*}
\EE(\hat{\thetab}_u-\thetab_u)(\hat{\thetab}_u-\thetab_u)^T&=
c^2\Big(c\Ib+\Vxx\Big)^{-1} \EE[\thetab_u\thetab^T _u] \Big(c\Ib+\Vxx\Big)^{-1}\nn\\ 
&+\Big(c\Ib+\Vxx\Big)^{-1}\sum_{e_i\in \HH}\phib^u_i\phib^{uT}_i
\frac{\EE(\epsilon^2(t_i))}{\sigma^4}\Big(c\Ib+\Vxx\Big)^{-1}\end{align*}
Note that, from the regularizer we have, $\thetab_u\sim\Ncal(0,\Ib/c)$. Furthermore $\EE(\epsilon^2(t_i))=\sigma^2$.
Using simple algebraic calculation,
we have the value for $\covm(\HH)$.

\subsection{Proof of Theorem~\ref{thm:subm}}\label{sec:proof2}
% \begin{proof}
\noindent (i) \textbf{Monotonicity of $f_X$ :}
To prove monotonicity, we need to show, $f_X(\HH\cup\{e_k\})-f_X(\HH)\geq 0$.
Recall that the set of users posting the messages is denoted by $\Vcal$. Assume the user $u$ has posted the event $e_k=\{u_k,m_k,t_k\}$, \ie\ $u_k=u$. 
We define an auxiliary function  of $p\in[0,1]$: 
%\vspace*{-0.2cm}
{\smaller\begin{align*}
g(p):=\sum_{u\in\Vcal}\tr\log\Big(c\Ib+\sigma^{-2}\sum_{e_i\in \Hcal_T}\phib^u _i\phib _i ^{uT}+p\sigma^{-2}\phib^u _k\phib _k ^{uT}\Big).
\vspace*{-0.2cm}
\end{align*}}
Monotonicity of $f_X$ in $\HH$ is equivalent to the condition: $g(1)\geq g(0)$ that can be shown by proving $\frac{d}{dp}g(p)\geq 0$.
For compactness, we define
$\Gb_u=\Big(c\Ib+\sigma^{-2}\sum_{e_i\in \Hcal_T}\phib^u _i\phib _i ^{uT}\Big)$.
Now we can show that: 
\begin{align}
\frac{d}{dp}g(p)
	  &=\sum_{u\in\Vcal}\tr\frac{d}{dp}\log(\Gb_u+p\phib^u _k\phib _k ^{uT})\nn\\
	       &=\sum_{u\in\Vcal}\tr\frac{d}{dp}\Big[\log(\Gb_u+p\phib^u _k\phib _k ^{uT})-\log\Gb_u\Big]\nn\\
	       &=\sum_{u\in\Vcal}\tr \big[(\Ib+p\phib^u _k\phib _k ^{uT} \Gb_u ^{-1})^{-1}\phib^u _k\phib _k ^{uT}\Gb_u ^{-1}\big]\nn\\
 	       &=\sum_{u\in\Vcal}\tr\phib _k ^{uT}\Gb_u ^{-1}(\Ib+p\phib^u _k\phib _k ^{uT} \Gb_u ^{-1})^{-1}\phib^u _k\nn\\
	       &
=\sum_{u\in\Vcal}\sigma^{-2}\phib _k ^{uT} (\Gb_u+ p \phib^u _k\phib _k ^{uT})^{-1}\phib _k ^{u}\geq 0	
\end{align}
Hence $f_X$ is monotone in $\HH$ $\forall X \in \{A,D,E,T\}$.

\noindent (ii) \textbf{Weak submodularity of $f_A$ and $f_E$ :} 
Since $f_A$ and $f_E$ are linear and monotone as proved above, their weak submodularity follows from the weak submodularity of the A and E optimality criteria which has been proved in \cite{bian2017guarantees,chamon2017approximate,hashemi2018randomized} for linear and monotone observation models.

\noindent (iii) \textbf{Submodularity of $f_D$ :} 
Since $f_D$ is linear and monotone as proved above, its submodularity follows from the submodularity of the D optimality criterion which has been proved in \cite{krause2008near,shamaiah2010greedy} for linear and monotone observation models.

\noindent (iv) \textbf{Modularity of $f_T$ :} 
Since $f_T$ is linear and monotone as proved above, its modularity follows from the modularity of the T optimality criterion which has been proved in \cite{krause2008near,summers2015submodularity} for linear and monotone observation models.

\section{Additional Results}
\label{sec:appen_expt}

\begin{table}[!h]
	\centering
%	\resizebox{0.7\textwidth}{!}{
		\begin{tabular}{|c|c|c|c|c|}
			\hline
			
			Datasets &  \ouragr & \ouraf & \ourdgr  & \ourdf  \\\hline
			&  \multicolumn{4}{c|}{ \textbf{Mean squared error:} $\EE(m-\hat{m})^2$ }\\\hline
			\barca  &  0.038 & 0.039 & 0.037 & 0.037 \\\hline
			\british  &  0.054 & 0.055 & 0.053 & 0.051 \\\hline
			\jaya  &  0.069 & 0.071 & 0.071 & 0.072 \\\hline
			\juv  &  0.066 & 0.067 & 0.056 & 0.057 \\\hline
			\twitter  &  0.035 & 0.036 & 0.039 & 0.039 \\\hline
			&  \multicolumn{4}{c|}{ \textbf{Failure Rate:} $\PP(\text{sign}(m)\neq \PP(\text{sign}(\hat{m}))$ }\\\hline
			\barca  &  0.120 & 0.124 & 0.113 & 0.116 \\\hline
			\british  &  0.169 & 0.176 & 0.176 & 0.182 \\\hline
			\jaya  &  0.207 & 0.204 & 0.210 & 0.213 \\\hline
			\juv  &  0.114 & 0.091 & 0.079 & 0.087 \\\hline
			\twitter  &  0.137 & 0.144 & 0.147 & 0.152 \\\hline			
	\end{tabular}

\caption{
%	Sentiment prediction performance for five real-world datasets for {\oura and \ourd} for a set of alternative algorithms for fixed 
	Comparative performance evaluation of \ouraf and \ourdf with \ouragr and \ourdgr for five real-world datasets for fixed $\gamma = 0.2$ and \T=4 hours. 
%	For \oura standard greedy is compared with random greedy. For \ourd standard greedy is compared with fast adaptive submodular maximization. 
	For each message $m$ in test set, we predict its sentiment value given the history up to \T=4 hours before the time of the message. 
	For the \T hours, we predict the opinion stream using a sampling algorithm. 
	Mean squared error and failure rate have been reported. 
%	We observe that the variants of \our generally perform better than the baselines on all datasets. Among the baselines, \ourR performs comparably with \our on some of the datasets presented here.
	We observe that, in general, greedy variants outperform adaptive variants by small margin. 
}
\label{tab:forecast_additional}
%\vspace*{-0.4cm}
\end{table}

\noindent 
For the experimental results, we apply standard greedy for maximizing $f_A$ and $f_D$ (see Section~\ref{sec:expt}). As we have mentioned earlier, there exist fast adaptive techniques as alternative to greedy for both maximization of $f_D$ as well as $f_A$. In this section, we repeat our primary experiments with fast adaptive techniques for \oura and \ourd and compare them with their greedy alternatives. 
For maximizing $f_D$, we use \textsc{Fast-Full}, a fast adaptive algorithm proposed in \cite{breuer2020fast}. We refer to this method as \ourdf. 
For maximizing $f_A$, we adopt DASH, a fast adaptive technique proposed by \cite{qian2019fast}, which we refer to as \ouraf . From now onwards, we refer to greedy techniques for $f_A$ and $f_D$ as \ouragr\ and \ourdgr\ respectively. 
We present their detailed comparative evaluation in our primary experiments on real datasets below.

\noindent 
\xhdr{Comparative analysis}\label{sec:perf_eval_additional} 
Here we compare the predictive performance of \ouraf and \ourdf with \ouragr and \ourdgr. 
We fix the fraction of endogenous chosen to $0.8$ and prediction time to 4 hours. We follow the same setup as in the corresponding experiment in Section~\ref{sec:expt}. 
Table~\ref{tab:forecast_additional} summarizes the results. 
In aggregate, we observe the trade-off that fast adaptive techniques either perform comparably with the greedy methods or standard greedy algorithms outperform adaptive methods by a small margin at the cost of speeding up the demarcation process.

\begin{figure}[h!]
	
	\centering
	\subfloat{ 	\includegraphics[scale=0.50]{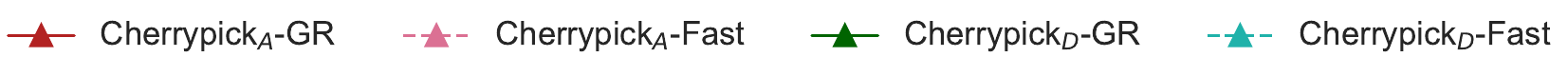}}
	\vspace{-3mm}
	
	\subfloat[\barca]{ 	\includegraphics[width=0.30\textwidth]{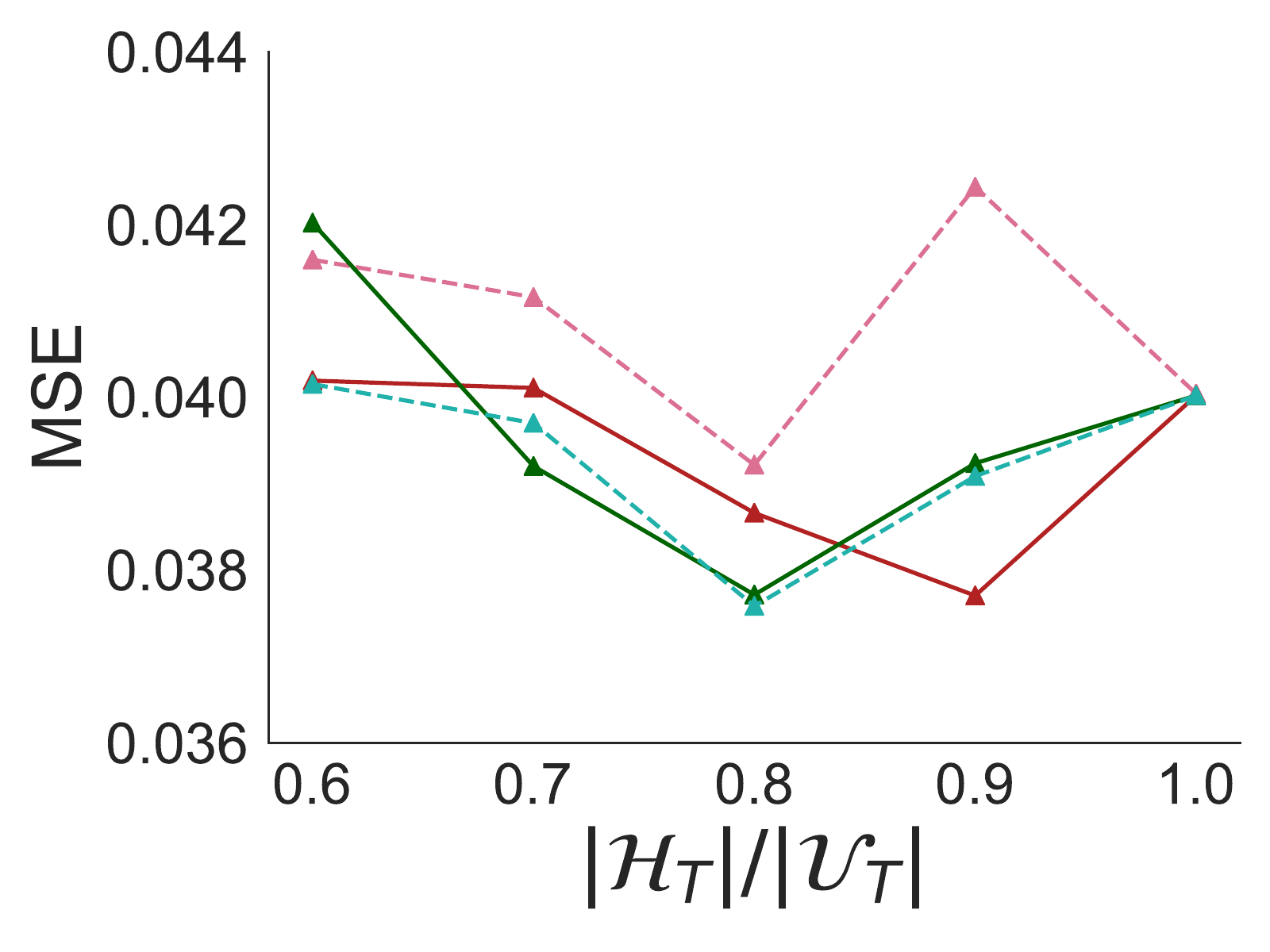}}\hspace*{0.2cm}
	%\subfloat[\british]{ 	\includegraphics[width=0.30\textwidth]{FIG_new/exp2_british_election_comparative_MSE.pdf}}\hspace*{0.2cm}
	%\subfloat{ 	\includegraphics[width=0.30\textwidth]{FIG_new/exp2_GTwitter_comparative_MSE.pdf}}
	\subfloat[\jaya]{ 	\includegraphics[width=0.30\textwidth]{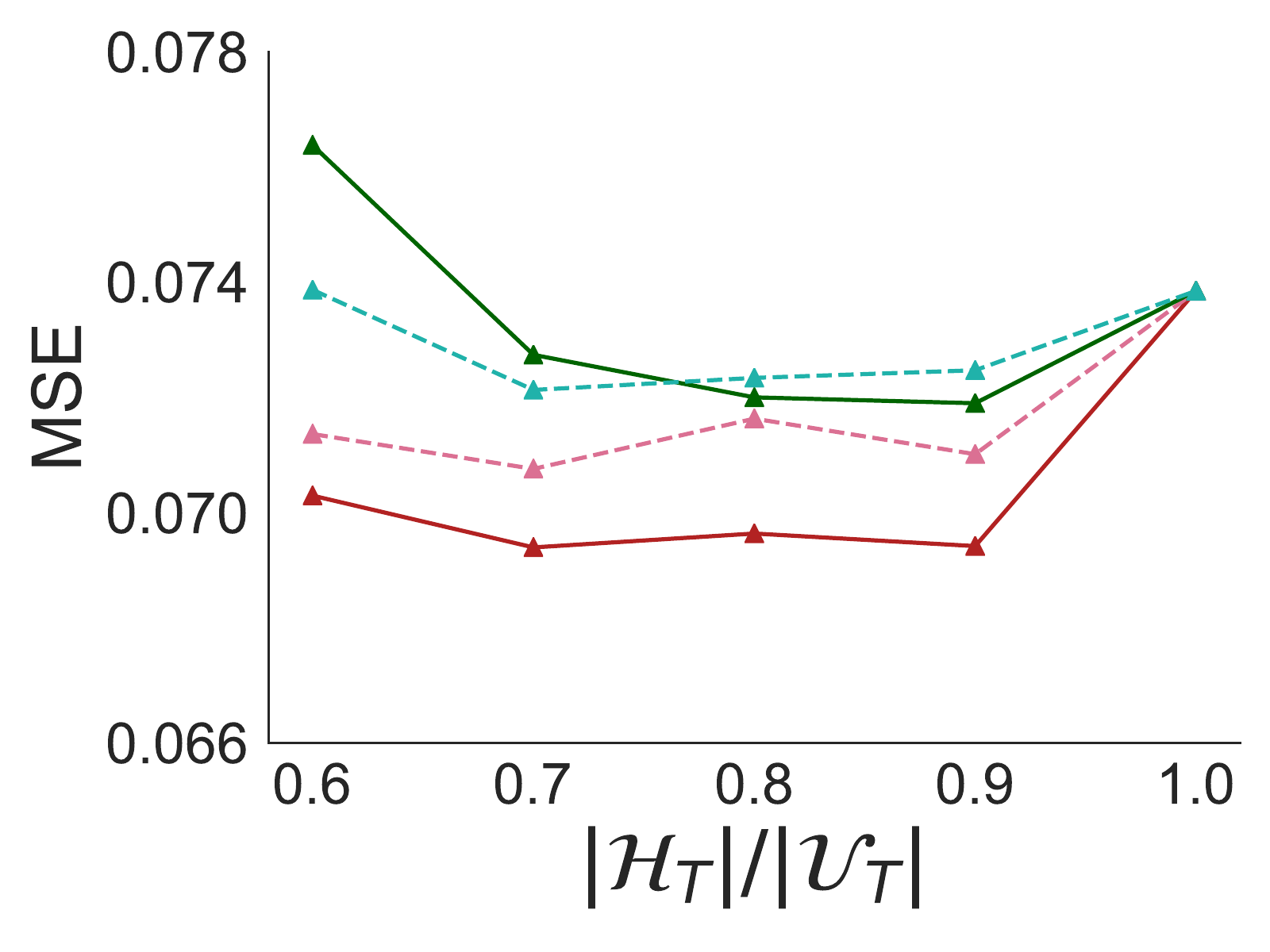}}\hspace*{0.2cm}
	\subfloat[\twitter]{ 	\includegraphics[width=0.30\textwidth]{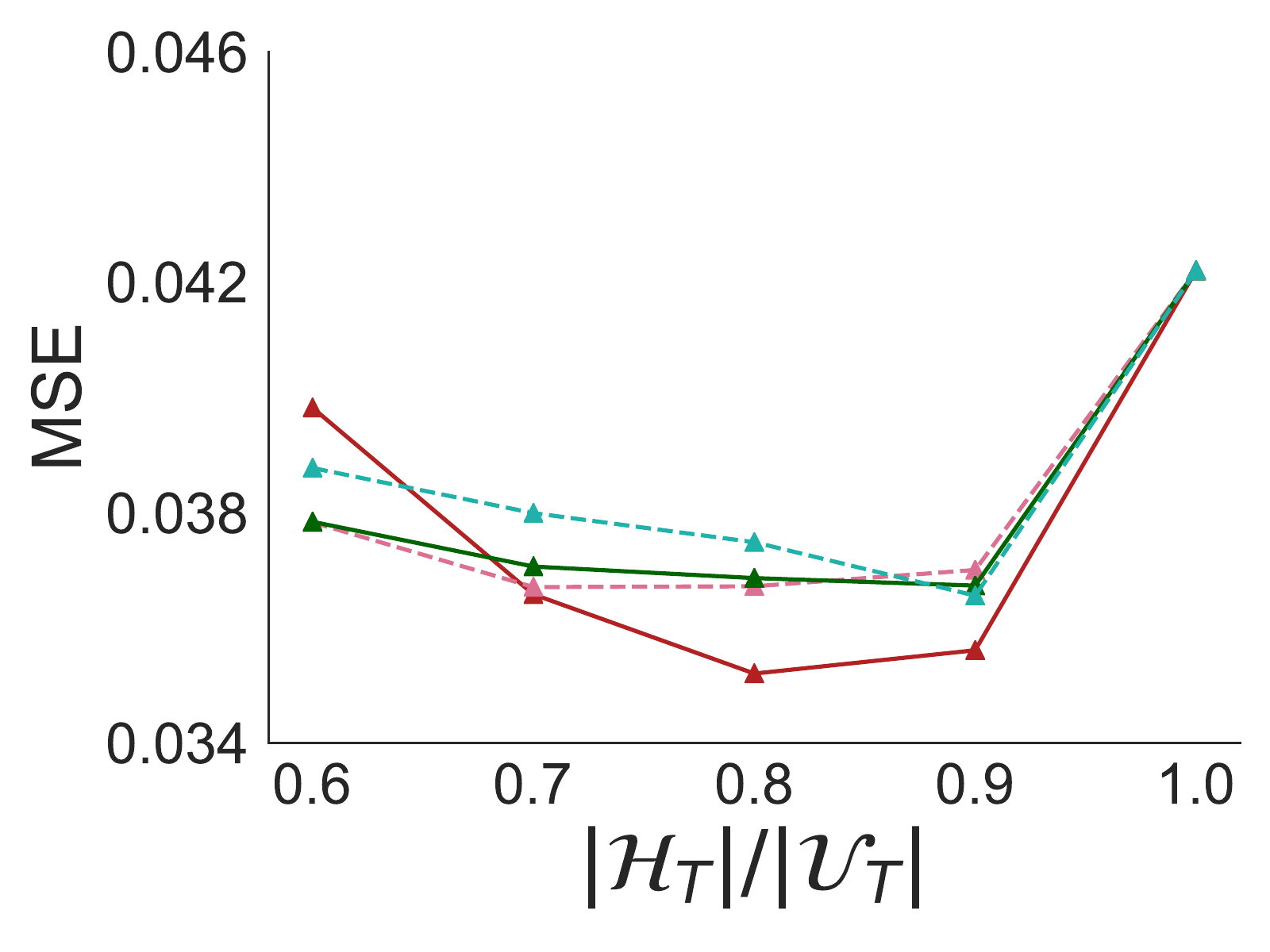}}\hspace*{0.2cm}

	%\subfloat[Juv]{ 	\includegraphics[width=0.30\textwidth]{FIG_new/exp2_JuvTwitter_MSE.pdf}}
	
	\vspace{-3mm}
	
	\subfloat[\barca]{ 	\includegraphics[width=0.30\textwidth]{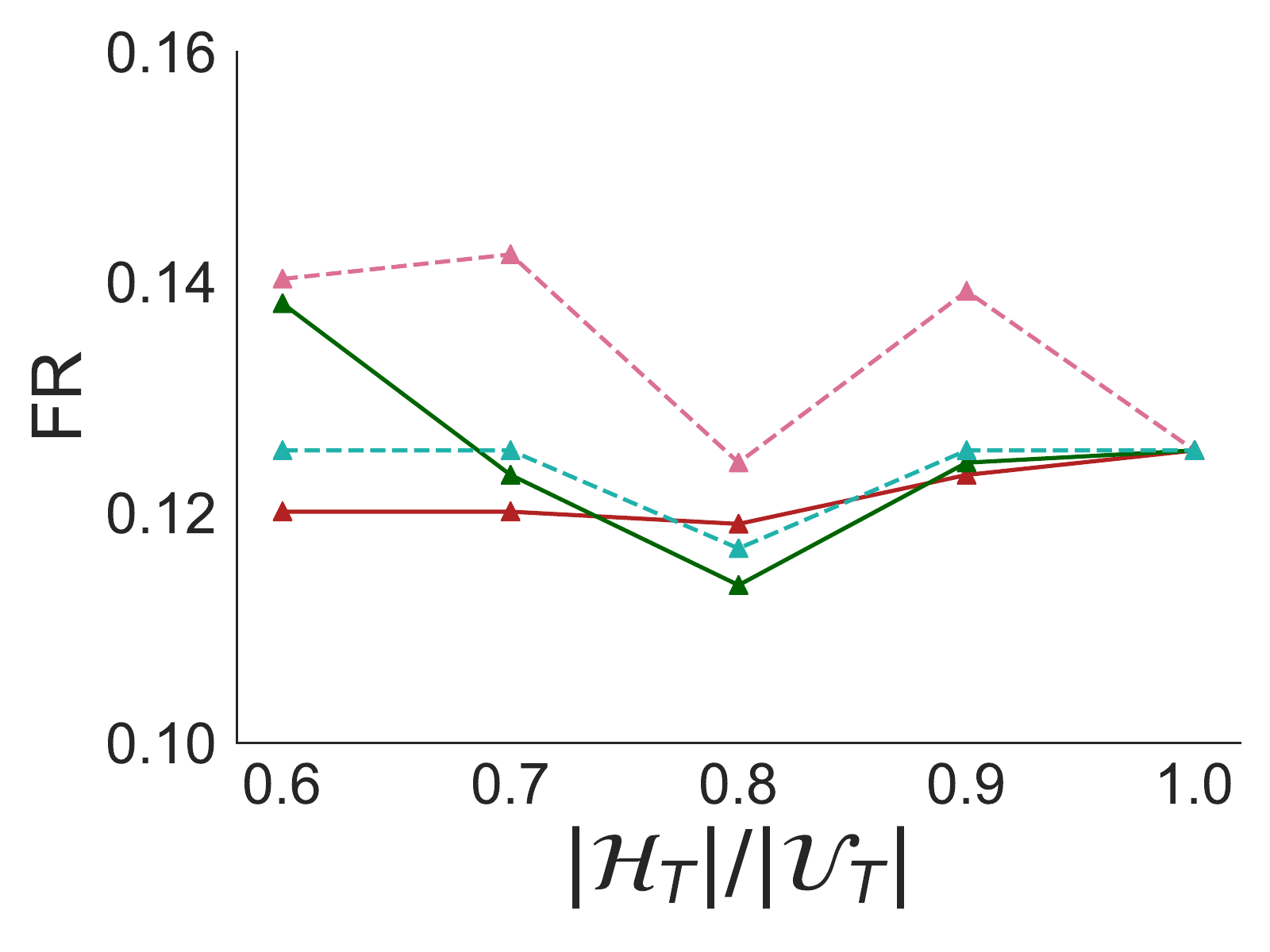}}\hspace*{0.2cm}
	%\subfloat[\british]{ 	\includegraphics[width=0.30\textwidth]{FIG_new/exp2_british_election_FR.pdf}}\hspace*{0.2cm}
	%\subfloat{ 	\includegraphics[width=0.30\textwidth]{FIG_new/exp2_GTwitter_FR.pdf}}
	\subfloat[\jaya]{ 	\includegraphics[width=0.30\textwidth]{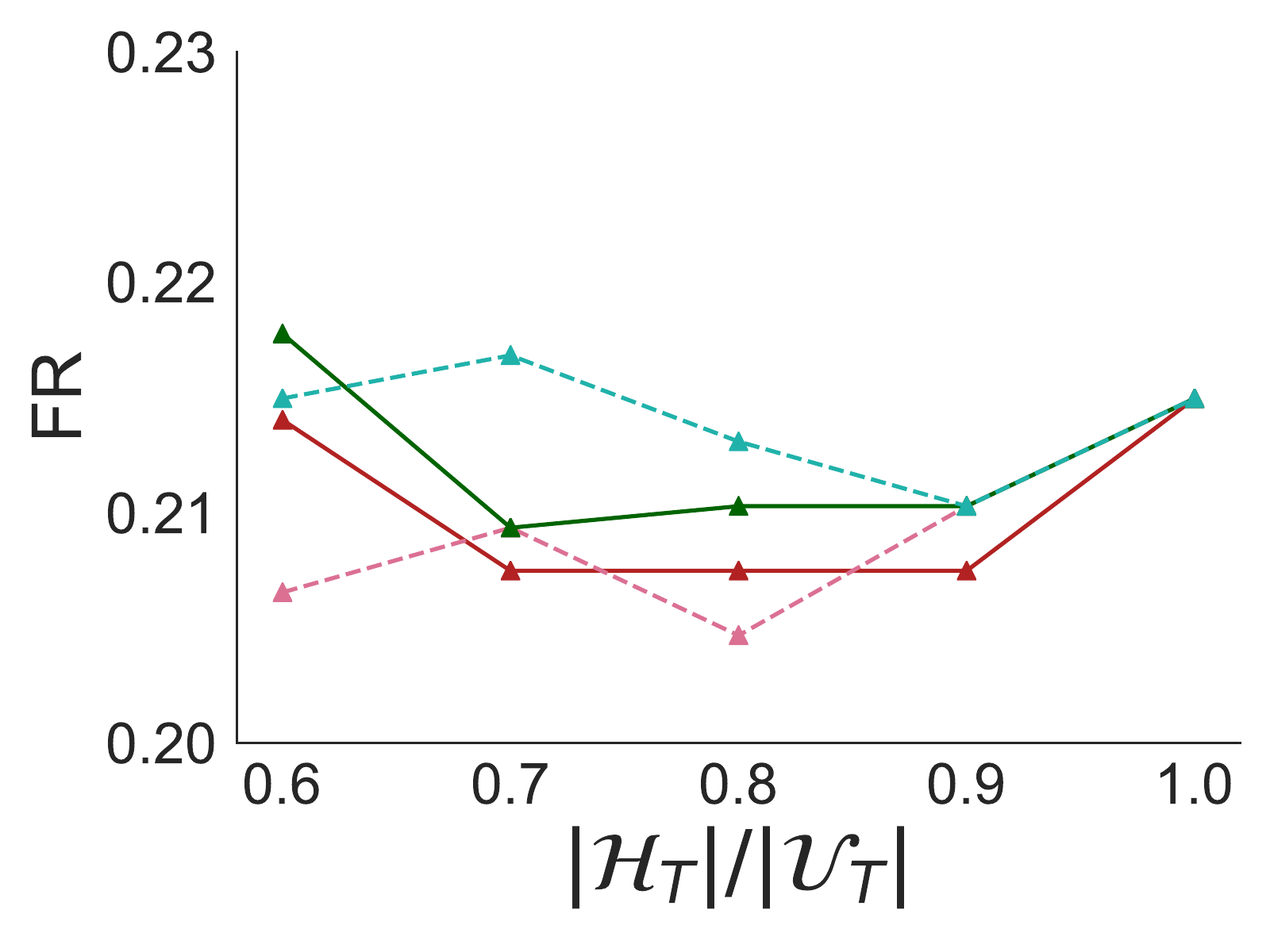}}\hspace*{0.2cm}
	%\vspace{-3mm}
	%\subfloat[Juv]{ 	\includegraphics[width=0.30\textwidth]{FIG_new/exp2_JuvTwitter_MSE.pdf}}
	%\subfloat{ 	\includegraphics[width=0.30\textwidth]{FIG_new/exp2_MsmallTwitter_MSE.pdf}}
	% \subfloat{ 	\includegraphics[width=0.30\textwidth]{FIG_new/real_vs_ju_703_MSE.pdf}}
	%\subfloat{ 	\includegraphics[width=0.30\textwidth]{FIG_new/exp2_VTwitter_MSE.pdf}}
	%\vspace{-3mm}
	%\subfloat[Juv]{ 	\includegraphics[width=0.30\textwidth]{FIG_new/exp2_JuvTwitter_FR.pdf}}
	%\subfloat{ 	\includegraphics[width=0.30\textwidth]{FIG_new/exp2_MsmallTwitter_FR.pdf}}
	% \subfloat{ 	\includegraphics[width=0.30\textwidth]{FIG_new/real_vs_ju_703_FR.pdf}}
	\subfloat[\twitter]{ 	\includegraphics[width=0.30\textwidth]{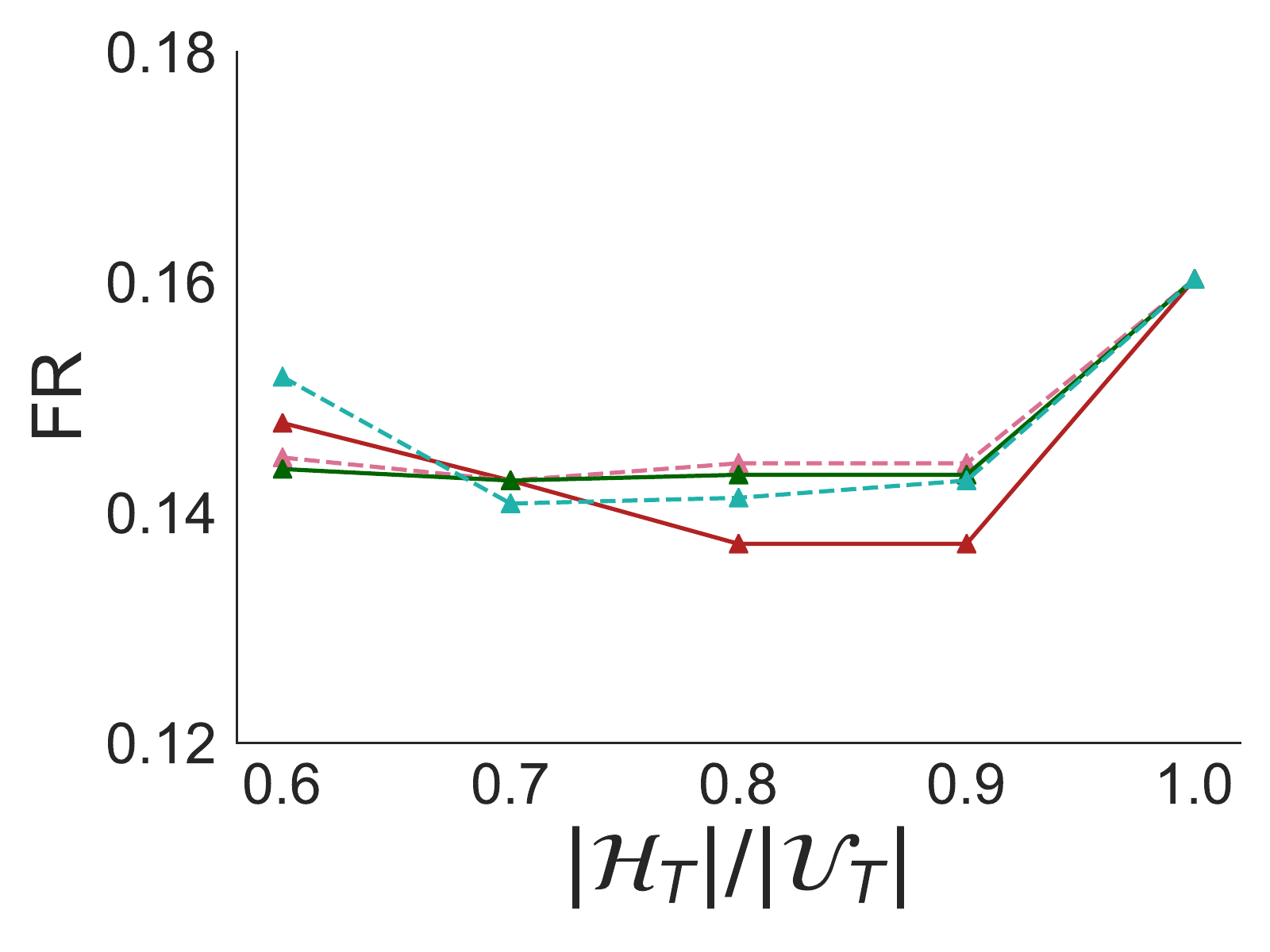}}\hspace*{0.2cm}
	%\subfloat{ 	\includegraphics[width=0.30\textwidth]{FIG_new/exp2_VTwitter_FR.pdf}}
	\vspace{-3mm}   
	\caption{Performance variation with size of endogenous subset for \barca, \jaya and \twitter datasets where the size of endogenous subset is varied from 60\% to 100\%. 
		Timespan \T has been set to 4 hours. 
		Mean squared error and failure rate have been reported. 
		Adaptive variants display almost identical behavior with greedy variants. 
		}
	\label{fig:varWithgammaAdditional}
\end{figure}

\noindent 
\xhdr{Variation of performance with the fraction of outliers ($\gamma$)}  
Figure \ref{fig:varWithgammaAdditional} compares greedy and adaptive variants of \oura and \ourd by observing the variation of forecasting performance for different values of $\gamma$ \ie\ the pre-specified fraction of outliers, for \barca, \jaya, and \twitter datasets. In this experiment, $\gamma$ is varied from $0.4$ to $0$, and the timespan has been fixed to  $4$ hours. Here we observe both \ouraf and \ourdf perform very similar to \ouragr and \ourdgr respectively. Moreover,  \ouraf and \ourdf preserve the same behavioral pattern as their greedy counterpart \ie\ both performing best for a fixed value of $\gamma$, with performance drop at both ends.

\noindent 
\xhdr{Forecasting performance}
Finally, we  compare the forecasting performance of \ouraf and \ourdf with their greedy counterparts  with respect to variation of \T, fixing $\gamma=0.2$. Figure \ref{fig:forecast_additional} summarizes the results, where we make following observations. (I) Generally, both \ouraf and \ourdf are outperformed by their greedy counterparts with very low margin as \T increases. (II) \ouraf and \ourdf display a similar pattern of performance drop with higher \T\ with small deviations from the greedy counterparts.

\begin{figure}[h!]
	
	\centering
	\subfloat{ 	\includegraphics[scale=0.5]{FIG_new/exp1_legend_short_comparative.pdf}}
	\vspace{-3mm}
	
	\hspace{-8mm}\subfloat{ 	\includegraphics[width=0.20\textwidth]{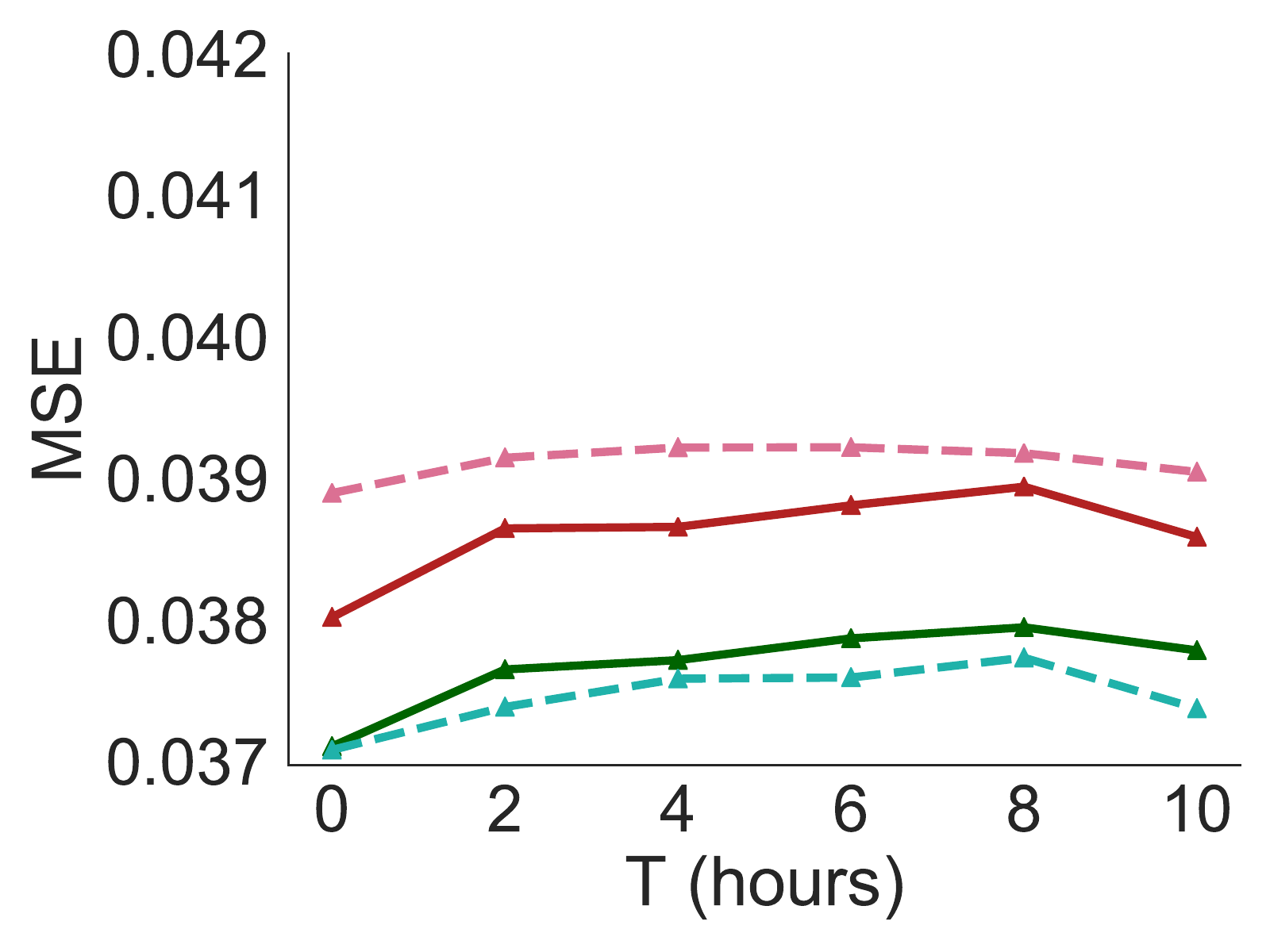}}
%	\subfloat{ 	\includegraphics[width=0.20\textwidth]{FIG_new/exp1_british_election_comparative_MSE.pdf}}
	%\subfloat{ 	\includegraphics[scale=0.20]{FIG_new/exp1_GTwitter_MSE.pdf}}
	\subfloat{ 	\includegraphics[width=0.20\textwidth]{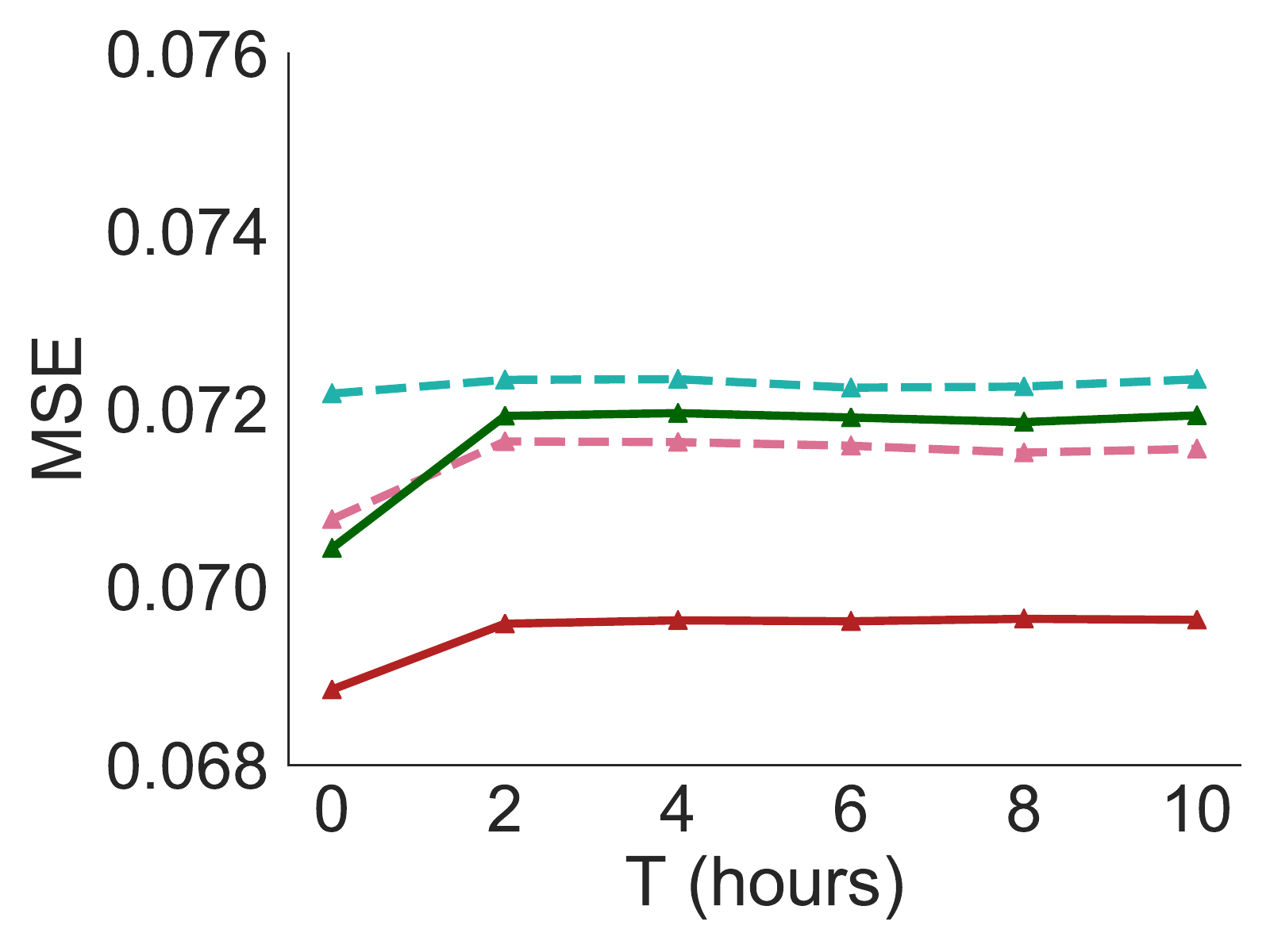}}
	\subfloat{ 	\includegraphics[width=0.20\textwidth]{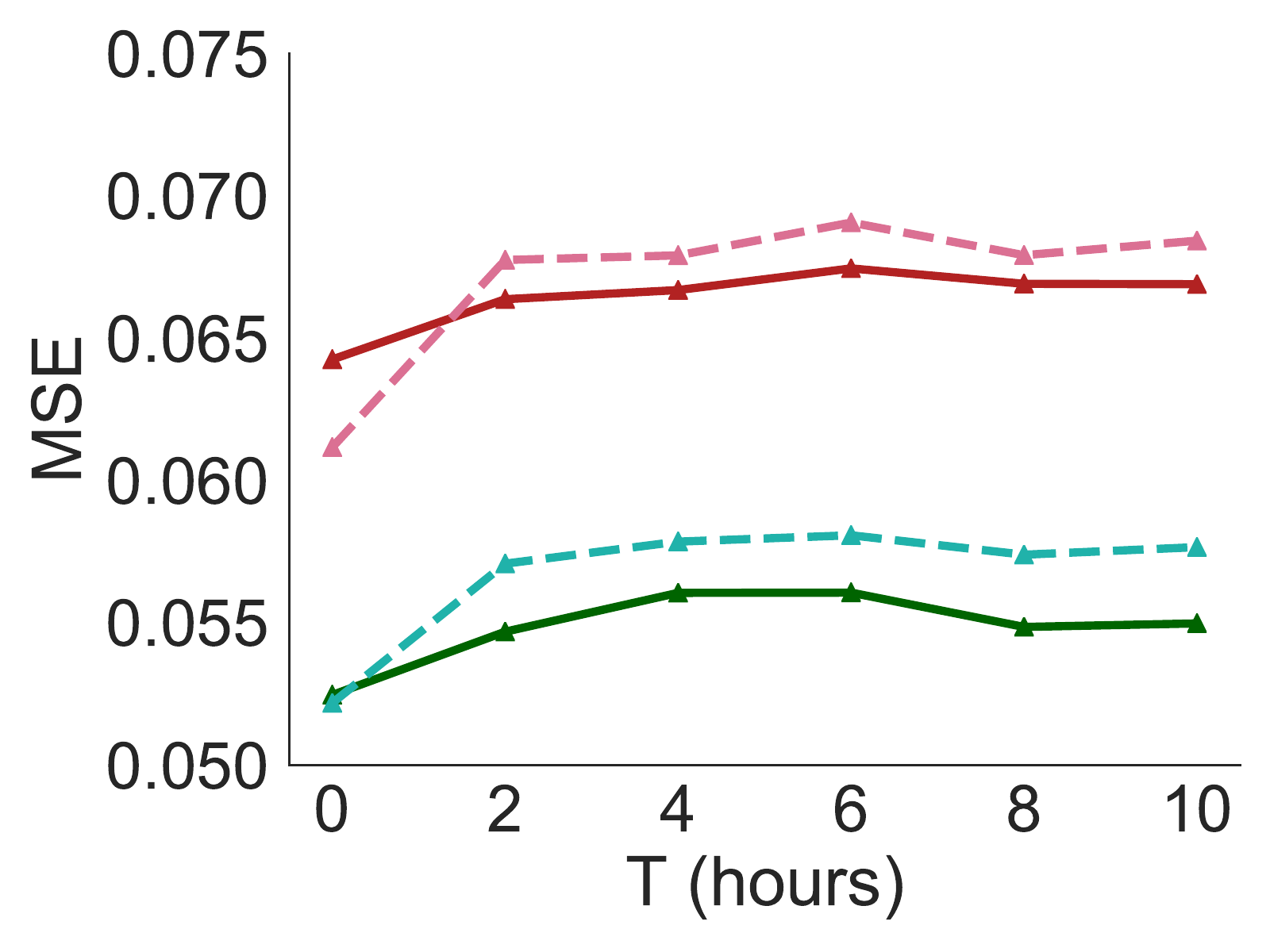}}
	%\subfloat{ 	\includegraphics[scale=0.20]{FIG_new/exp1_MsmallTwitter_MSE.pdf}}
	\subfloat{ 	\includegraphics[width=0.20\textwidth]{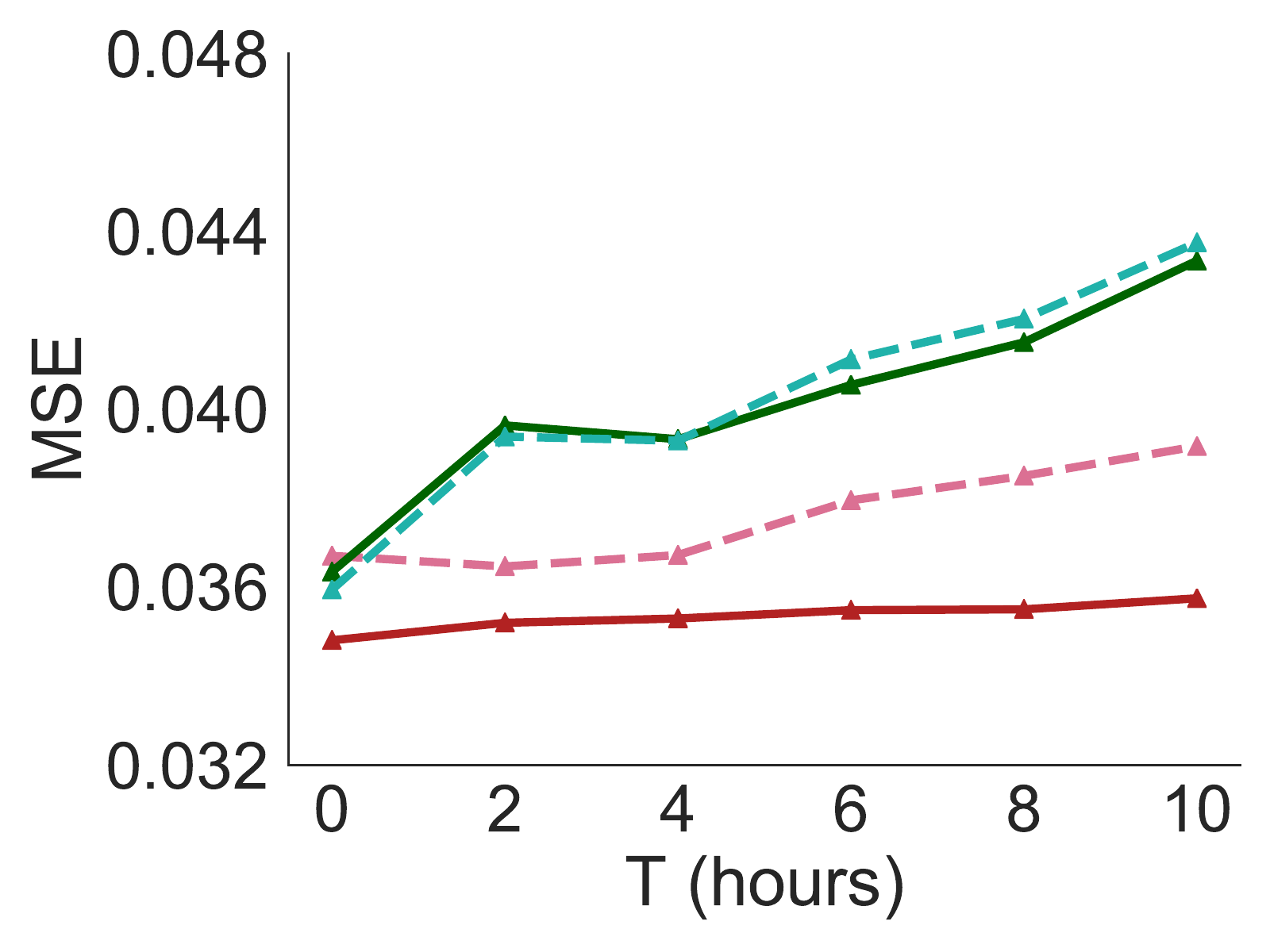}}
	%\subfloat{ 	\includegraphics[scale=0.20]{FIG_new/exp1_VTwitter_MSE.pdf}}
	\\	
	\hspace{-8mm}	\subfloat[\barca]{ 	\includegraphics[width=0.20\textwidth]{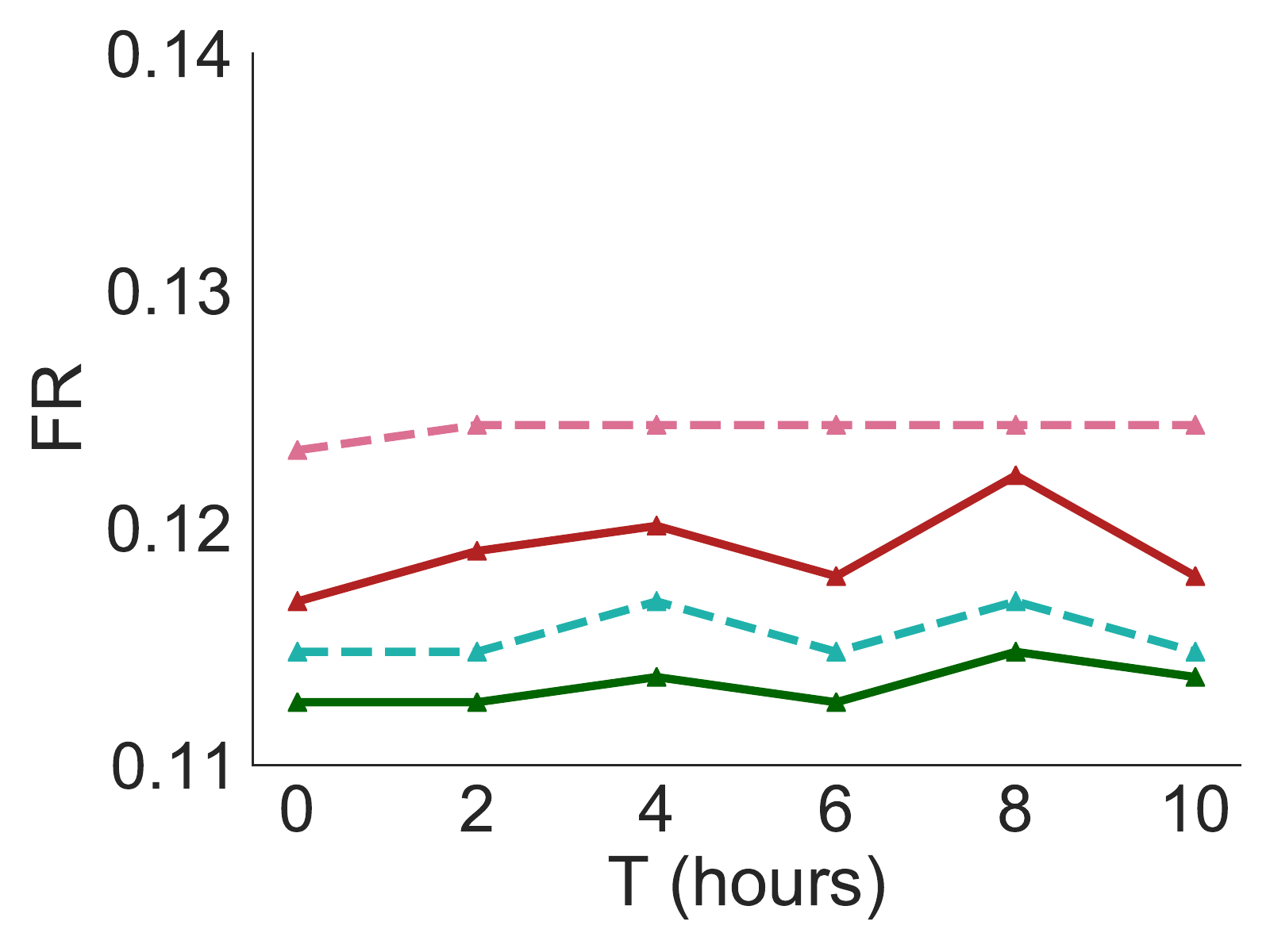}}
%	\subfloat[\british]{ 	\includegraphics[width=0.20\textwidth]{FIG_new/exp1_british_election_comparative_FR.pdf}}
	%\subfloat{ 	\includegraphics[scale=0.20]{FIG_new/exp1_GTwitter_FR.pdf}}
	\subfloat[\jaya]{ 	\includegraphics[width=0.20\textwidth]{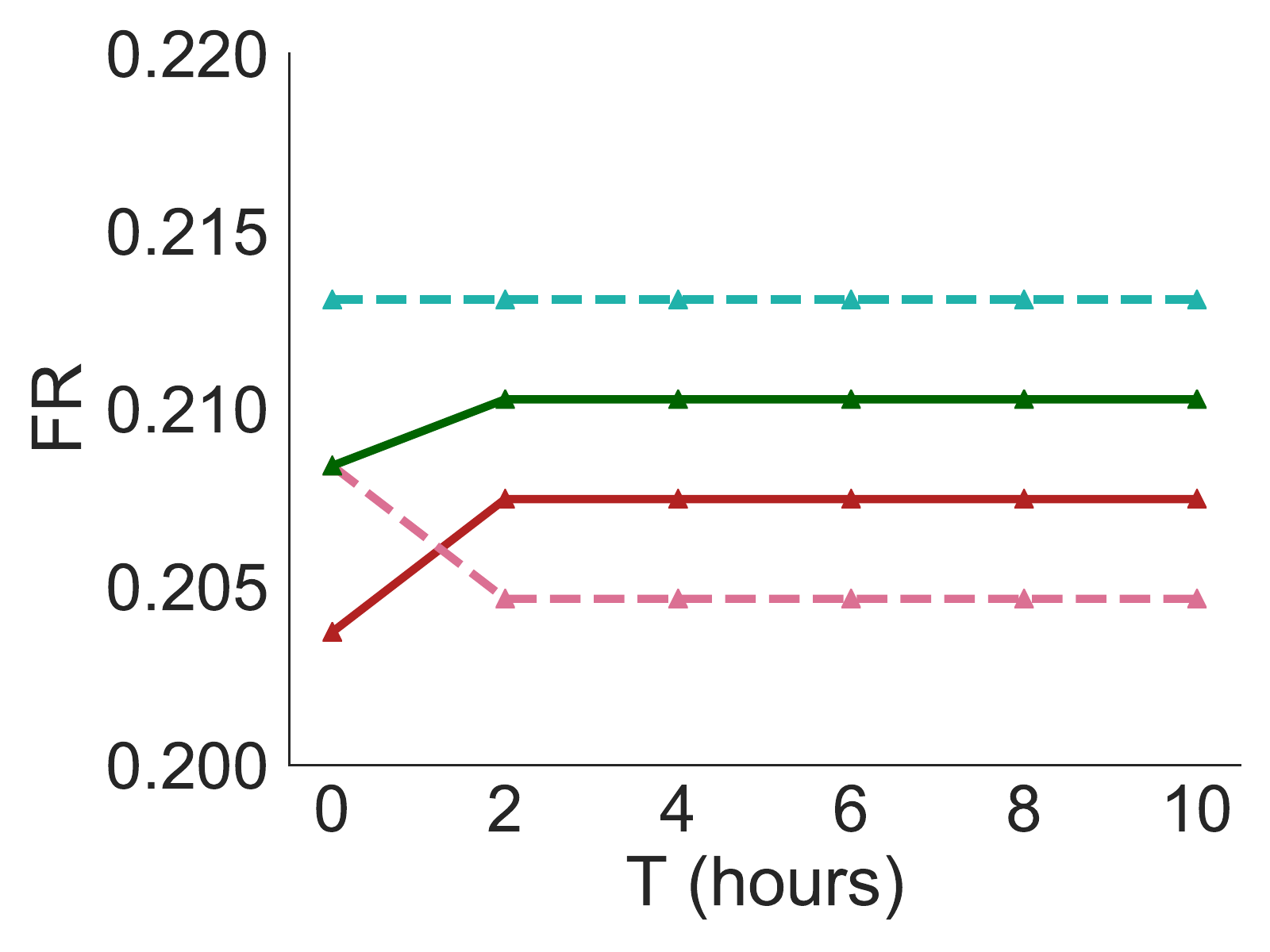}}
	%\vspace{-3mm}
	%\vspace{-3mm}
	\subfloat[\juv]{ 	\includegraphics[width=0.20\textwidth]{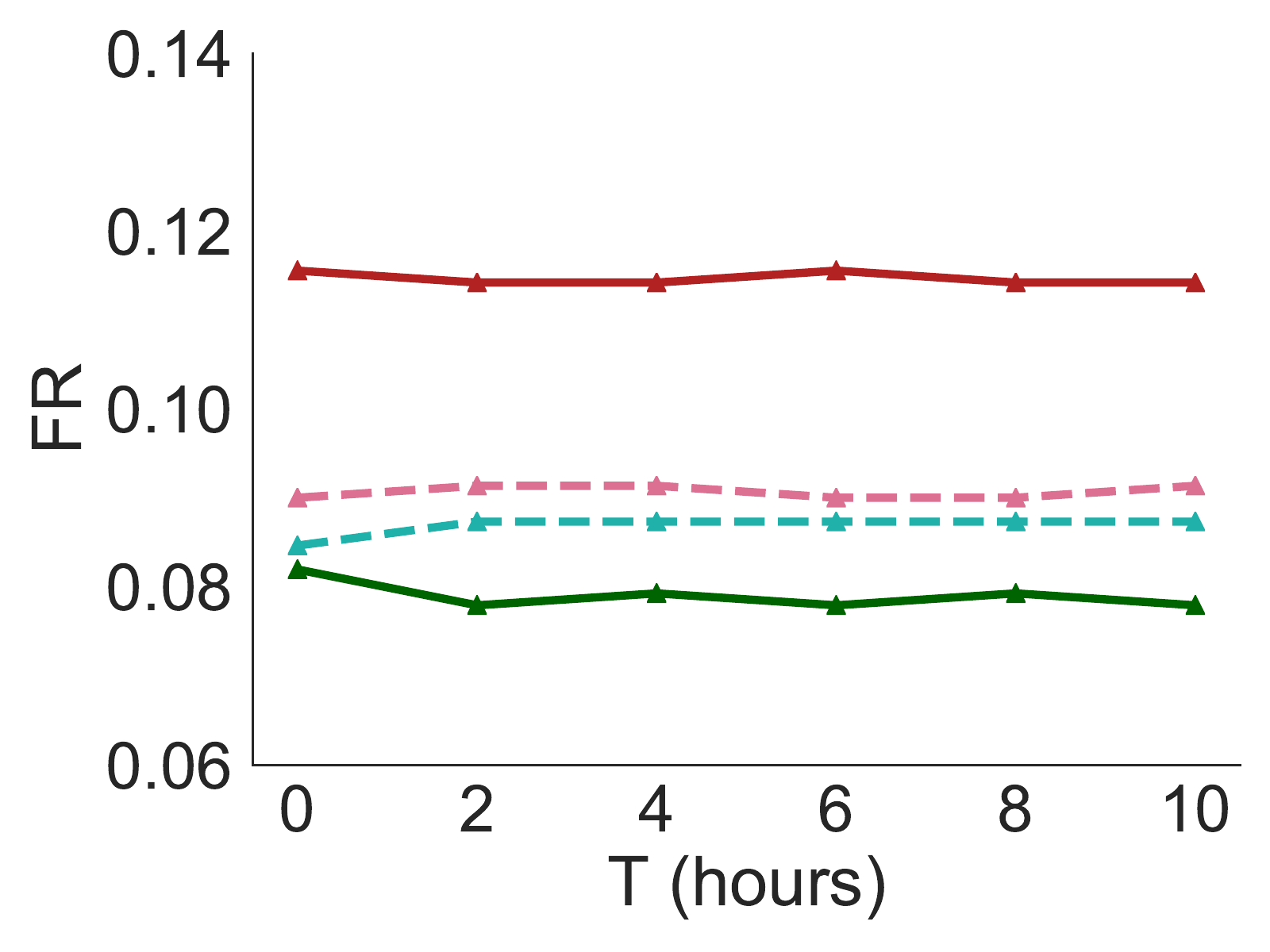}}
	%\subfloat{ 	\includegraphics[scale=0.20]{FIG_new/exp1_MsmallTwitter_FR.pdf}}
	\subfloat[\twitter]{ 	\includegraphics[width=0.20\textwidth]{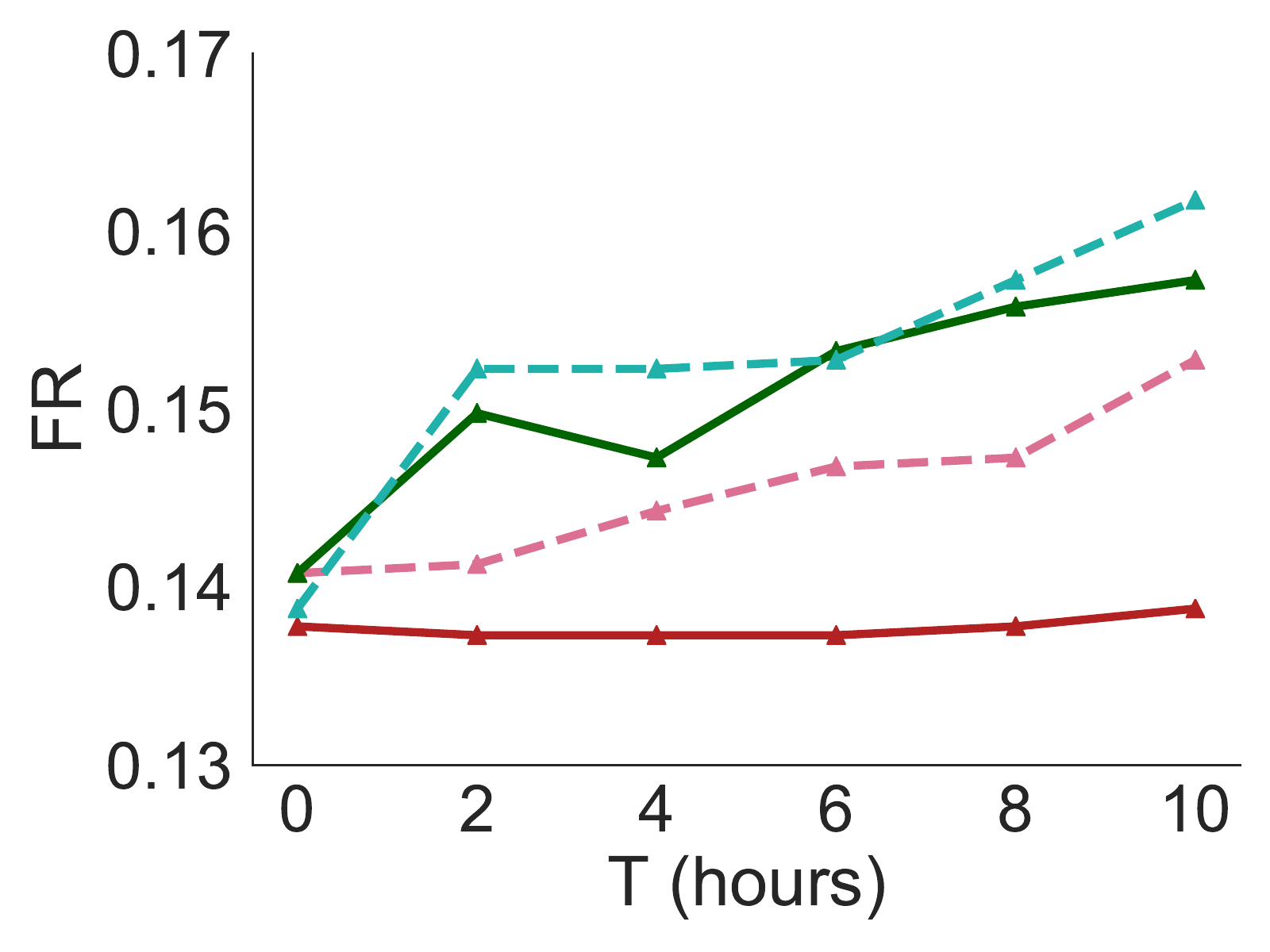}}
	%\subfloat{ 	\includegraphics[scale=0.20]{FIG_new/exp1_VTwitter_FR.pdf}}
	\vspace{3mm}
	
	% \subfloat{ 	\includegraphics[scale=0.30]{FIG_new/legend.png}}
	\caption{Forecasting performance for greedy and adaptive variants of \oura and \ourd for a fixed $\gamma = 0.2$ for four major real datasets. 
%		For \oura standard greedy is compared with random greedy. For \ourd standard greedy is compared with fast adaptive submodular maximization. 
		For each message $m$ in the test set, we predict its sentiment value given the history up to \T hours before the time of the message. 
		For the \T hours, we predict the opinion stream using a sampling algorithm. Mean squared error and failure rate have been reported.
		We observe almost identical behavior of greedy and adaptive variants, with adaptive methods outperformed by greedy methods with negligible margin. 
	}
	\label{fig:forecast_additional}
\end{figure}

\end{document}